%% file: main.tex
\documentclass[aps,prr,reprint,twocolumn,superscriptaddress,floatfix,amsmath, amssymb, longbibliography]{revtex4-2}
\usepackage[english]{babel}
\usepackage[T1]{fontenc}
\usepackage[utf8]{inputenc}

\usepackage{graphicx}% Include Fig. files
\usepackage{dcolumn}% Align table columns on decimal point
\usepackage{bm}% bold math
\usepackage{amssymb}
\usepackage{amsmath}
\usepackage{multirow}
\usepackage{color}
\usepackage{hyperref}
\usepackage{float}
\usepackage{comment}

\usepackage{xr}
\begin{document}

\input{definitions}

\input{title}
\title{\manuscripttitle}
\input{authors}

\date{\today}

\begin{abstract}
Harnessing the topology of ring polymers as a design motif in functional nanomaterials is becoming a promising direction in the field of soft matter. For example, the ring topology of DNA plasmids prevents the relaxation of excess twist introduced to the polymer, instead resulting in helical supercoiled structures. In equilibrium semi-dilute solutions, tightly supercoiled rings relax faster than their torsionally relaxed counterparts, since the looser conformations of the latter allow for rings to thread through each other and entrain through entanglements. Here we use molecular simulations to explore a non-equilibrium scenario, in which a supercoiling agent, akin to gyrase enzymes, rapidly induces supercoiling in the suspensions of relaxed plasmids. The activity of the agent not only alters the conformational topology from open to branched, but also locks-in threaded rings into supramolecular clusters, which relax very slowly. Ultimately, our work shows how the polymer topology under non-equilibrium conditions can be leveraged to tune dynamic behavior of macromolecular systems, suggesting a method to create a class of driven materials glassified by activity.

\vspace{0.5cm} % Adjust the space as needed
\end{abstract}

\maketitle

\makeatletter
\def\@startsection#1#2#3#4#5#6{%
  \if@noskipsec \leavevmode \fi
  \par
  \@tempskipa #4\relax
  \@afterindenttrue
  \ifdim \@tempskipa <\z@
    \@tempskipa -\@tempskipa \@afterindentfalse
  \fi
  \if@nobreak
    \everypar{}%
  \else
    \addpenalty\@secpenalty\addvspace\@tempskipa
  \fi
  \@ifstar
    {\@ssect{#3}{#4}{#5}{#6}}%
    {\@dblarg{\@sect{#1}{#2}{#3}{#4}{#5}{#6}}}}
\makeatother

% introduction
\section{Introduction}
\label{sec:intro}
Understanding of the topological constraints and entanglements in polymer solutions and melts under diverse conditions is among the biggest open problems in the field of soft matter \cite{Tubiana2024a, Qu2021a, Hart2021a, Shankar2022a}.
Different forms of entanglement and entrainment can enhance or diminish certain relaxation modes, in turn affecting the viscoelastic properties of polymer materials \cite{Michieletto2022a, Kong2022a, Peddireddy2020b, Rosa2020a, Kapnistos2008a, Neill2024a, Liu2022a, Ito2011a, Li2024a}, but also the dynamics of other classes of fibrous matter such as weaved molecular sheets \cite{Zhang2022a}, macroscopic worms \cite{Deblais2020a, Patil2023a, OzkanAydin2021a}, ropes or soft robots \cite{Patil2020a, Becker2022a}. 
The ability to select the relaxation pathways would allow to tune the properties of the solution without the need to change the polymer chemistry. 
The entanglements and the ensuing dynamics are many-body in nature and crucially depend on the chain topology \cite{Huang2019a} and external \cite{Parisi2021b} or internal driving \cite{Breoni2025a}. 
These reasons make a theoretical prediction of the properties challenging, but concomitantly present vast possibilities for diverse modulation of the behavior.

%rings+noneq conditions 
In particular, polymers of \emph{ring} topology that are unknotted and mutually nonconcatenated have been studied extensively as a prototypical example \cite{Kruteva2023a, Tubiana2024a}. The well-studied equilibrium relaxation of linear polymers (reptation) proceeds by one-dimensional diffusion along the chains contour, while the transverse motions are restricted by the entanglements with the other chains. 
While rings can also exhibit linear-like entanglements \cite{Wang2021a}, the rings have no ends, hence the reptation mechanism is altered \cite{Smrek2015a,Rubinstein_FLG_MAMOL16,Ghobadpour2021a,Ghobadpour2025a}.
Additionally, rings allow for \emph{threading} entanglements when one chain enters the opening of another one, but the exact contribution of threadings to the viscoelasticity is not fully understood.
To satisfy the constraints of mutual ring nonconcatenation, at semidilute concentrations the rings adopt branched configurations \cite{Rosa2014a,Smrek2016a} that are however not tightly double-folded and include mutual threadings \cite{Ubertini2022a}. 
On the one hand, treating rings in melts as branched annealed trees without threading \cite{Grosberg2014a, Kapnistos2008a}, or other models that neglect threadings \cite{Rubinstein_FLG_MAMOL16}, reproduce the experimental moduli almost quantitatively, suggesting that threading relaxation does not contribute significantly to viscoelastic properties. %Similarly, the Fractal Loopy Globule model that does not consider threadings explicitly reproduces power-law stress relaxation modulus \cite{Rubinstein_FLG_MAMOL16}. 
On the other hand, the theories without threadings predict faster ring diffusion \cite{Smrek2015a,Rubinstein_FLG_MAMOL16,Grosberg2014a,Ghobadpour2021a} than observed \cite{Halverson2011b} and molecular simulations \cite{Jung_Macromol15,Tsalikis2016a,Smrek2016a,Jung_Lee_MC_Polymers19,Michieletto2017a} show that threading is present in polymer melts, and the threaded rings do diffuse slower than their unthreaded counterparts. 
Recent studies show that stiffer rings can exhibit prolonged subdiffusive regime and glassy dynamical features \cite{Goto2023a,Tu2023a,Roy2024a,Stano2023c,Michieletto2017a}. The latter is strongly enhanced also in certain ring systems out of equilibrium, either driven externally by extensional flows \cite{OConnor2020a,Huang2019a} or internally, for example in active topological glass of block co-polymers with hot and cold blocks \cite{Smrek2020a, Micheletti2024a}. 
In both of the above non-equilibrium systems, the dynamics effects have been directly linked to threadings, as the flow field or activity can turn threadings into deadlocks \cite{Micheletti2024a} -- geometrically entangled constructs with very long relaxation times, causing either large increase in viscosity or even vitrification of the whole system. 
Hence active induction of threadings can slow down the system and impact the viscoelasticity, but is difficult to reverse or regulate to achieve a control.

A promising candidate for such a regulation is the supercoiling \cite{Fathizadeh2015a, Krajina2016a, Segers2025a} that can be present in ring polymers with both, bending and torsional elasticity, readily realized with circular DNA plasmids or other ribbon-like polymers. 
In such \emph{equilibrium} solution, the abundance of linear-like and threading entanglements can be controlled by the degree of supercoiling. 
The enhanced supercoiling decreases the threadable rings area and stiffens their effective backbone, leading to a decrease of both, the threadings and the entanglements, and thereby speeds up the system \cite{Smrek2021a}. 
The supercoiling can be also induced after the polymer synthesis providing the sought-for control over the properties.
In the case of DNA, this can be achieved by molecular motors such as gyrase \cite{Gellert1976a, Menzel1994a} or RNA polymerase \cite{Fosado2021a} or intercalators \cite{Wu1989a, Wu1988a, Kolbeck2023a, Ganji2016a}. 
The detailed action of these supercoiling agents can be different. 
For example, while the intercalator is inserted in the base pairs and thereby changing the bending-torsion balance, gyrase temporarily cuts DNA strand, reposition the chain under- or over-crossings and then binds the cut parts together, thereby inducing excess torsion \cite{Bates2005a}. For other ribbon-like polymers the supercoiling can be induced by the incorporation of artificial molecular motors in the polymers backbone \cite{Kassem2017a,Yao2025a}.
While the equilibrium consequences of the supercoiling are known, the impact of the active induction of supercoiling on the threadings and dynamic properties of such ribbon-like ring solutions remains unexplored. 
We do not aim at modeling a specific supercoiling agent, but rather focus on the universal properties that arise from the coarse-grained perspective of the supercoiling induction.

The crucial parameter in considering the role of the active process is the time to reach the supercoiling steady state, when the active forces are balanced by the interactions, in comparison to the polymer relaxation time. Taking gyrase as an example of the supercoiling agent, it operates on the scale of seconds \cite{Gore2006a,Basu2018a,Galvin2023a} providing enough time for short plasmids ($<100$kbp) to equilibrate on the scale of the diffusion time in aqueous solution as measured from diffusion experiments \cite{Robertson2006a,Hammermann1997a,Liu1993a}. 
Yet, the gyrase operation rate is energy- and tension-dependent, and for longer plasmids or solvents of higher viscosity the relaxation times can be orders of magnitude longer. Under these conditions, the supercoiling process operates far from equilibrium, potentially leading to diverse dynamical properties.
To understand the impact of active induction of supercoiling and to explore its potential use for the control of polymer solution dynamics, we focus on this regime where the supercoiling is induced faster than the diffusional relaxation. 
We investigate a solution of ring polymers actively driven by external torques applying excess twist on the contour and we assess its impact on the conformations, dynamics and the entanglement properties. 
We show that depending on the active torque, the generated conformational changes, from open ring through branched to linear-like supercoils, can exhibit metastable global states affecting the dynamics akin topological glasses \cite{Michieletto2017a,Smrek2020a}. 
The plasmid system we focus on is not only biologically interesting but also falls within a broader field of prospective DNA nanotechnology augmented by enzyme activity or external fields \cite{Stano2025a,Neill2024a,Conforto2024a,Michieletto2022a,Bonato2022a}. 

% model and single ring results
\section{Results and Discussion}
\label{sec:results}
\begin{figure*}[!ht]
	\centering
	\includegraphics[scale=0.95]{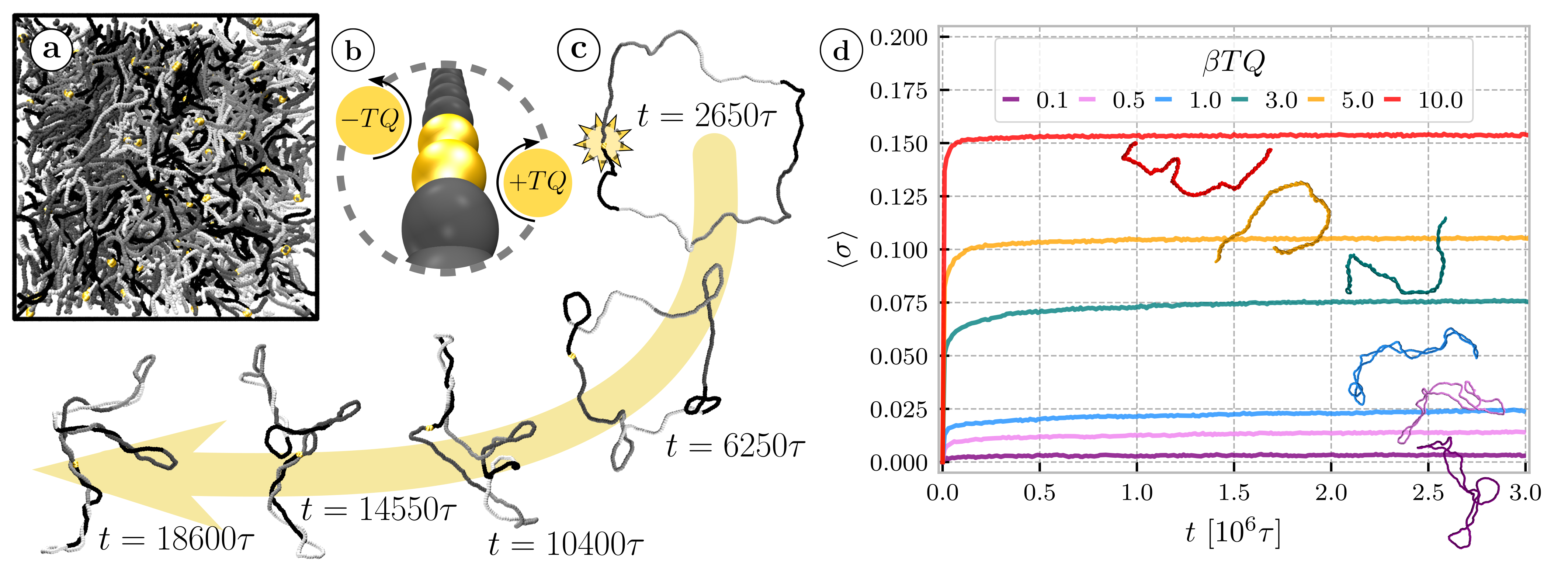}
	\caption{
    \textbf{a} Snapshot of the equilibrium system with fixed $\sigma = 0$ with different rings in shades of gray and gold beads representing the pair of monomers where the active torque is subsequently applied. 
    \textbf{b} Scheme depicting the application of the active torque
    \textbf{c} Time evolution of conformations of a single ring of $N = 400$ in infinite dilution with $\beta TQ = 3$ applied.
    \textbf{d} Mean supercoiling density $\sigma$ as a function of time for different values of active torque $TQ$.
    Snapshots show typical polymer conformations in the dense systems at $2\cdot10^6\tau$ at different torques respectively.
    }
	\label{fig:1}
\end{figure*}

We represent the ring polymers using bead-spring models with bending and torsional stiffnesses. 
We use the models \cite{Brackley2014a,racko_molecular_2017}  for completeness detailed in Methods (\refsec{esi:sec:model:details}) and ESI (\refsec{esi:sec:model:alternative}) where we also show that the fine model details affecting microscopic dynamics keep the large-scale properties and phenomena we report here largely unaffected. The models are characterized by torsion between consecutive pairs of monomeric units modeled using a dihedral potential, controlling the preferred pitch along the chain $p = 2\pi / \psi_{0}$, where $\psi_{0}$ is the angle for which the dihedral potential has the minimum. In equilibrium, by choosing appropriate $\psi_0$, we can control the supercoiling density $\sigma$, which is a measure of the total torsional-bending stress contained within the polymer, defined as $\sigma = 1/p = |\Lk|/N$, where $\Lk$ is the (excess) linking number $\Lk = N/p$ \cite{Smrek2021a}. 
We first simulate $200$ equilibrium flat ring ribbons ($\psi_{0}=0$) of length $N = 400$ in solution with monomer density $\rho s^3 = 0.08$ (\reffig{fig:1}a), corresponding to the overlap parameter $\rho R_\mathrm{g}^3/N$ of about $4.4$, signifying semidilute regime with threading entanglements \cite{Smrek2021a}. 
Subsequently, we apply the activity on each ring through a supercoiling agent, such as gyrase. Our simulations do not follow the exact microscopic pathway of this enzymatic process \cite{Bates2005a} that involves double strand passage and occurs in bursts with waiting times due to structural changes of the motor \cite{Gore2006a}. As mentioned, we rather explore its large-scale and coarse-grained consequence, being the change of the supercoiling density $\sigma$. %coarse-grained in the sense that we do not resolve the strand passage, and also multiple acting gyrases. 
Such models with active swivels introducing supercoiling by gyrase or transcription were used previously to address biological problems like unknotting\cite{racko_generation_2015},  formation of topologically associating domains \cite{10.1093/nar/gkx716} and cohesin loop extrusion \cite{Racko2017a}.
To apply the motor enzyme, we select a pair of consecutive monomeric units, we remove the dihedral potentials constraining the pitch of this pair, and we apply oppositely oriented torques on the monomers (\reffig{fig:1}b and Methods \refsec{esi:sec:model:gyrase}), rotating the monomers around the bond vector. The accumulated angle $\psi_{a}$, defines the pitch $p=2\pi / (\psi_{a}/N)$ and supercoiling density $\sigma=1/p$ in the nonequilibrium situation. 
The activity of the motor is controlled by the value of the fixed applied torque $TQ$ (see Methods \refsec{esi:sec:model:gyrase}).

At low $TQ$ the activity is not strong enough to significantly overcome the thermal fluctuations and the supercoiling density $\sigma$ fluctuates around zero, since the ribbon has zero excess supercoiling ($\psi_{0}=0$). %Fig. 1 [MODEL]: sketch with snapshot + gyraze effect: a) dense system with supercoiling, b) polymer model + gyraze
At higher $TQ$ the supercoiling initially grows in time (\reffig{fig:1}c,d) until it reaches a steady state at which the applied torques are balanced by the bending and torsional stiffness, not allowing further supercoiling.
Eventually, for sufficiently large $TQ$, we reach values $|\sigma| \approx 6\%$ (or even larger) corresponding to the maximal supercoiling density commonly observed in DNA \cite{Bates2005a}, with typical plectonemic structures with a high writhe resembling double folded chains or trees with small number of branches (see snapshots in \reffig{fig:1}d). 
The highest torques and the resulting highest induced supercoilings we find are on the limit of structural stability of the DNA \cite{Bryant2003a}, but our results apply to other ribbon-like polymers beyond DNA.
The driven transition from the flat ribbon to the supercoiled one is rather fast and, even at low values of $TQ$, takes less than the self-diffusion time in dilute conditions $\tau_{R} = R_{g}^{2}/D \simeq 6 \cdot 10^{4}\tau$, which we measured in independent simulations. 

\begin{figure}[!ht]
	\centering
    \includegraphics[width=0.47\textwidth]{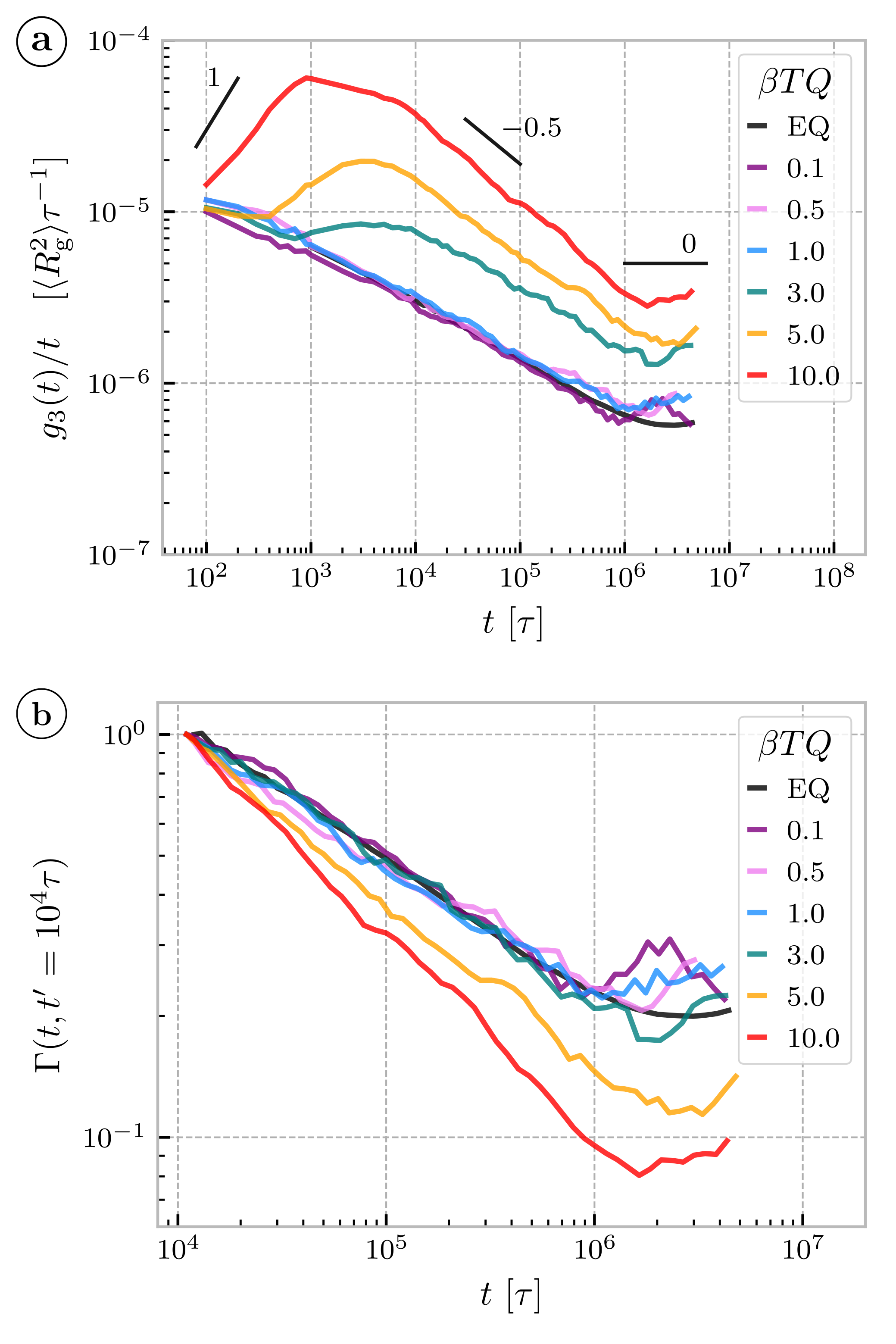}
	\caption{
    \textbf{a} Mean square displacement ~\eqref{eq:msd} divided by time since the onset of the activity in the units of mean square radius of gyration over monomer relaxation time for different torques. 
    EQ stands for the equilibrium system with no driving.
    \textbf{b} Relaxation functions ~\eqref{eq:gamma} for $t^{\prime} = 10^4\tau$, probing the dynamics after saturation of supercoiling and after the sharp initial drop in threadings.
    The equilibrium system has rings in flat ribbon configuration ($\sigma = 0$). The data of the equilibrium system are also time-averaged and active systems with $\beta TQ\leq 1$ are averaged over four system replicas with different initial conditions.
	}
	\label{writhing:fig:msd}
\end{figure}

% results for melts
In equilibrium semi-dilute solutions, the flat ribbon rings ($\sigma=0$) are rather swollen and of elongated prolate shape (\reffig{fig:1} and \reffig{esi:fig:results:shape}) with large opening and numerous threadings and interpenetration. The equilibrium dynamics, characterized by the center of mass mean-squared displacement $g_{3}(t)$, 
\begin{equation}\label{eq:msd}
    g_3(t) = \langle [\mathbf{r}_{\mathrm{C}}(t) - \mathbf{r}_{\mathrm{C}}(0)]^{2} 
    %\cdot [\mathbf{r}_{\mathrm{C}}(t) - \mathbf{r}_{\mathrm{C}}(0)] 
    \rangle
\end{equation}
is sub-diffusive for short times and diffusive one for longer times, with the crossover time $\tau_{D}\simeq 1\cdot 10^{6}\tau$ (\reffig{writhing:fig:msd}a). In comparison to \cite{Smrek2021a}, our present estimate of $\tau_{D}$ is less impacted by finite system size effects, as our systems are factor of four larger.

In contrast to equilibrium, we find diverse dynamical behavior for actively supercoiled rings depending on the value of the active torque, as measured by $g_{3}(t)$ from the point of activity onset, without time averaging for the active systems (\reffig{writhing:fig:msd}). Systems with low active torques $\beta TQ\leq 1.0$ are essentially identical to the equilibrium system with $\sigma=0$. The seemingly larger fluctuations at late times in comparison to equilibrium are caused by the time-averaging that we applied to the equilibrium system only. Systems with high active torques $\beta TQ>1$ exhibit early superdiffusive behavior that crosses over to distinct subdiffusive regimes.

%Explanations - Short time
The short-time behavior ($t<10^{5}\tau$) is readily explained by the induced level of supercoiling (\reffig{fig:1}d). In equilibrium, the conformations are weakly doubly-folded and branched due to the constraint that each ring must remain nonconcatented with the others \cite{Ubertini2022a}. High active torques induce strong supercoiling that propagates along the chain and creates ring-scale rearrangements from weakly to tightly doubly-folded conformations (\reffig{fig:1}d). Additionally, the enhanced intra-chain interactions induce rapid stiffening of the effective chain \cite{Smrek2021a} that although does not manifest itself strongly in the overall size $R_g$, but does so in the shape anisotropy of the conformation (\reffig{esi:fig:results:shape}). In contrast, low torque systems are only weakly supercoiled and as being already partly doubly-folded their shape transformation is hence less significant.

%Explanations - Long time
In the long-time regime ($t>10^{5}\tau$) the supercoiling degree almost already reaches the steady state but the overall dynamics contrasts with that of the equilibrium simulations. In equilibrium melts highly supercoiled plasmids diffuse \emph{faster} than their less supercoiled counterparts \cite{Smrek2021a}.
Here we find that the active melts exhibit a torque-dependent subdiffusive regimes. We observe (\reffig{writhing:fig:msd}b) the rings with higher active torque ($\beta TQ>3$) show stronger slowdown as measured by the relaxation function $\Gamma(t,t')$, 
\begin{equation}\label{eq:gamma}
    \Gamma(t,t') = \dfrac{g_3(t)}{t} \cdot \dfrac{t'}{g_3(t')},
\end{equation}
a normalized version of the $g_{3}(t)/t$ to the value at time $t'=10^{4}\tau$ corresponding to the end of the regime of large-scale structural relocations. The stronger relaxation slowdown for the rings with high torques (supercoiling) is in sharp contrast to the relaxation process of equilibrium rings, with corresponding matching $\sigma$, exhibiting \emph{speedup} for rings of higher supercoiling (see the equilibrium relaxation function in \reffig{esi:fig:results:dynamics3}).

\begin{figure}[!ht]
	\centering
	\includegraphics[width=0.47\textwidth]{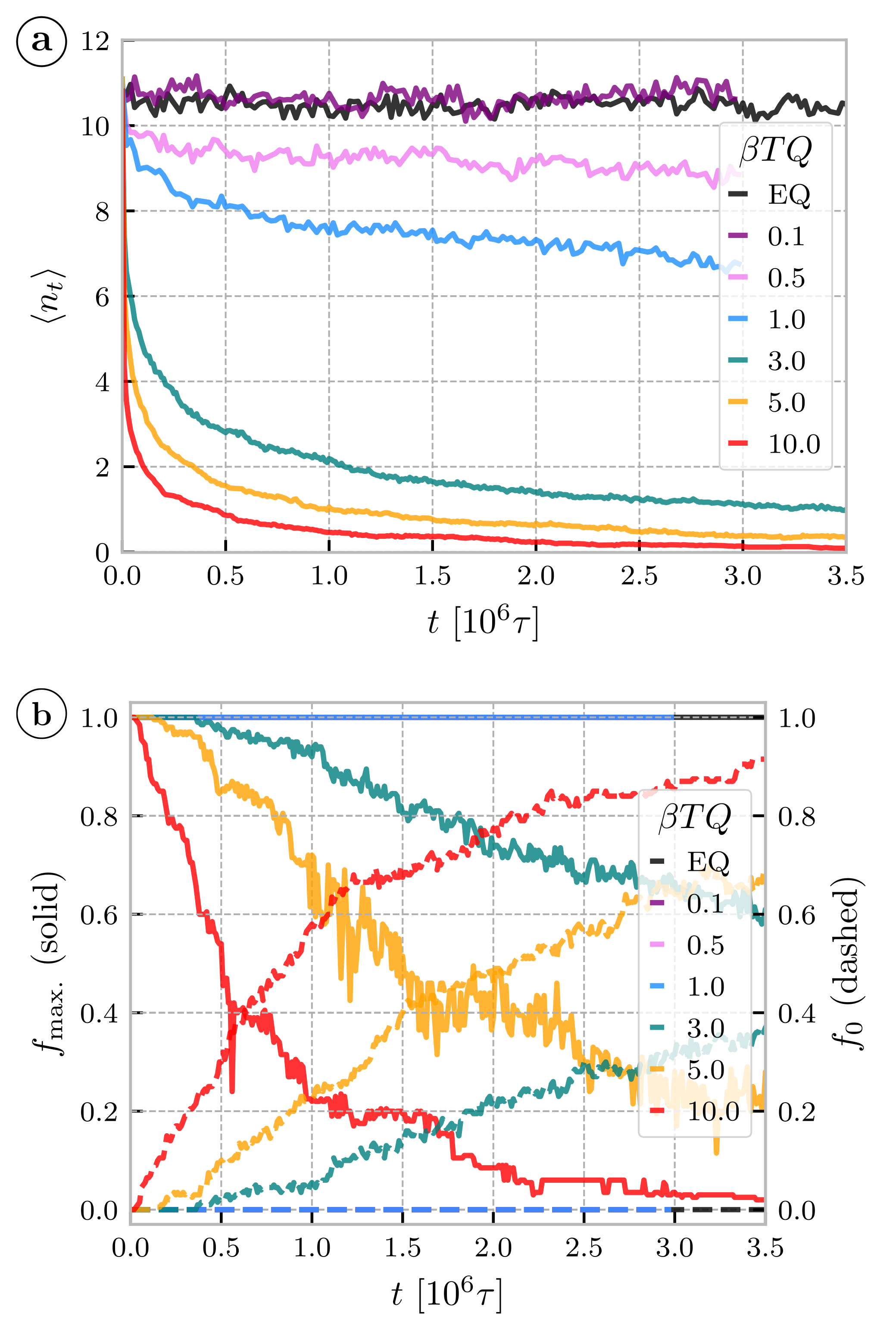}
	\caption{
    \textbf{a} The mean number of threadings per ring as a function of time after the onset of the activity, shown for different torques.
    EQ stands for the equilibrium system with no driving.
    \textbf{b} Fraction $f_{\rm max.}$ of the rings belonging to the largest continuous networks of threadings present in the system (solid, left axis) as a function of time.
    Fraction $f_{0}$ of dangling rings (dashed, right axis), rings with no threading, as a function of time.
    In both cases, the criterion for threading is non-zero threading depth (separation length), defined by eq.~\eqref{eq:Lsep}, $L_\mathrm{sep}^\mathrm{thr}$, the effect of this threshold is explored in the ESI (Fig.~\ref{esi:fig:results:threshold}).
	}
	\label{fig:threading}
\end{figure}

\begin{figure*}[!ht]
	\centering
    \includegraphics[scale = 0.95]{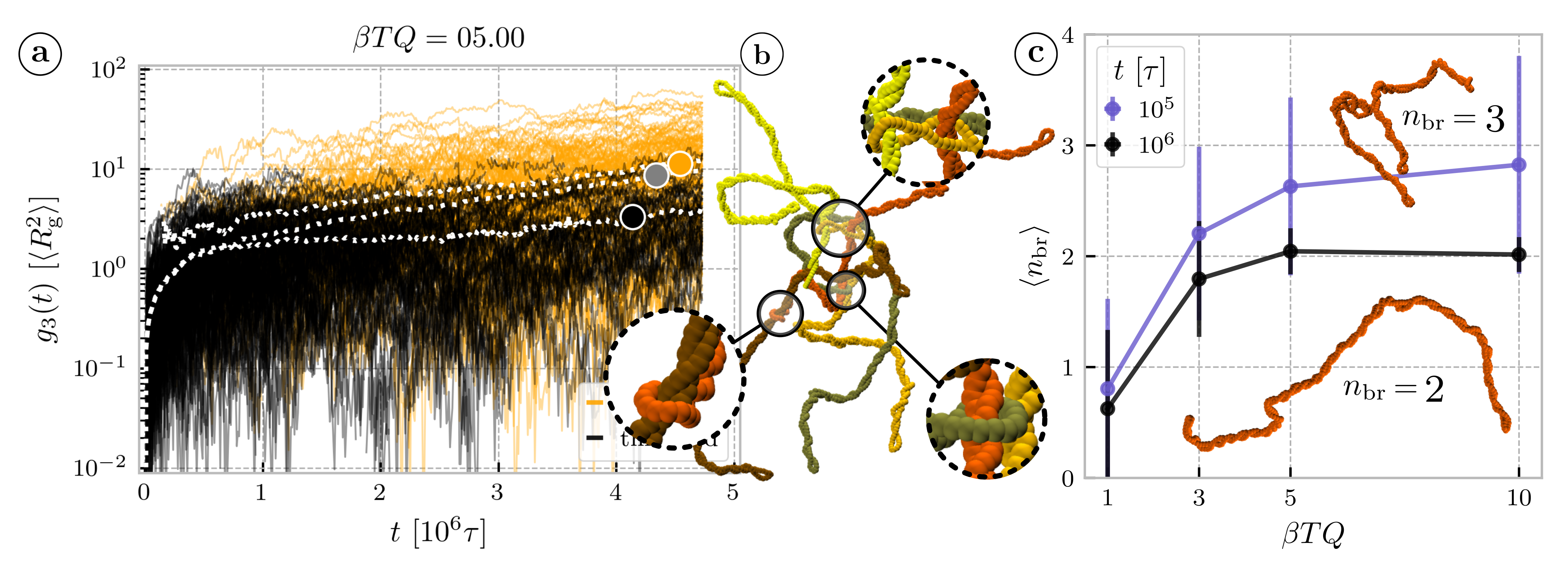}
	\caption{
    \textbf{a} The mean square displacement \eqref{eq:msd} of all individual rings, with the point at time $t$ colored black if the ring is threaded or orange if it is dangling.
    Three white dotted lines show means over three subpopulations of rings -- over all dangling rings, over all threaded rings and over all of the rings, marked with orange, black and gray circular marker respectively.
    \textbf{b} Snapshot showing long-lived and deep locked-in threadings, magnified in the insets, in a cluster of rings (different shades of orange) at $2 \cdot 10^6 \tau$ for $\beta TQ = 5$.
    \textbf{c} Mean number of branches per polymer as a function of torque for two different times -- after establishing the steady state of supercoiling and during the stage of slow aging.
    The errorbars are the standard deviations over the ensemble of ring conformations.
    The inset snapshots show a conformation of a selected ring at $\beta TQ = 5$ in the time $t = 10^5\tau$ where it has three branches and the same ring at $t = 10^6\tau$ where it relaxes to a doubly folded, linear-like chain.
	}
	\label{fig:deadlock}
\end{figure*}

To explain the contrast we investigate the inter-chain threading constraints by means of evaluating intersections of a rings' contour with another rings' disk-like minimal surface (see Methods \ref{esi:sec:model:threading} and \cite{Ubertini2022a}). In \reffig{fig:threading}a, the mean number of rings threading a ring as a function of time exhibits two regimes. The initial ($t<10^{5}\tau$) sharp drop coincides with the fast structural rearrangements after the activity onset and, as expected, affects mostly shallow threadings that vanish in this regime (\reffig{esi:fig:results:lsep:histogram}). The subsequent slow threading relaxation indicates that larger displacements are necessary to release the remaining threadings but such rearrangements are likely restricted by the threadings themselves. To explore such collective constraints we categorize rings into threading clusters, if ring A threads ring B they belong to a common cluster. 
As expected from the large number of threadings per ring, the systems at $\beta TQ \leq 1.0$ form a fully connected network -- all rings belong to the largest cluster at all times \reffig{fig:threading}b. A percolating threading cluster would indicate a possible vitrification or gel-like response if the threadings were hindering the rings motion significantly 
as found in \cite{Micheletti2024a,Smrek2020a,Michieletto2017a,Michieletto2016a} for different systems, but not here at low torques. At higher active torques, where the threadings are tight due to supercoiling and restrict the ring motion significantly, we expected that a high fraction of rings belonging to the largest cluster, $f_{\rm max.}$ would be predictive of the system slowdown. 
Contrary to these expectations, the initially fully connected system disintegrates into individual dangling rings and smaller clusters. 
Moreover, the higher the torque the \emph{sooner} the largest cluster disintegrates (\reffig{fig:threading}b) but, surprisingly, the stronger the subdiffusive slowdown is (\reffig{writhing:fig:msd}b). We checked that the cluster structure and the order of the disintegration, remains qualitatively the same even if the criterion of a cluster is based on threadings of a certain critical depth (\reffig{esi:fig:results:threshold}). Nevertheless, we do find that the membership to a cluster does impact the ring dynamics, even at later stages when the clusters are certainly not percolating. To prove that we plot the $g_{3}(t)$ for each ring separately and at a given time $t$ we compute separately the mean over rings in clusters and the mean over dangling rings. 
As shown in \reffig{fig:deadlock}a (also Fig.~\ref{esi:fig:results:deadlock} and Fig.~\ref{esi:fig:results:dynamics2} for all torques) the resulting $g_{3}(t)$ of the rings from threading clusters is consistently below the total $g_{3}(t)$ and the dangling rings constitute the more mobile component. Moreover, not only is the $g_{3}(t)$ of the rings in clusters lower, it also grows in time more slowly than the $g_{3}(t)$ of the free counterpart. Comparison across different torques (Fig.~\ref{esi:fig:results:dynamics2}) shows that both, the dangling and the threaded classes are slower for higher torques.

%\section{Discussion}
%\label{sec:discussion}
While clearly the rings in threading clusters are slower, why are then the systems with higher torques and hence more dangling rings slower? We identified two distinct mechanisms underlying this slowdown, distinguished by whether the effect originates from the topology of the threading constraint network or from the topology of the chain itself. First, the threading cluster network topology can constrain the unthreading process to follow a sequential pathway, with this effect becoming more pronounced at higher torques where the network topology is more restrictive. The snapshot in \reffig{fig:deadlock}b illustrates such a cluster, which due to mutual threading constraints can disassemble only sequentially (brown chain first).
Second, in contrast to the network-based constraints, the chain topology can limit the relaxation. As active torque increases, the resulting higher supercoiling not only tightens the threadings, thereby increasing effective friction, but more importantly can induce geometrical/topological changes of the conformation, constraining the available relaxation modes.

To investigate the impact of the threading network topology on the dynamics we considered an adapted version of the effective 1D topological glass model proposed by Lo and Turner \cite{Lo2013a}. There, each ring is represented by a doubly-folded chain of linear-like topology, an accurate description of highly supercoiled equilibrium chains of this length \cite{Smrek2021a}. Each chain can move along its contour and thread or be threaded by another chain. The latter, passive threading, at the end of a chain, such as the threading of orange by the brown chain in \reffig{fig:deadlock}b, prohibits the curvilinear motion of the threaded chain and hence its disengagement from the cluster. We constructed the corresponding Monte Carlo (MC) simulation (detailed in Methods \refsec{sec:effective_simulation})), where in a single step each chain attempts to move in a curvilinear direction and the move is rejected if the chain contains such a passive threading. If the threading is not at the end of the chain, the move is accepted and the location of the threading is updated in the direction corresponding to the move. If actively threading chain moves out of the threaded chain, the mutual threading vanishes. In contrast to \cite{Lo2013a} that was focusing on equilibrium, we do not allow the creation of new threadings, as these are highly unlikely for the present highly supercoiled case \cite{Smrek2021a}, and we track only the relaxation from the given threading network conformation that we have from our Molecular Dynamics simulations. Our MC simulations show that the threading networks are potentially vitrifying the system, but again, with trends contrasting with our observations. The effective MC model predicts slower dynamics for the lower torques (ESI sec.~\ref{esi:sec:effective_simulation}), consistent with the Lo and Turner model where the relaxation time grows with the number of threadings, the latter being indeed higher for lower torques in our MD systems (\reffig{fig:threading}a).

We find that the Lo and Turner model is not applicable in our case, because of the induced \emph{internal} geometrical and topological changes of the conformations. The 1D contour diffusional relaxation (reptation) modes would be restricted in case of a branched conformations because the branches would have to ``squeeze" through the threading openings in order to reach the chain end. We therefore measured, using local Writhe (Methods \refsec{esi:sec:model:threading}), the topology of the supercoiled chains, characterized by the number of branches $n_\mathrm{br}$, at the onset of the slow relaxation regime ($t=10^{5}\tau$). We find that not only the mean number of branches grows with the torque $\beta TQ$, but the width of their distribution does so too due to highly branched outliers (\reffig{fig:deadlock}c and \reffig{esi:fig:results:branching_histogram} for the distribution details). In contrast, at late stages ($t=10^{6}\tau$) when many of the initial threadings are relaxed (\reffig{fig:threading}a), the conformations are mostly linear-like, consistent with the equilibrium observations \cite{Smrek2021a}. Similarly, the system with $\beta TQ = 3$, separating the behavior of low and high-torque systems, confirms that number of branches close to two already at early times $10^{5}\tau$, translates into equilibrium-like dynamics later (see also \reffig{esi:fig:results:g3_vs_branching} for finer time resolution of branching statistics, separately for threaded and dangling rings, and its effect on the individual ring dynamics).

These results together show that the stronger active torques couple the threading and branching features, characterizing the inter- and intra-chain topologies of the supercoiled systems. Not only are some of the pre-existing threadings locked-in, but their subsequent relaxation is hindered by the induced branching causing a slow down of the global system dynamics.

%Fig. 3 [THREADING]: a) threading vs. time (different torques), b) threading "network" vs. time

\section{Conclusions}
\label{sec:conclusions}
Our investigated systems induce supercoiling much faster than the relaxation time resulting in a nonequilibrium conformations and partial locking of threadings. Whether such effects survive for slower supercoiling dynamics is not clear at the moment, despite the observation that highly supercoiled rings have little threadings in equilibrium \cite{Smrek2021a}, because the threading nature and the resulting threading constraint network can be complex functions of the supercoiling rate. Moreover, going beyond a simple constant rate or torque, and inducing the supercoiling with a time-dependent protocol, can have different impact on the local threading and branching statistics, indicating a possible way to tune the global dynamics by the non-equilibrium driving. The present work is hence the first step in this direction.

In the limit of long rings and high densities we conjecture even stronger slowdown than the one observed here for moderately long rings. Although the rings in a melt adopt compact conformations \cite{Halverson2011a}, meaning a saturating number of (threading) neighbors in the asymptotic limit, the threading depth statistics is a power-law indicating the existence of many deep threadings \cite{Ubertini2022a,Smrek2016a}. If these threadings are locked-in by strong induced supercoiling they might possibly lead to a complete vitrification of the system.
Moreover, highly supercoiled conformations of long rings are branched for entropic reasons \cite{Vologodskii1992a} and the branch retraction and contour diffusion are slow processes themselves \cite{vanLoenhout2012a}. The question however remains if sufficiently many threadings with branched-enough conformations on either side of the threaded chain can be trapped. While there are possibly many ways the threading or branching network can be manipulated, such as use of different stiffness or multiple supercoiling agents (e.g. enzymes) on a single ring, the future theoretical crux of the problem is to understand the connection of the branching and threading statistics with their dynamical consequences. 

Some works on equilibrium ring melts indicate that the slow and glassy dynamics emerging from threadings is a nontrivial function of the rings stiffness and concentration \cite{Goto2023a,Tu2023a,Roy2024a}. 
For the same reason, it would be interesting to investigate also blends of rings with different stiffness, fraction of active rings and higher concentrations. The concentration, in particular, can be a crucial parameter in tuning the behaviour. While below the overlap concentration, there can be hardly any threadings locked-in and hence no slowdown, one can expect a stronger slowdown as a function of overlap, as the number of threadings is expected to be proportional to the overlap parameter. Yet, higher concentrations at the same chain stiffness might induce nematic ordering of the supercoiled conformations \cite{Smrek2021a} and hence the behavior can be more complex. 

Older works \cite{Halverson2011b} indicate that even in melts of semiflexible rings exist ring pairs that remain spatially proximal for times significantly exceeding $\tau_{D}$. Recent simulations \cite{Giesinger2025a} link this observation to ring pairs with geometrically deadlocked threading conformations that relax on time scales surpassing the diffusion time. Inducing supercoiling in these pairs would likely prolong the relaxation even more. Our present work shows how the actively induced supercoiling control the dynamic response of the polymeric material and it indicates that quickly supercoiled long rings might be a path to create an active topological glass in experiment. Whether it could be achieved also with a gyrase operating at longer time scales or with other faster supercoiling agents remains currently unknown.

\section{Methods}
%%%%%%%%%%%%%%%%%%%%%%%%%%%%%%%%%%%%%%%%%%%%%%%%%%
\subsection{Microscopic Model of Plasmid Suspension}
\label{esi:sec:model:details}
%%%%%%%%%%%%%%%%%%%%%%%%%%%%%%%%%%%%%%%%%%%%%%%%%%

We model the DNA plasmids using the standard bead-spring coarse-grained model of a polymer \cite{Grest1986a, Kremer1990a} with elements of elastic twistable ribbon \cite{Brackley2014a, Smrek2021a}.
A single plasmid consists of $N = 400$ connected monomeric units, forming an unknotted ring polymer.
Each pair of monomeric units at instantaneous distance $r$ interacts with an isotropic pair (WCA) potential
\begin{equation}\label{eq:model:wca}
	U_{\mathrm{WCA}}(r) = 4 \varepsilon \left[ 
    \left( \dfrac{s}{r} \right)^{12} - \left( \dfrac{s}{r} \right)^{6} 
    + \dfrac{1}{4} \right] 
    \cdot H \left( 2^{1/6} - \dfrac{r}{s} \right),
\end{equation}
where $H(\cdot)$ is the Heavside step-function, $\varepsilon = 1/\kT = \beta = 1$ sets the energy scale and $s = 2.5\ \mathrm{nm} \approx 7.4\ \mathrm{bp}$ sets the length scale of the model, yielding $\approx 3.0\ \mathrm{kbp}$ for the whole plasmid.
The bonds between monomeric units are emulated using finitely extensible non-elastic (FENE) bonds
\begin{equation}\label{eq:model:fene}
	U_{\mathrm{FENE}}(r) = -\dfrac{1}{2} K_{\mathrm{FENE}} R_0^2 \ln \left[ 1 - \left( \dfrac{r}{R_0} \right)^2 \right],
\end{equation}
where $K_{\mathrm{FENE}} = 40\kT / s^2$ and $R_0 = 1.6s$ is the maximal allowed bond length beyond which the potential diverges.
To simulate the polymer contour bending, every triplet of consecutive monomeric units interacts with Kratky-Porod angular potential
\begin{equation}\label{eq:model:bending}
	U_{\mathrm{bend}}(r) = K_{\mathrm{bend}} \left(1 - \dfrac{\boldt_i \cdot \boldt_{i+1}}{|\boldt_i| |\boldt_{i+1}|} \right),
\end{equation}
where $K_{\mathrm{bend}} = 20 \kT$ and the bond vector $\boldt_i = \mathbf{r}_{i+1} - \mathbf{r}_i$, where $\mathbf{r}_i$ is the position of the monomeric unit $0 \leq i < N$, and further we also assume $0 \equiv N$ to simplify the notation in summations running over the whole ring.
The persistence length resulting from the above potentials is $l_\mathrm{p} \approx 20s \approx 50\ \mathrm{nm} \approx 150 \ \mathrm{bp}$.

\begin{figure}
    \centering
    \includegraphics[scale = 0.90]{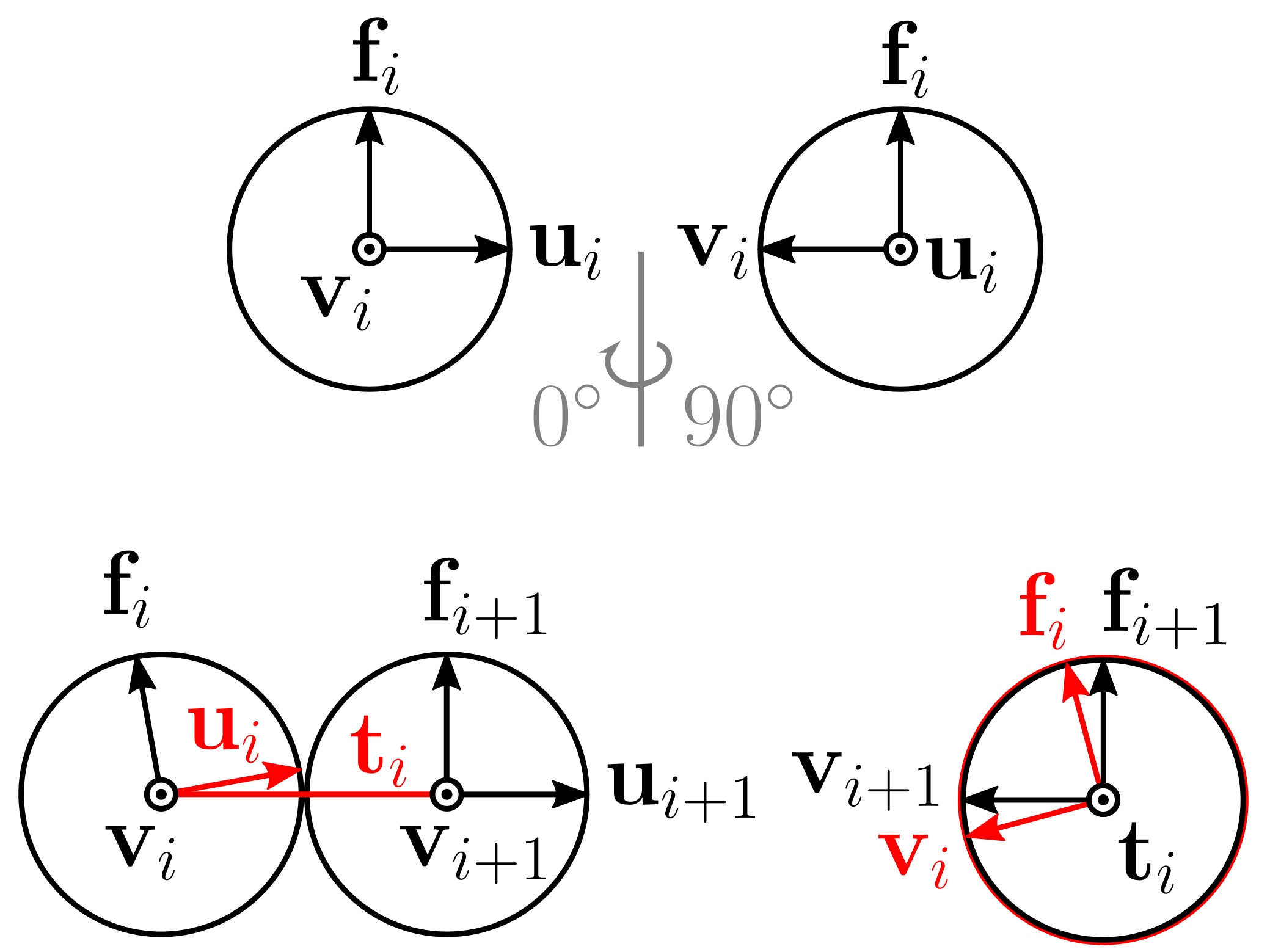}
    \caption{Top: monomeric unit with the set of orientation vectors, bottom left: scheme showing the vectors involved in the tilting potential \eqref{eq:model:tilt}, bottom right: scheme showing the vectors involved in the dihedral potential \eqref{eq:model:torsion:1} and \eqref{eq:model:torsion:2}}
    \label{esi:fig:model}
\end{figure}

To embed the torsional elasticity into the DNA chain, each monomeric unit is inscribed three vectors $[\boldu, \boldf, \boldv]$ forming a right-handed orthonormal set of axes \cite{Brackley2014a}, giving the orientation to the otherwise spherically symmetric particle as shown in \reffig{esi:fig:model}.
For each pair of consecutive monomeric units, we apply a stiff tilting potential
\begin{equation}\label{eq:model:tilt}
    U_{\mathrm{tilt}} = K_{\mathrm{tilt}} \left( 1 - \boldu_i \cdot \dfrac{(\boldu_i - \boldt_i)}{|\boldt_i|} \right),
\end{equation}
with $K_{\mathrm{tilt}} = 200\kT$ locally aligning $\boldu_i$ with the bond vector $\boldt_i$ of the pair as shown in \reffig{esi:fig:model}.
Finally, we apply two dihedral potentials on each pair of consecutive monomeric units
\begin{equation}\label{eq:model:torsion:1}
    U_{\mathrm{torsion}} = K_{\mathrm{torsion}} \left( 1 - 
    \dfrac{(\boldt_i \times \boldf_i) \cdot (\boldt_i \times \boldf_{i+1})}{|\boldt_i \times \boldf_i| |\boldt_i \times \boldf_{i+1}|}  
    \right),
\end{equation}
\begin{equation}\label{eq:model:torsion:2}
    U_{\mathrm{torsion}} = K_{\mathrm{torsion}} \left( 1 -  
    \dfrac{(\boldt_i \times \boldv_i) \cdot (\boldt_i \times \boldv_{i+1})}{|\boldt_i \times \boldv_i| |\boldt_i \times \boldv_{i+1}|}
    \right),
\end{equation}
with $K_{\mathrm{torsion}} = 50\kT$ giving the chain the torsional stiffness as shown in \reffig{esi:fig:model}.
Since the monomeric units have internal orientation, they also possess the rotational degrees of freedom controlled by the angular forces originating from the tilting and dihedral potentials.
The supercoiling density within the DNA chain is defined as $\sigma = \Lk / N$, where $\Lk$ is the linking number \cite{Clauvelin2012a, Klenin2000a} calculated as
%\begin{widetext}
\begin{equation}\label{eq:model:lk}
    \Lk = \dfrac{1}{4\pi} \oint_{C_1} \oint_{C_2} 
    \dfrac{\boldr_1 - \boldr_2}{|\boldr_1 - \boldr_2|^3} 
    \left( \dot{\boldr}_1 \times \dot{\boldr}_2 \right) \dd s_1 \dd s_2,
\end{equation}
%\end{widetext}
where $\boldr_i = \boldr_i(s_i)$ is a parametrization of curve $C_i$ with arc length $s_i$ and $\dot{\boldr}_i=\dot{\boldr}_i(s_i) = {\dd \boldr_i(s_i)}/{\dd s_i}$ \cite{Tubiana2024a}.
The curve $C_1: \boldr_1(s_1)$ follows the polymer contour and $C_2: \boldr_2(s_1) = \boldr_1(s_1) + \epsilon \boldu_1(s_1)$ where $\epsilon \ll 1$.

The solution of $M = 200$ plasmids is enclosed in a cubical box with periodic boundary conditions and with dimension $L = 100 s \approx 250\ \mathrm{nm}$, resulting in the monomer concentration $\rho s^3 = 0.08$ corresponding to $\rho/\rho^* \approx 4.4$ in terms of polymer concentration relative to the overlap one, herein defined as $\rho^* = N / R_\mathrm{g}^3$ with $R_\mathrm{g}$ being the radius of gyration of a ring with $\sigma = 0$, or corresponding to weight concentration of $c_{\mathrm{m}} \approx 40 \mathrm{mg}/\mathrm{ml}$ of DNA.

%%%%%%%%%%%%%%%%%%%%%%%%%%%%%%%%%%%%%%%%%%%%%%%%%%
\subsection{Supercoiling Agent Activity}
%%%%%%%%%%%%%%%%%%%%%%%%%%%%%%%%%%%%%%%%%%%%%%%%%%
\label{esi:sec:model:gyrase}
To simulate the non-equilibrium effect of the supercoiling agent, for each of the plasmids, we select a pair of consecutive monomeric units and remove the dihedral potentials \eqref{eq:model:torsion:1} and \eqref{eq:model:torsion:2} from this pair.
Simultaneously, we apply an external torque $TQ$ on the two above monomers, rotating one of them clockwise and the other one counter-clockwise along the bond vector between them.
The forces stemming from the applied torque are applied at the distance $0.80\sigma$ from the center of the particle, which considering $s \approx 2.5\ \mathrm{nm}$ and $T = 300\ \mathrm{K}$, results in typical torsional forces $\approx 2.07\ \mathrm{pN}$ for $TQ = \kT$.
The excess energy pumped into the system by the means of activity is dissipated through the Langevin thermostat as detailed later.
We note that in the limit of diminishing activity $\beta TQ \approx 0$, the polymer behaves as a \emph{nicked} DNA ring, which can fully relax the torsional stress within.
For the high values of $\beta TQ$, torque applied to the two monomers excites the elastic response of the sequence of the dihedral potentials along the contour, propagating excess torsional stress over the polymer, until reaching a steady-state as explained in detail in the discussion of the results.
In summary, the supercoiling agent (e.g.~gyrase) increases the absolute value of supercoiling density contained within the ring, with higher value of $\beta TQ$ resulting in a higher absolute value of supercoiling.

%%%%%%%%%%%%%%%%%%%%%%%%%%%%%%%%%%%%%%%%%%%%%%%%%%
\subsection{Simulation Method}
%%%%%%%%%%%%%%%%%%%%%%%%%%%%%%%%%%%%%%%%%%%%%%%%%%

To sample the configurations we propagate the particles using the LAMMPS (\texttt{29Oct2020} version) \cite{Plimpton1995a, Brackley2014a} implementation of molecular dynamics with Langevin thermostat with friction $\gamma$.
Translational degrees of freedom are propagated using the equation of motion
\begin{equation}\label{esi:eq:langevin}
    m\ddot{\boldr_i}(t) = \mathbf{F}_i - m\gamma\dot{\boldr}_i(t) + \mathbf{Y}(t), 
\end{equation}
where $m$ is particle mass, $\mathbf{F}$ is the force originating from the potentials and active forces, $\mathbf{Y}(t)$ is a random force obeying fluctuation-dissipation theorem such that $\langle \mathbf{Y}(t) \rangle = 0$ and $\langle \mathbf{Y}_i(t)\mathbf{Y}_i(t') \rangle = 2\gamma m\kT \delta_{ij}\delta(t-t')$, where $\delta$ is the Kronecker delta.
Simultaneously, rotational degrees of freedom are propagated under the action of the active torques and torques from the angular potentials. 
We use $\gamma = 1.0/\tau$, where $\tau = (m\sigma^2/\kT) = 1/\gamma = 1 = 1000\Delta t$ is the time unit and $\Delta t = 0.001\tau$ is the integration time step.

To prepare the initial configurations for the non-equilibrium simulations, we first simulate a solution of flat ribbons ($\sigma = 0$) in equilibrium.
We initialize the ring polymers as circles, placed on a well spaced primitive cubic lattice, making sure that the rings are unknotted and unlinked.
We then iteratively thermalize the system, and compress it by rescaling the coordinates by a factor $(L_0-s)/L_0$, where $L_0$ is the box length before the rescaling.
We repeat the sequence of the two steps until reaching the desired concentration of $\rho s^3 = 0.08$, in our case over a cumulative simulation length of $10^6 \tau$, exceeding the Rouse time $\tau_{\mathrm{RS}} \sim (400)^2 \sim 10^5 \tau$.
Upon reaching the final density, we equilibrate the system for another $4\cdot 10^6 \tau$ and then collect the equilibrium properties over the course of final time interval $8\cdot 10^6 \tau$.
Each ring traverses a distance of several of its radii of gyration during the simulation.
For the non-equilibrium simulations, we start from an equilibrium snapshot, apply the active torques and then run the simulation for $4 \cdot 10^6\tau$.
For each value of torque, we simulated four independent replicas with varying initial configurations and different random seeds, and the presented ensemble averages of total mean square displacements are averaged over all four runs.

%%%%%%%%%%%%%%%%%%%%%%%%%%%%%%%%%%%%%%%%%%%%%%%%%%
\subsection{Threading and Branching Analysis}
\label{esi:sec:model:threading}
%%%%%%%%%%%%%%%%%%%%%%%%%%%%%%%%%%%%%%%%%%%%%%%%%%

To detect the threadings we use the same method as detailed in \refref{Smrek2019a}. In short, at every analyzed snapshot a disk-like surface, triangulated to $4N$ triangles, is spanned on every ring and the surfaces are independently minimized by moving of the triangle vertices that do not belong to the boundary (the ring). In rare cases where the minimization does not converge \cite{Smrek2019a}, we do not analyze the threading of that snapshot.
Subsequently, we detect the intersections of the contour of one ring with another ring's minimal surface, signifying a threading of the latter by the former. The minimal surface of a threaded ring splits the threading ring in consecutive segments of lengths $L_{1}$, $L_{2}$, \ldots $L_{p}$, where $p$ is the number of piercings through the surface. As the rings are nonconcatenated $p$ is an even number and the subsequent segments are located at the opposite sides of the minimal surface \cite{Smrek2019a}. We quantify the depth of the threading by the so-called separation length $L_{\rm sep}$ defined by
\begin{equation}
\label{eq:Lsep}
    L_{\rm sep} = \textrm{min} \left( \sum_{i \,\textrm{even}}L_{i},\sum_{i\,\textrm{odd}}L_{i} \right).
\end{equation}
The separation length $L_{\rm sep} \in [0,N/2]$ quantifies the amount of the polymer material of the threading ring on one of the two sides of the minimal surface of the threaded ring.

To describe the threading network, for each configuration, we construct a connectivity matrix $\mathbb{A}$ of size $M\times M$, such that if ring $0 \leq i < M$ threads ring $0 \leq j < M$ with $L_\mathrm{sep} \geq L^\mathrm{thr}_\mathrm{sep}$, then $\mathbb{A}[i,j] = 1$, otherwise $\mathbb{A}[i,j] = 0$.
This matrix in turn defines a graph of $M$ nodes, where a threading event corresponds to a directed edge.
Within this graph, we localize the maximal connected components corresponding to threaded clusters.
We consider $L^\mathrm{thr}_\mathrm{sep} = 0$ unless stated otherwise.

To estimate the number of branches, we detect the tips of plectonemes by calculating the profile of local segmental writhe, $W(i)$, \cite{Smrek2021a, Sleiman2022a} for each monomer $i$ as
\begin{equation}
    W(i) = \dfrac{1}{4\pi} \int_{C_1} \int_{C_2}
    \dfrac{\boldr_1 - \boldr_2}{|\boldr_1 - \boldr_2|^3} 
    \left( \dot{\boldr}_1 \times \dot{\boldr}_2 \right) \dd s_1 \dd s_2,
\end{equation}
where $\boldr_i = \boldr_i(s_i)$ is a parametrization of curve $C_i$ with arc length $s_i$ and curves $C_1: \boldr_1(s_1)$ and $C_2: \boldr_2(s_2)$ follow the polymer contour between the monomers $[i-w,i]$ and $[i,i+w]$ respectively, where $w = 20$ covers the segment length of approximately one persistence length.
The local maxima of $W(i)$ correspond to the tips of branches, and we also require that $W(i) \geq W_\mathrm{thr}$ at these maxima, where $W_\mathrm{thr}$ is the threshold set to avoid the false branch recognition, representing only  topologically insignificant local segmental bending due to fluctuations. 
Number of branches $n_{\mathrm{br}}$ on a polymer is then defined as the number of local maxima on the segmental writhe profile. In essence the threshold defines what is considered a branch. Branch configuration with a locally open (flat non-supercoiled) tip, for example due to a threading, might exhibit smaller value of $W$ and hence can lead to undercounting of the number of branches. 
In contrast, a long branch, if wrapped around itself, can locally exhibit higher $W$ even at segments that are not at its tip, that can lead to overcounting of the number of branches. The latter effect can be reduced by the use of relatively short segment length $w$, as we do, that is relatively stiff and reduces such bending. 
We show the mean number of branches and the standard deviation of their distribution as a function of the threshold in \reffig{esi:fig:results:branching} showing that our two conclusions from the Results section, namely $(i)$ branching being the increasing function of the torque at time $t=10^{5}\tau$ and $(ii)$ the branching is strongly reduced to the extent of linear-like chains at $t=10^{6}\tau$, hold irrespective of the specific value of the threshold ($W_\mathrm{thr}\geq 0.2$). We confirmed visually in a number of cases that the branching is correctly captured. To produce the figure \reffig{fig:deadlock}c we used the threshold $W_{\rm{thr}}=0.35$ as in \cite{Smrek2021a}.

\subsection{Effective Monte Carlo Simulations}
\label{sec:effective_simulation}
%%%%%%%%%%%%%%%%%%%%%%%%%%%%%%%%%%%%%%%%%%%%%%%%%%
The effective 1D Monte Carlo simulation based on the work of Lo and Turner \cite{Lo2013a} treats each ring as a linear doubly-folded chain, of effective length $N_{\rm{eff}}$, that can be passively threaded or can actively thread other chains at some contour position $x_{a/p} \in [0,N_{\rm{eff}}-1]$. At every step each chain is selected in random order and a random move of unit length left/right is attempted. If a passive threading exists at a contour coordinate $x_{p}=0$ or $x_{p}=N_{\rm{eff}}-1$ the left/right move is rejected. If move is accepted the chain is shifted, we update the position of the chain, the locations of all the passive threadings $x_{p}$ of the given chain and the locations of all the active threadings $x_{a}$. If an active threading location $x_{a}$ moves out of bounds $[0,N_{\rm{eff}}-1]$ the threading is relaxed and vanishes. The initial threading network for the effective simulation is from our MD simulations at time $t=10^{5}\tau$. We use the information of ring pairs that are in threading configuration, which of them is passively/actively threaded and we use the value of $L_{\rm{sep}}$ as a proxy for the $x_{a}$. Specifically, $x_{a} = L_{sep} N_{\rm{eff}}/(N/2)$. Unfortunately from the threading analysis \refsec{esi:sec:model:threading} we do not have access to the locations of passive threadings because these correspond to the locations of the intersections of the minimal surfaces and not of the ring contours, hence are more difficult to define. Therefore we tested our simulations for two setups. First, the locations of the passive threadings are chosen uniformly randomly $x_{p}\in [0,N_{\rm{eff}}-1]$. Such a choice can however lead to stalled global conformations, in particular when there exist many mutual threadings and the effective ring lengths are short, because many rings can have mutually (or in cycles) $x_{p}=0$ and $x_{p}=N_{\rm{eff}}-1$ making such conformations impossible to relax. More freedom can be obtained by either increasing $N_{\rm{eff}}$ or by restricting the initial $x_{p}$'s to be located away from the chain ends. We comment on these options in ESI (\refsec{esi:sec:effective_simulation}).

\section{Associated Content}
\subsection{Supporting Information} 
The Supporting Information (ESI) is available below the references. It contains details on Effective Monte Carlo Simulations, comparison with different microscopic model and additional simulation results: polymer shape parameters, branching analysis for threaded and dangling ring classes, separation length distributions, network topology analysis, dynamics of deadlocked rings and dynamics of rings in steady state.

 \section{Acknowledgements}
This research was funded by the Grant Agency of the Ministry of Education, Science, Research and Sport of the Slovak Republic, Grant VEGA 2/0038/24 (Polymers with Active Chiral Topology and Nanotechnology, PACT\&NANOTEC) and Slovak Research and Development Agency SRDA 21-0346, SK-AT 20-0011. 
This research was supported by the OeAD Austria-Slovakia Scientific and Technological Cooperation Grant No: SK 05/2021. 
The computational results presented have been achieved in part using the Vienna Scientific Cluster of Austrian Scientific Computing.
Part of the research results was obtained using the computational resources procured in the
national project National competence centre for high performance computing (project code:
311070AKF2) funded by European Regional Development Fund, EU Structural Funds Informatization
of society, Operational Program Integrated Infrastructure.
R.S. acknowledges the financial support by the Doctoral College Advanced Functional Materials-Hierarchical Design of Hybrid Systems DOC 85 doc.funds funded by the Austrian Science Fund (FWF).
R.S. is funded by the UK Research and Innovation (UKRI) Engineering and Physical Sciences Research Council (EPSRC) under the UK Government’s guarantee scheme (EP/Z002028/1), following funding by the European Research Council (Consolidator Grant) under the European Union’s Horizon Europe research and innovation programme.

\makeatletter
\def\selectlanguage#1{%
  \@ifundefined{date#1}{}{%
    \@nameuse{date#1}%
  }%
}
\makeatother

\let\oldaddcontentsline\addcontentsline% Store \addcontentsline
\renewcommand{\addcontentsline}[3]{}% Make \addcontentsline a no-op
\bibliography{gyrase,bibjs} % Produces the bibliography via BibTeX.
\let\addcontentsline\oldaddcontentsline% Restore \addcontentsline

% Switch to single column for supplementary
\clearpage
\onecolumngrid
\begin{center}
    \centerline{\Large \textbf{Supplementary Information}}
\end{center}

\renewcommand{\theequation}{S\arabic{equation}}
\renewcommand{\thefigure}{S\arabic{figure}}
\renewcommand{\thetable}{S\arabic{table}}

\input{ESI_content.tex}

\end{document}

%% file: definitions.tex
% abbreviations and references
\newcommand*{\reffig}[1]{Fig.~\ref{#1}}
\newcommand*{\refref}[1]{Ref.~\cite{#1}}
\newcommand*{\refsec}[1]{Sec.~\ref{#1}}

% physics
\newcommand*{\dd}{\mathrm{d}}
\newcommand*{\kB}{k_\mathrm{B}}
\newcommand*{\kT}{\kB T}
\newcommand*{\Lk}{\mathrm{Lk}}

% vectors for angular potentials
\newcommand*{\boldf}{\mathbf{{f}}}
\newcommand*{\boldr}{\mathbf{{r}}}
\newcommand*{\boldt}{\mathbf{{t}}}
\newcommand*{\boldu}{\mathbf{{u}}}
\newcommand*{\boldv}{\mathbf{{v}}}

% labels and tags
\newcommand*{\ready}{\colorbox{green}{\textbf{! READY FOR REVISIONS !}}}
\newcommand*{\todo}{\colorbox{orange}{\textbf{! TODO !}}}
\newcommand*{\finished}{\colorbox{teal}{\textbf{! FINISHED !}}}
\newcommand*{\unknown}{\colorbox{red}{\textbf{! STATUS UNKNOWN !}}}
\newcommand*{\wip}{\colorbox{yellow}{\textbf{! WORK IN PROGRESS !}}}
\newcommand*{\doneJS}[1]{\colorbox{green}{\textbf{Done by JS - see {#1} comments.}}}

%% file: title.tex
\newcommand*{\manuscripttitle}{Actively induced supercoiling can slow down plasmid solutions by trapping the threading entanglements}

%% file: authors.tex
% affiliations
\newcommand*{\affUniVie}{Faculty of Physics, University of Vienna, Boltzmanngasse 5, 1090 Vienna, Austria}
\newcommand*{\affVDSP}{Vienna Doctoral School in Physics, University of Vienna, Boltzmangasse 5, 1090 Vienna, Austria}
\newcommand*{\affSAV}{Polymer Institute, Slovak Academy of Sciences, Dúbravská cesta 9, 845 41 Bratislava, Slovakia}
\newcommand*{\affCam}{Yusuf Hamied Department of Chemistry, University of Cambridge, Lensfield Road, Cambridge CB2 1EW, UK}
\newcommand*{\affSap}{Department of Physics, Sapienza University of Rome, Piazzale Aldo Moro 5, 00185 Rome, Italy}

% authors
%\author{\textbf{*authors alphabetically for now}}
%\affiliation{placeholders}

\author{Roman Staňo}
\affiliation{\affUniVie}
%\affiliation{\affVDSP}
\affiliation{\affCam}

\author{Renáta Rusková}
\affiliation{\affSAV}
\affiliation{\affSap}

\author{Dušan Račko}
\email{dusan.racko@savba.sk}
\affiliation{\affSAV}

\author{Jan Smrek}
\email{jan.smrek@univie.ac.at}
\affiliation{\affUniVie}

%% file: ESI_content.tex
%\tableofcontents

%%%%%%%%%%%%%%%%%%%%%%%%%%%%%%%%%%%%%%%%%%%%%%%%%%
%%%%%%%%%%%%%%%%%%%%%%%%%%%%%%%%%%%%%%%%%%%%%%%%%%
%%%%%%%%%%%%%%%%%%%%%%%%%%%%%%%%%%%%%%%%%%%%%%%%%%
\section{Models \& Methods}
\label{esi:sec:model}
%%%%%%%%%%%%%%%%%%%%%%%%%%%%%%%%%%%%%%%%%%%%%%%%%%
%%%%%%%%%%%%%%%%%%%%%%%%%%%%%%%%%%%%%%%%%%%%%%%%%%
%%%%%%%%%%%%%%%%%%%%%%%%%%%%%%%%%%%%%%%%%%%%%%%%%%

%%%%%%%%%%%%%%%%%%%%%%%%%%%%%%%%%%%%%%%%%%%%%%%%%%
\subsection{Effective Monte Carlo Simulations}
\label{esi:sec:effective_simulation}
\begin{figure}[htbp]
    \centering
    \includegraphics[scale = 0.90]{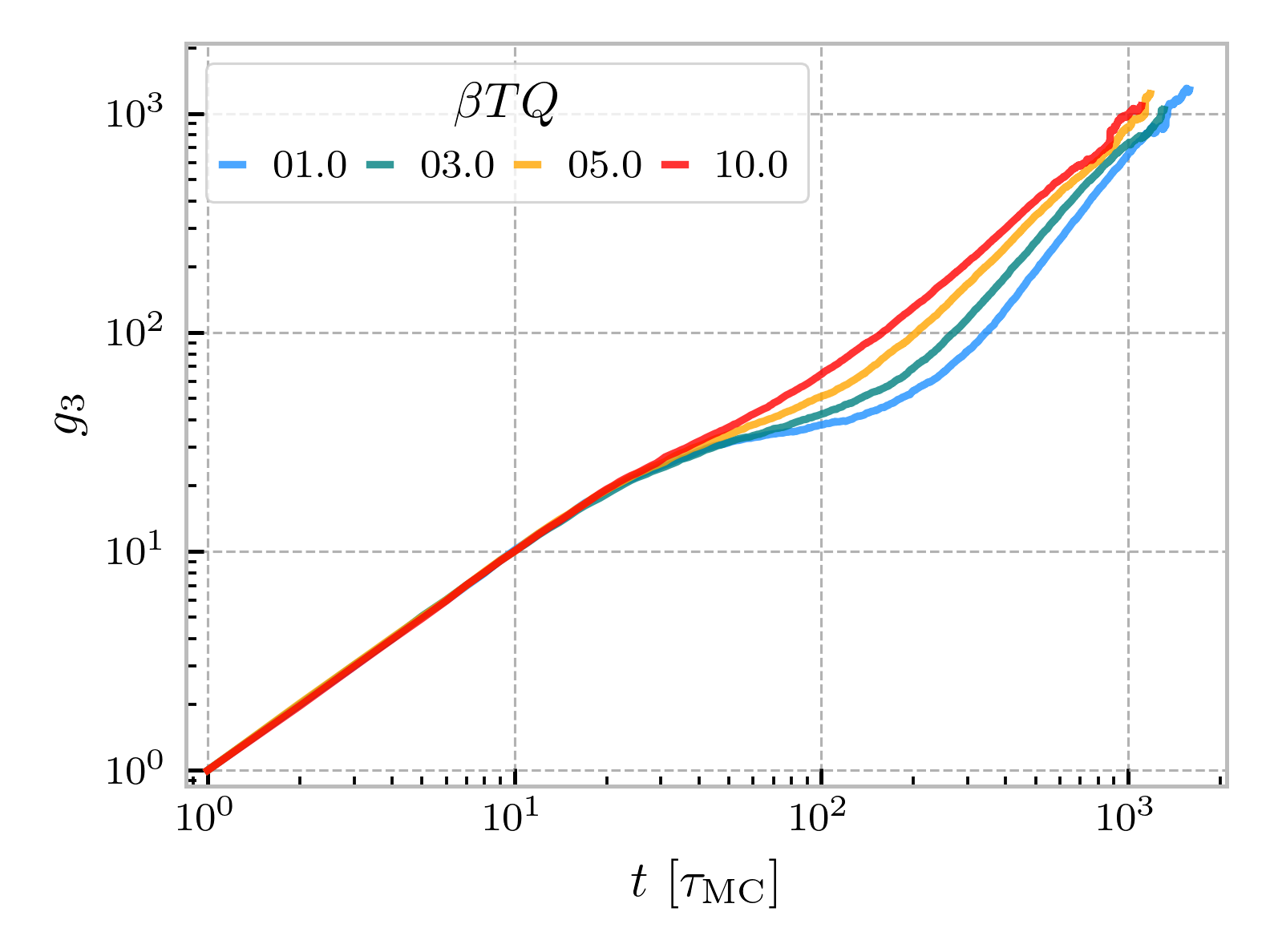}
    \includegraphics[scale = 0.90]{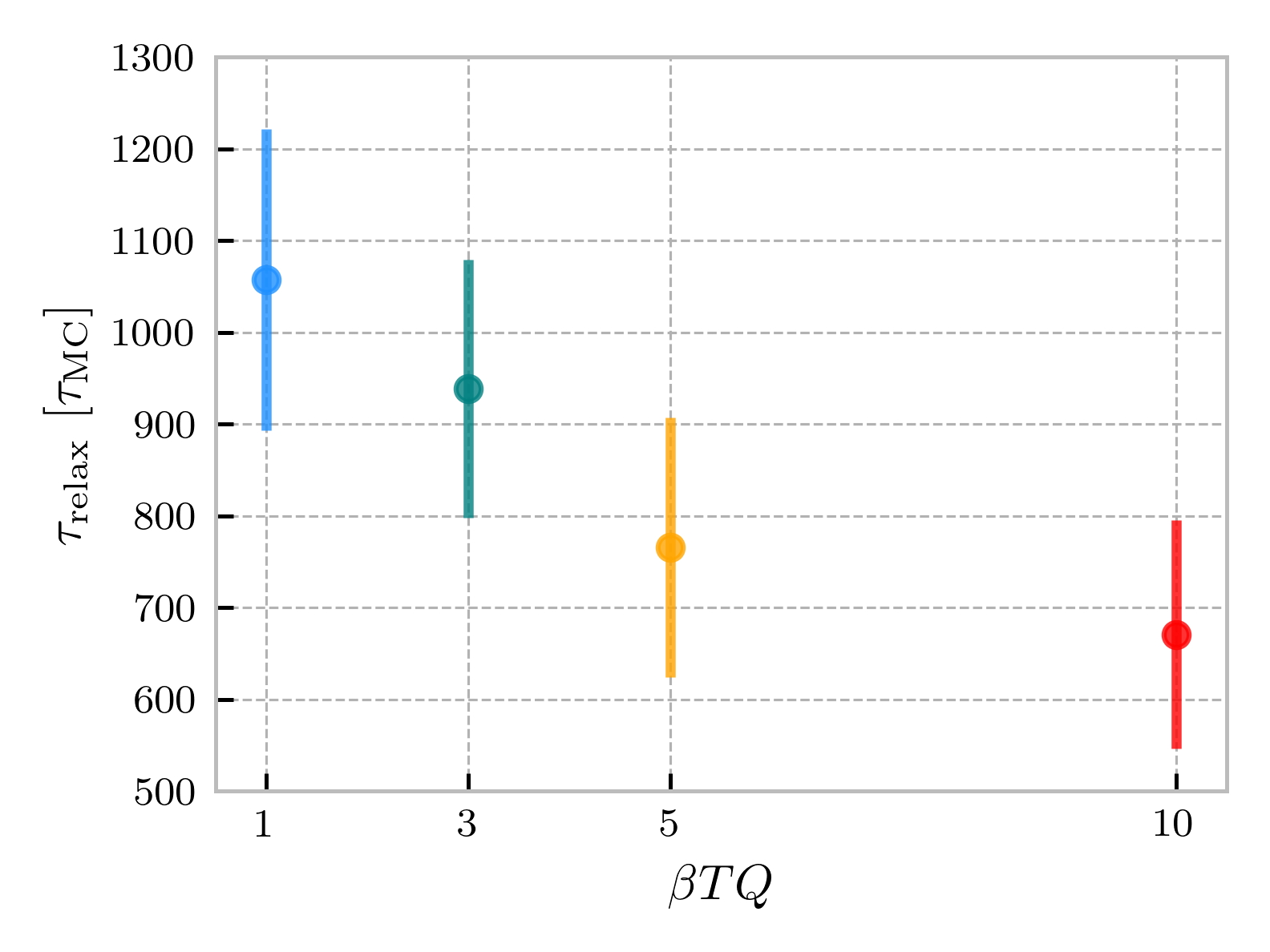}
    \caption{
    Mean-squared displacement $g_{3}$ (left) and relaxation time (right) $\tau_{\rm{relax}}$ of rings from the effective MC simulations as specified in \refsec{esi:sec:effective_simulation}.
    The presented data are the average of $50$ samples where each sample had the same initial configuration specified in the text, but likely different sequence of random numbers. 
    The error-bars of $\tau_{\rm{relax}}$ represent the standard deviation of the sample.
    $\tau_\mathrm{MC}$ is the Monte Carlo time (number of steps) and $g_3$ is dimensioneless as it refers to the MSD in the space of particle indices.
    }
    \label{esi:fig:results:effective_simulations}
\end{figure}

%%%%%%%%%%%%%%%%%%%%%%%%%%%%%%%%%%%%%%%%%%%%%%%%%%
In the effective 1D Monte Carlo simulation we track the mean-squared displacement of the chains and the system relaxation time defined as the time until the last threaded ring pair unthreads. In \reffig{esi:fig:results:effective_simulations} 
we show the results for systems with $N_{\rm{eff}}=20$ and initial $x_{p}=N_{\rm{eff}}/2$ for all the threadings. The mean-squared displacement shows a prolonged subdiffusion and longer relaxation times for systems with smaller active torques that exhibit relatively more threadings per ring. These observations are consistent with equilibrium simulations \cite{Lo2013a} that show significant increase of relaxation time with the number of threadings. Another set of simulations with uniformly distributed $x_{p}$ typically stall for lower torques, as explained above, as a fraction of the rings remain trapped in a threading cluster. The fraction is higher for lower torques consistent with a higher number of threadings. In the last set of simulations we used $N_{\rm{eff}}=100$ and uniformly distributed $x_{p}$, leading to qualitatively similar results to the ones in \reffig{esi:fig:results:effective_simulations} (not shown) but poor statistics due to long simulation times.

%%%%%%%%%%%%%%%%%%%%%%%%%%%%%%%%%%%%%%%%%%%%%%%%%%
\subsection{Original Reference Model}
\label{esi:sec:model:alternative}
%%%%%%%%%%%%%%%%%%%%%%%%%%%%%%%%%%%%%%%%%%%%%%%%%%

Herein, we compare our current polymer model (\refsec{esi:sec:model:details}) with the original one, which was used in earlier studies by some of us \cite{racko_molecular_2017} and designed
to quantitatively reproduce the topological aspects of DNA \cite{benedetti_introducing_2019,10.1093/nar/gky1091,biology10020130}. The implementation of the model we have at hand (in a non-maintained version ESPResSo-3.1 \cite{arnold_espresso_2013, limbach_espressoextensible_2006}) appears slower than the implementation of the model in \refsec{esi:sec:model:details}, presumably because of the larger number of particles and older software of the former, but no direct detailed comparison of the performance was done.
Although the current model \cite{Brackley2014a} appears %, is 
more efficient,
it was not tuned to quantitatively emulate all of the properties of DNA. 
The point of comparison was %mostly technical -- we wanted 
to demonstrate, that the observed phenomena are robust to the used model and they agree with the previously published results.

In the %\sout{alternative} 
original model, a monomeric unit, $i$, is composed of a single bead with $s = 2.5\ \mathrm{nm}$, and five virtual sites, labeled as ${p_{1i}, p_{2i}, p_{3i}, p_{4i}, a_i}$.
All of the virtual sites lie on the same plane, having positions $\{p_{1i}: [-x,0]; p_{2i}: [x,0]; p_{3i}: [0,-x]; p_{4i}: [0,x]; a_i: [0,0]\}$ and this plane is placed perpendicularly to the bond vector connecting bead $i$ and $i+1$, where virtual beads form a cross with the arm length $x=0.9s$ (see below the potential details). 
Beads of consecutive monomeric units are bonded with a harmonic bond
\begin{equation}
    U_{\mathrm{h}}(r) = \dfrac{1}{2} K_{\mathrm{h}} (r - r_0)^2,
\end{equation}
with $K_{\mathrm{h}} = 100\kT/s^{2}$ and $r_0 = s$. %according to the tcl script
The non-bonded interactions between beads of any two monomeric units are modeled using WCA with the form Eq.~\eqref{eq:model:wca} with $\varepsilon = \kT$, while virtual sites have none non-bonded interactions.
The polymer bending is emulated with harmonic potential:
\begin{equation}
    U_{\mathrm{b}}(r) = \dfrac{1}{2} K_{\mathrm{b}} (\theta - \theta_0)^2,
\end{equation}
with $K_{\mathrm{b}} = 20\kT $ and $\theta_0 = \pi$, and $\theta$ being the angle between two consequent bond vectors, all together yielding the persistence length of $l_\mathrm{p} \approx 50 \ \mathrm{nm}$.
To model the torsion, we utilize harmonic dihedral potentials
\begin{equation}
    U_{\mathrm{d}}(r) = \dfrac{1}{2} K_{\mathrm{d}} (\phi - \phi_0)^2,
\end{equation}
with $K_{\mathrm{d}} = 25\kT$, %according to tcl script
 $\phi_0 = 0$ %according to biology10020130
and $\phi$ being the angle between two planes -- one defined by the triplet of particles $\{ a_i, a_{i+1}, p_{1i} \}$, the other by the triplet $\{ a_i, a_{i+1}, p_{1i+1} \}$ (see also the model scheme in \cite{racko_molecular_2017}). 
The choice of $K_{\mathrm{d}}$ should result to the mean excess writhe-to-twist ratio of $\approx 7:3$ \cite{Bednar1994a}. 
Finally, the inner geometry of the virtual sites is fixed by a series of harmonic potentials.
First a bonding potential fixing the arm length of the cross, having the stiffness $80\kT$ %$1000\kT$ 
with the potential well minimum at $0.9s$ %$\sigma/2$ 
applied between pairs $(a_i, p_{1i}), (a_i, p_{2i}), (a_i, p_{3i}), (a_i, p_{4i})$. 
Second a bonding potential fixing the position of the virtual sites relative the monomeric units, having the stiffness $80\kT$ %according to tcl script
with the potential well minimum at $s/2$ applied between pairs $(i,a_i), (i+1,a_i)$. 
Third, a bending angular potential controlling the relative orientation of the neighboring virtual sites, having the stiffness $50\kT$ %according to tcl script
with the potential well minimum at $\pi/2$ applied at triplets $(p_{1i},a_i,p_{3i},), (p_{1i},a_i,p_{4i},), (p_{2i},a_i,p_{3i},), (p_{2i},a_i,p_{4i})$. 
Finally, a bending angular potential for keeping the virtual sites in the single plane, having the stiffness 
$50\kT$ %according to tcl script%$800\kT$ 
with the potential well minimum at $\pi$ applied at triplets $(p_{1i},a_i,p_{2i}), (p_{3i},a_i,p_{4i})$.

To simulate the effect of the supercoiling agent, we remove the dihedral potential from a pair of consecutive monomeric units, 
or rather their virtual sites, and instead apply a constant torque $TQ$ in the form of constant force $F$ acting on the cross arms, rotating them around the axis of the bond

We used systems with $M = 50$ rings of length $N = 400$ ($\approx 3\mathrm{kbp}$) 
in a cubic box with periodic boundary conditions at monomer density $\rho\sigma^3 = 0.08$. 
We use the ESPResSo-3.1 implementation of Langevin dynamics as formulated in Eq.~\eqref{esi:eq:langevin}, but with $\Delta t = 0.0012 \tau$ and $\gamma\tau = 3.9$, and with total length of $\sim 10^5 \tau$ just slightly above the Rouse time of a single ring. 

%%%%%%%%%%%%%%%%%%%%%%%%%%%%%%%%%%%%%%%%%%%%%%%%%%
%\subsection{A Brief Comparison of the Models}
%\label{esi:sec:comparison}
%%%%%%%%%%%%%%%%%%%%%%%%%%%%%%%%%%%%%%%%%%%%%%%%%%

\begin{figure}
    \centering
    \includegraphics[scale=0.90]{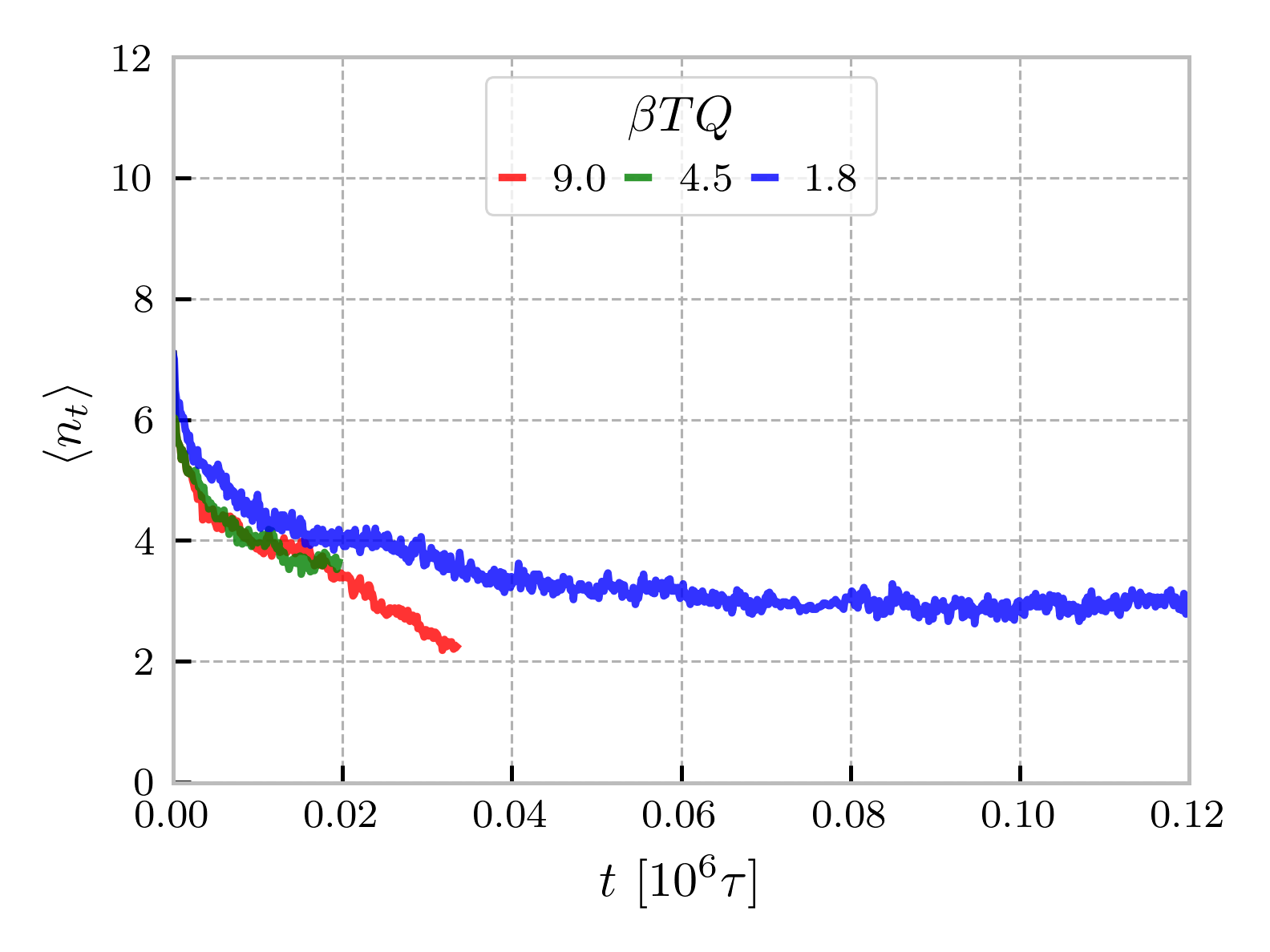}
    \caption{
        Time series of the number of threadings per ring for different torques for the original model from Section~\ref{esi:sec:model:alternative}.
        We use $\beta TQ = Fl$, where $F$ is the input force of the model and $l = 0.9s$ is the arm length, as described in Section~\ref{esi:sec:model:alternative}.
    }
    \label{esi:fig:comparison}
\end{figure}

In \reffig{esi:fig:comparison} we show the number of threadings in time for the original 
model. 
The comparison with the \reffig{fig:threading} in the main text shows that the observed phenomena (two regimes of the $n_{t}$ decay) are universal and robust to the choice of a model.
The small quantitative disparities (somewhat smaller $n_{t}$) are caused by the smaller system size, where a ring could in principle interact and entangle with itself through the boundary conditions. Such an events decrease the measured $n_{t}$ because we do not detect self-threadings. 
Further differences between the current and the original model constitute the interaction potentials, namely different effective torsional stiffness and tilting potential combined with presence of additional degrees of freedom in the original 
model -- patchy monomeric unit vs. rigid bodies with inscribed directionality. 
Note that the longest simulations in \reffig{esi:fig:comparison} are only of the order of the Rouse time of the rings. Although longer simulations are  possible in principle, we have not used the original model for production simulations. 

%%%%%%%%%%%%%%%%%%%%%%%%%%%%%%%%%%%%%%%%%%%%%%%%%%
%%%%%%%%%%%%%%%%%%%%%%%%%%%%%%%%%%%%%%%%%%%%%%%%%%
%%%%%%%%%%%%%%%%%%%%%%%%%%%%%%%%%%%%%%%%%%%%%%%%%%
\clearpage
\section{Additional Simulation Results}
\label{esi:sec:results}
%%%%%%%%%%%%%%%%%%%%%%%%%%%%%%%%%%%%%%%%%%%%%%%%%%
%%%%%%%%%%%%%%%%%%%%%%%%%%%%%%%%%%%%%%%%%%%%%%%%%%
%%%%%%%%%%%%%%%%%%%%%%%%%%%%%%%%%%%%%%%%%%%%%%%%%%

%%%%%%%%%%%%%%%%%%%%%%%%%%%%%%%%%%%%%%%%%%%%%%%%%%
\subsection{Shape Parameters}
\label{esi:sec:results:shape}
%%%%%%%%%%%%%%%%%%%%%%%%%%%%%%%%%%%%%%%%%%%%%%%%%%

\begin{figure}[htbp]
    \centering
    \includegraphics[scale = 0.90]{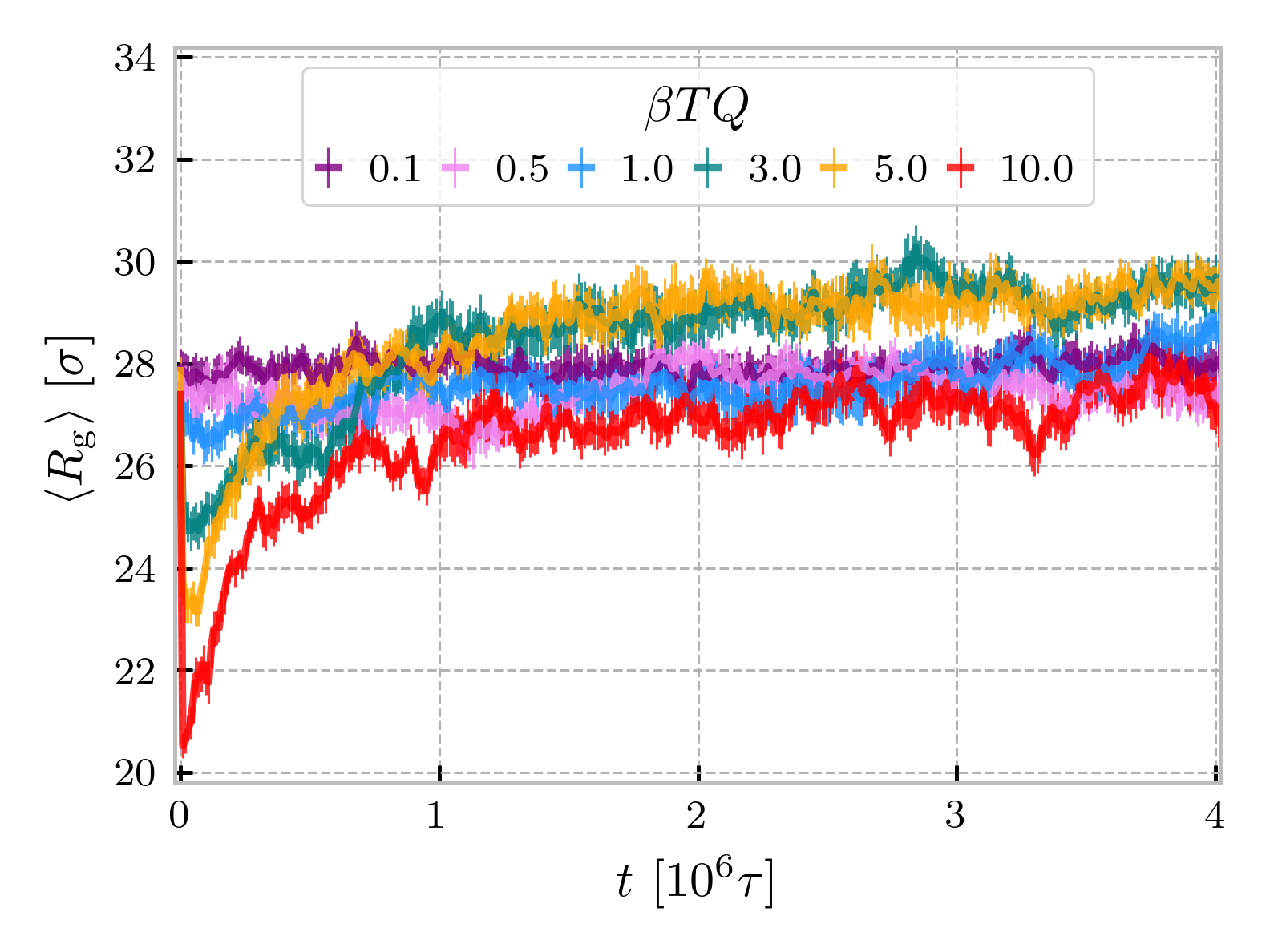}
    \includegraphics[scale = 0.90]{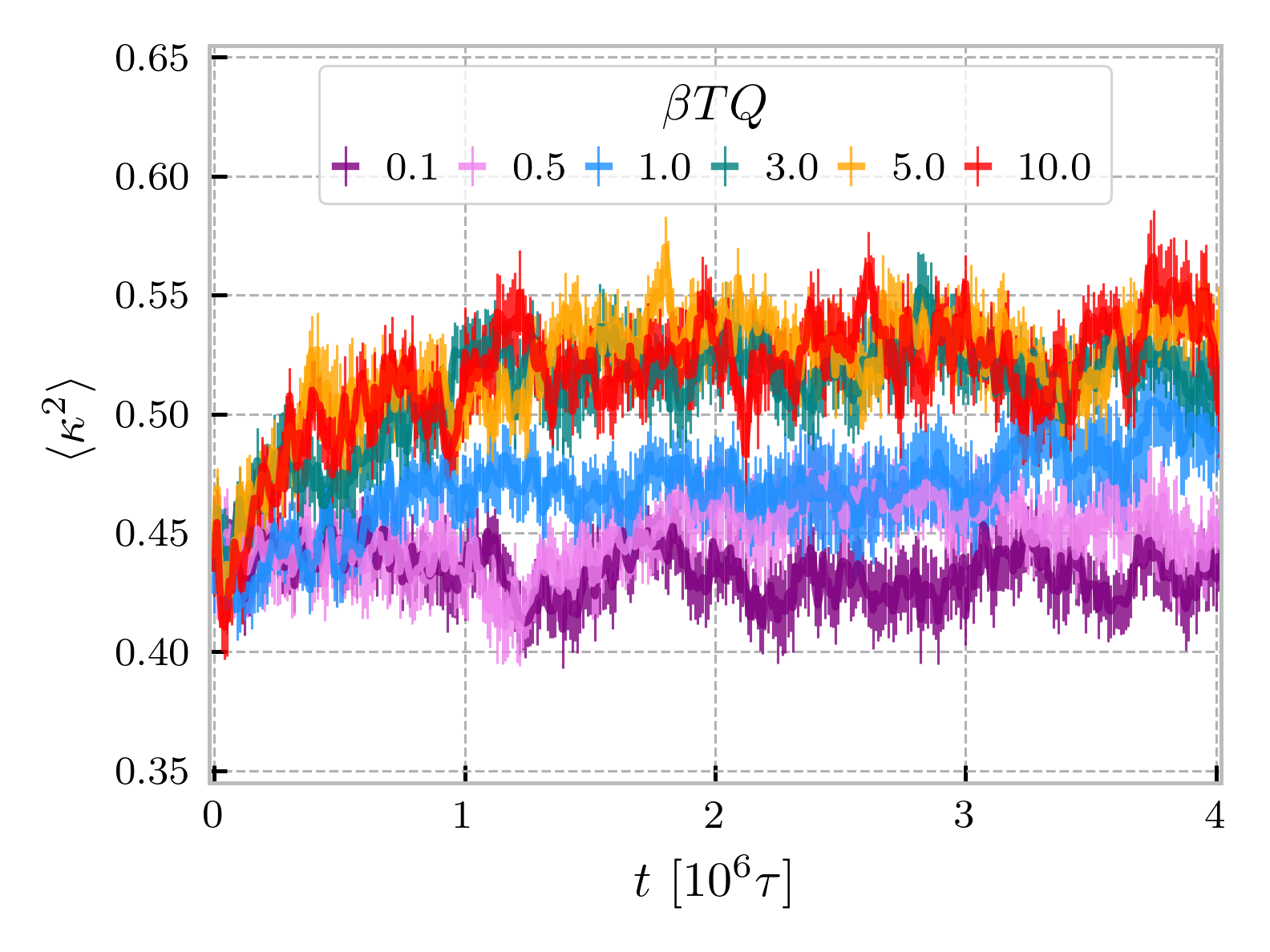}
    \caption{
    Time series of the mean radius of gyration $\langle R_{\rm g} \rangle$ 
    and the mean shape anisotropy $\langle \kappa^{2} \rangle$ averaged over all rings, plotted for different values of the active torque.
    The error bars are the error of the mean.
    }
    \label{esi:fig:results:shape}
\end{figure}

To describe the size and shape of the rings, we use the radius of gyration, $R_\mathrm{g}$ and relative shape anisotropy $\kappa^2$ respectively.
For each of the rings in a given time, we first construct the gyration tensor, $\mathcal{G}$ with $3\times 3$ components
\begin{equation}
    \mathcal{G}_{xy} = \dfrac{1}{2N^2} \sum_{i=1}^N \sum_{j=1}^N 
    (x_i - x_j) (y_i - y_j),
\end{equation}
where $x_i$ (or $x_j$) is the cartesian $x$ coordinate of the monomeric unit $i$ (or $j$), analogically defined for axes $[x,y,z]$.
Diagonalization of this gyration tensor yields three eigenvalues $\lambda_1^2 \leq \lambda_2^2 \leq \lambda_3^2$ such that $\mathrm{Tr}(G) = \lambda_1^2 + \lambda_2^2 + \lambda_3^2 = R_\mathrm{g}^2$.
We define the relative shape anisotropy \cite{Narros2013a} as
\begin{equation}
    \kappa^2 = \dfrac{3}{2} \dfrac{\lambda_1^4 + \lambda_2^4 + \lambda_3^4}{(\lambda_1^2 + \lambda_2^2 + \lambda_3^2)^2} -\dfrac{1}{2},
\end{equation}
which takes values between zero and one, with zero being attained only for spherically symmetric objects and one occurring only for linear objects.
In \reffig{esi:fig:results:shape} we then present expectations values $\langle \cdot \rangle$ of these single-ring properties, obtained by averaging over all of the rings in the system.

%%%%%%%%%%%%%%%%%%%%%%%%%%%%%%%%%%%%%%%%%%%%%%%%%%
\clearpage
\subsection{Branching Analysis}
\label{esi:sec:results:branching}
%%%%%%%%%%%%%%%%%%%%%%%%%%%%%%%%%%%%%%%%%%%%%%%%%%

\begin{figure}[htbp]
    \centering
    \includegraphics[scale = 0.90]{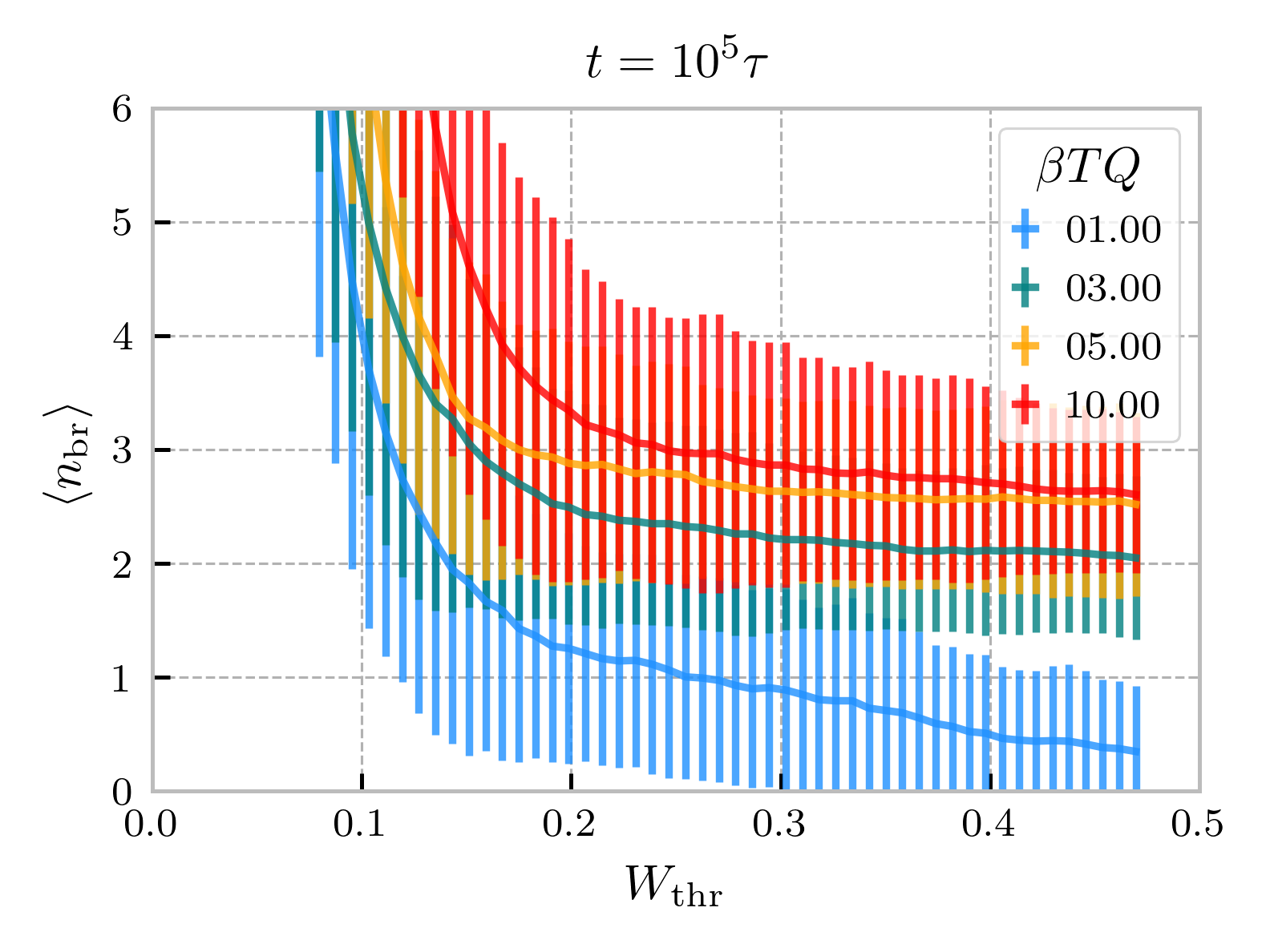}
    \includegraphics[scale = 0.90]{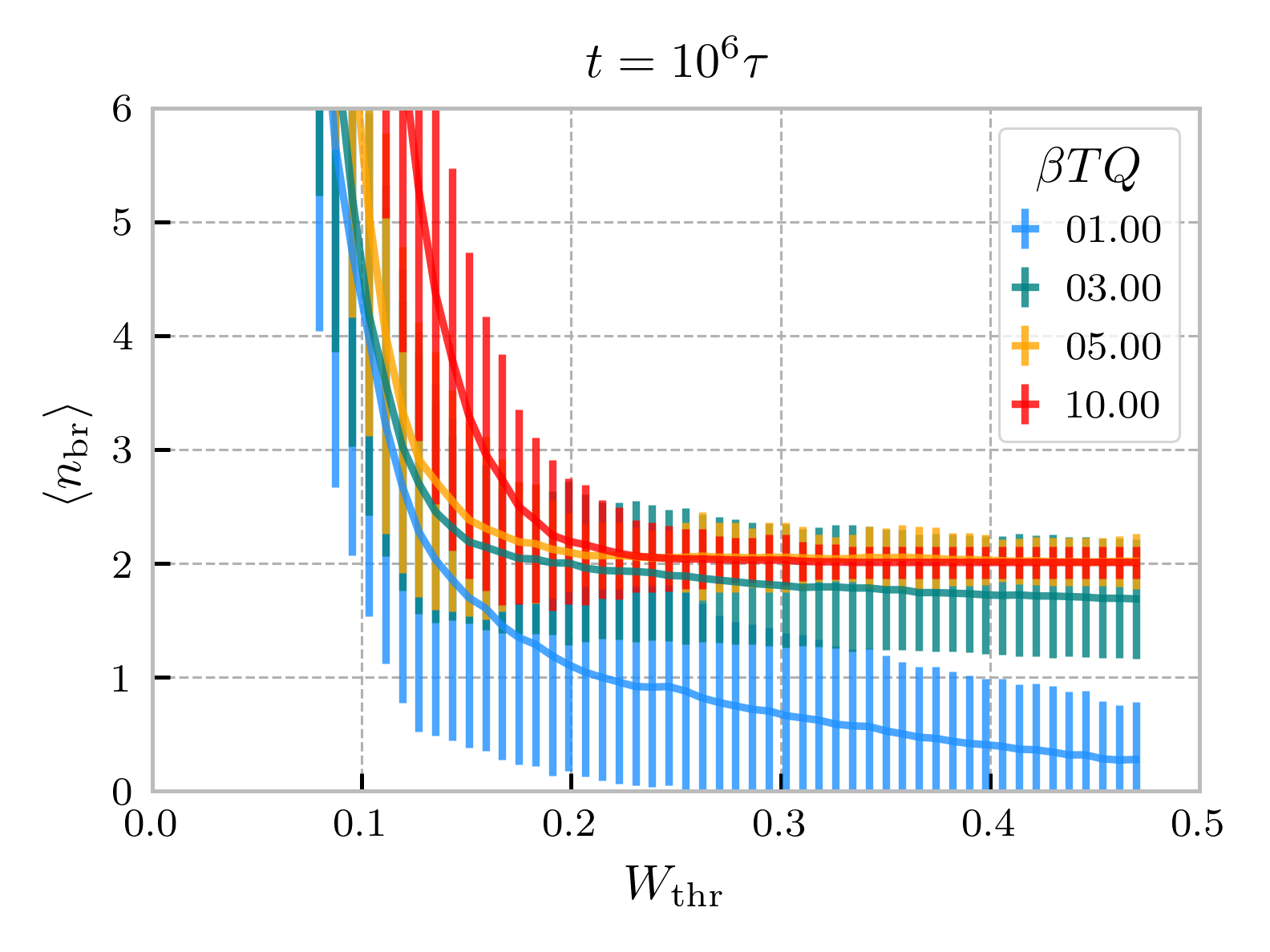}
    \caption{
    Mean number of branches as a function of the threshold value of segmental writhe as detailed in \refsec{esi:sec:model:threading}, plotted for different torques at two different times -- at the end of the initial stage of supercoiling tightening and during the slow aging regime respectively.
    The error bars are the standard deviation over the ensemble of rings.
    }
    \label{esi:fig:results:branching}
\end{figure}

\begin{figure}[htbp]
    \centering
    \includegraphics[scale = 0.90]{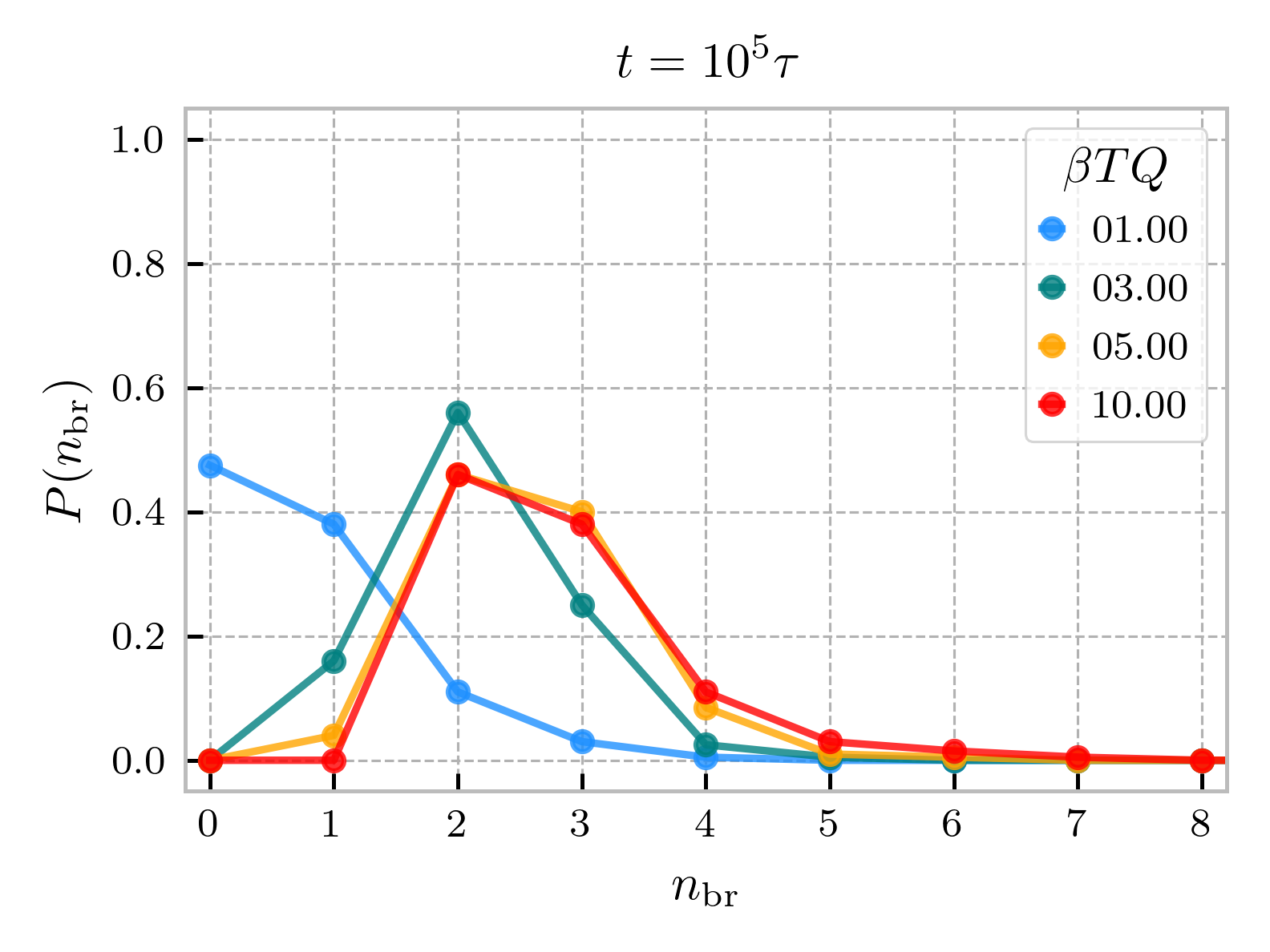}
    \includegraphics[scale = 0.90]{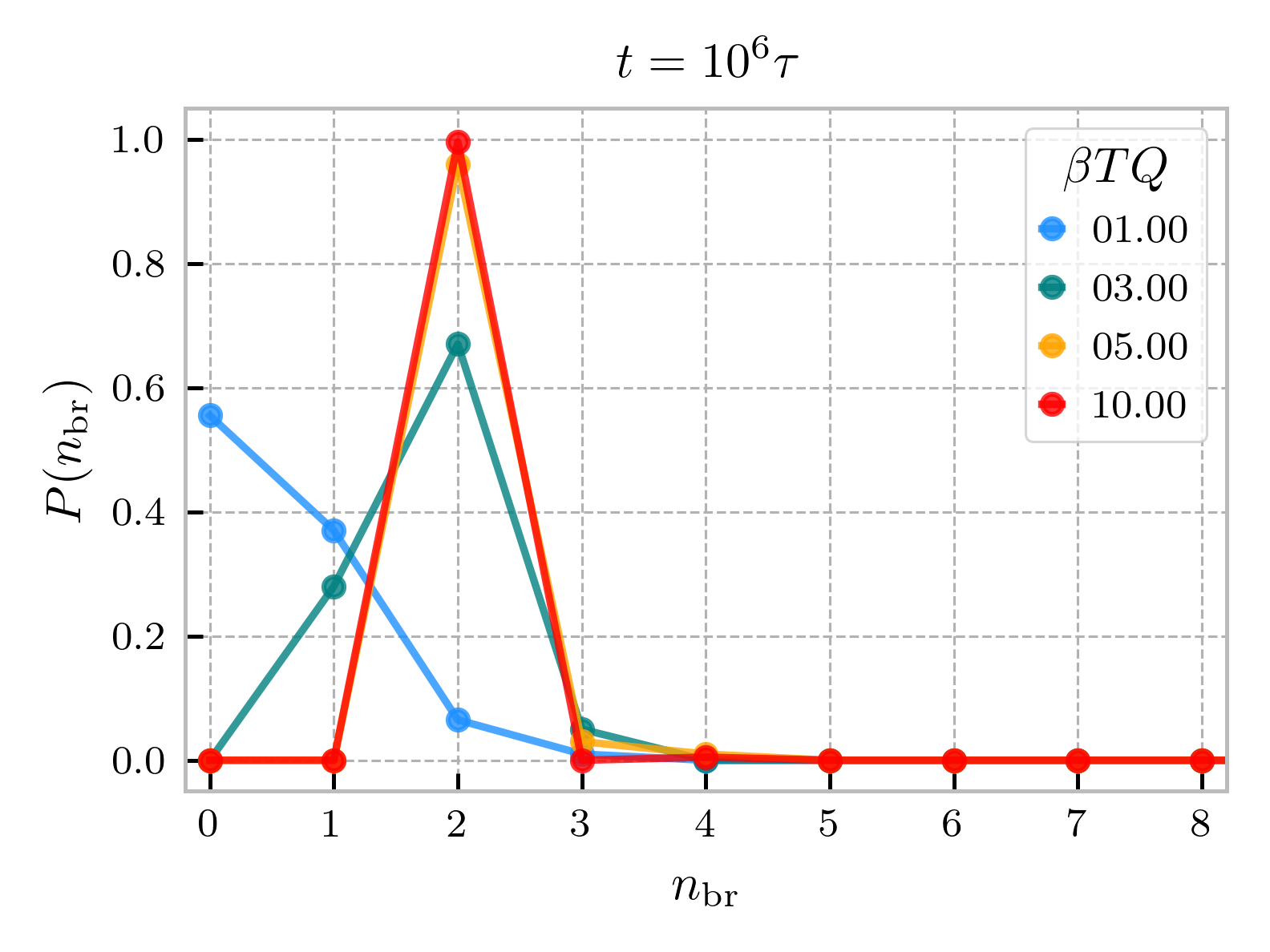}
    \caption{
    Normalized histograms of the number of branches for different torques at two different times -- at the end of the initial stage of supercoiling tightening and during the slow aging regime respectively.
    }
    \label{esi:fig:results:branching_histogram}
\end{figure}

\begin{figure}[htbp]
    \centering
    \includegraphics[scale=0.80]{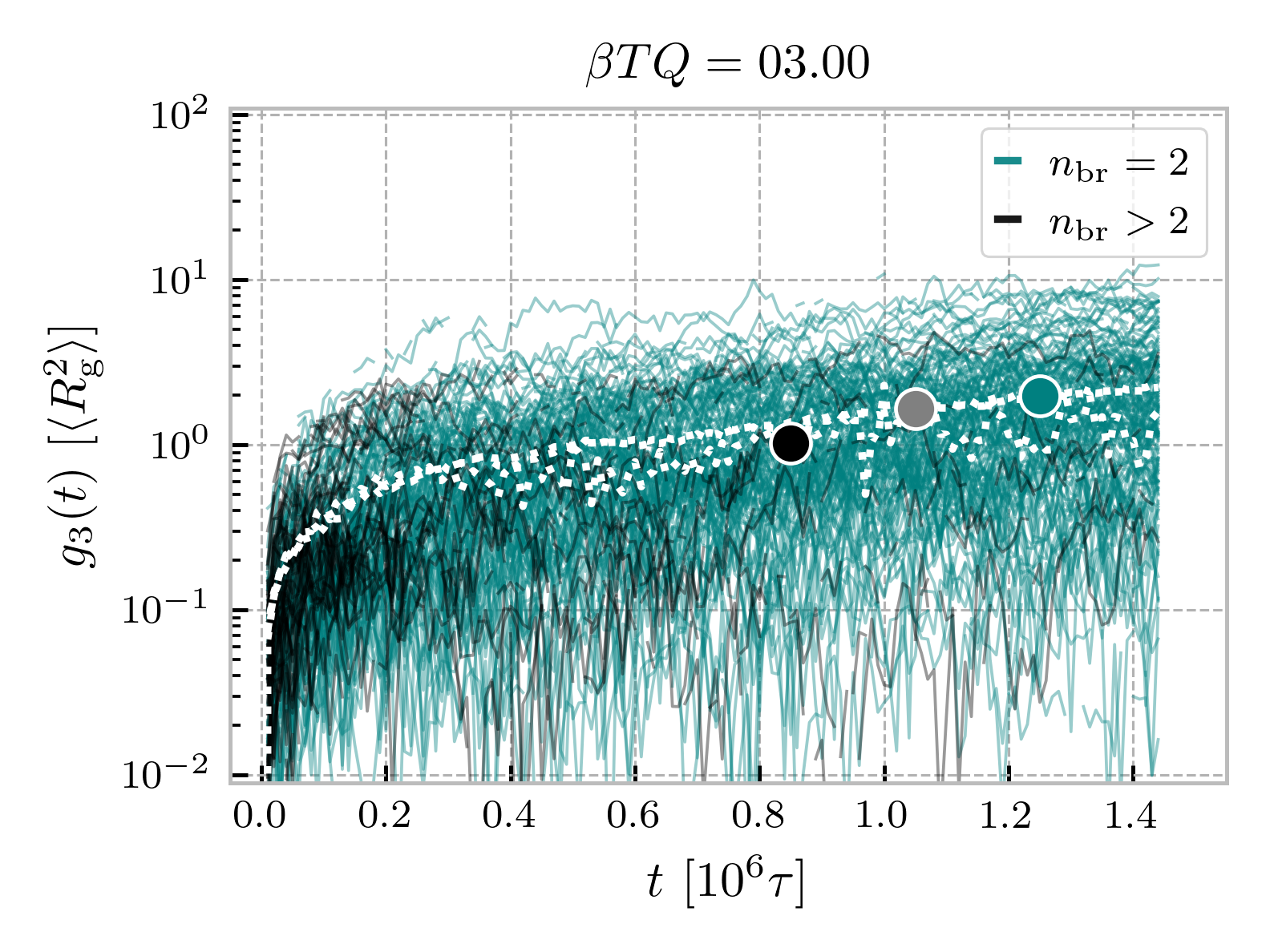} \hfill \includegraphics[scale=0.80]{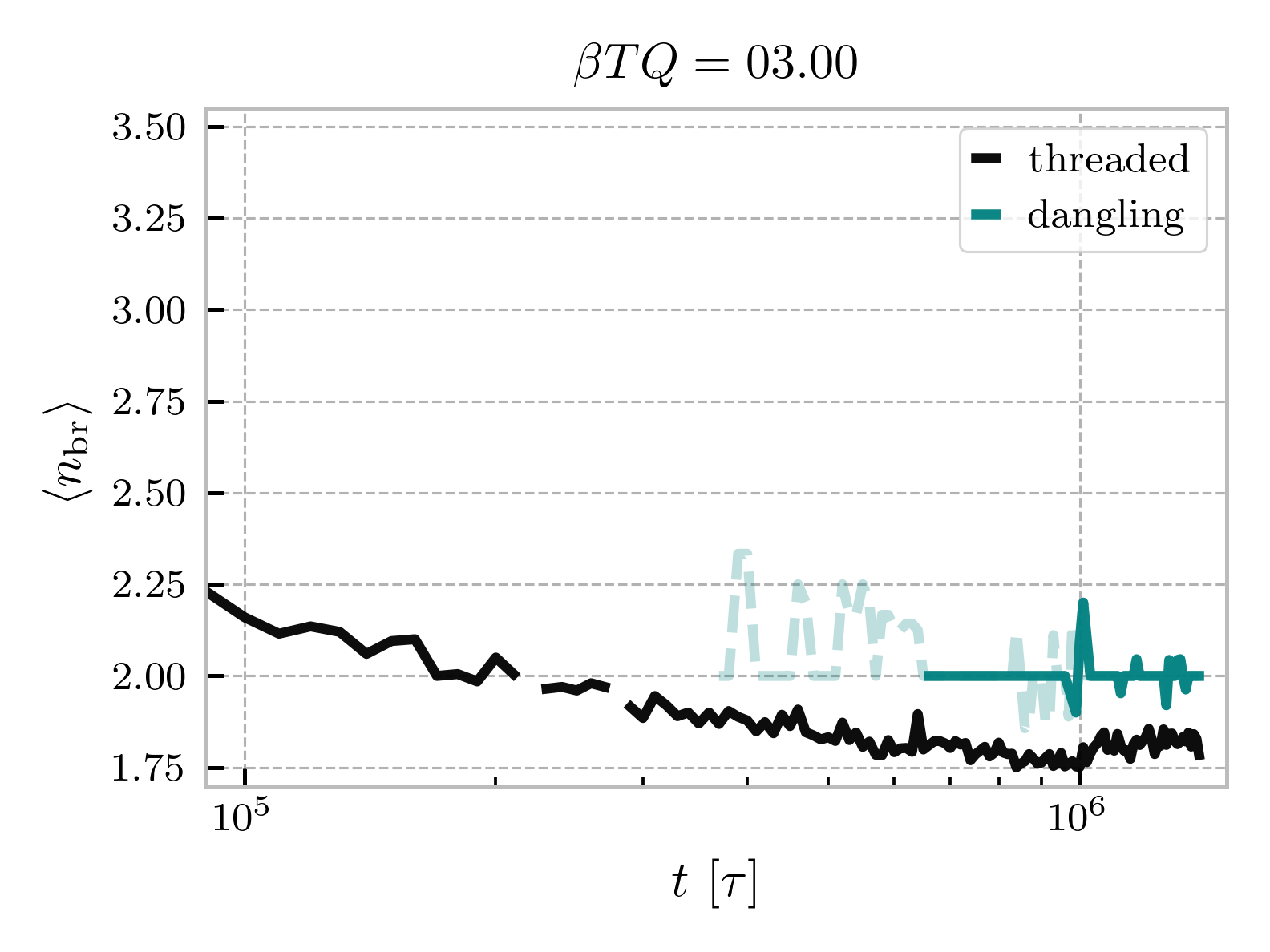} \\
    \includegraphics[scale=0.80]{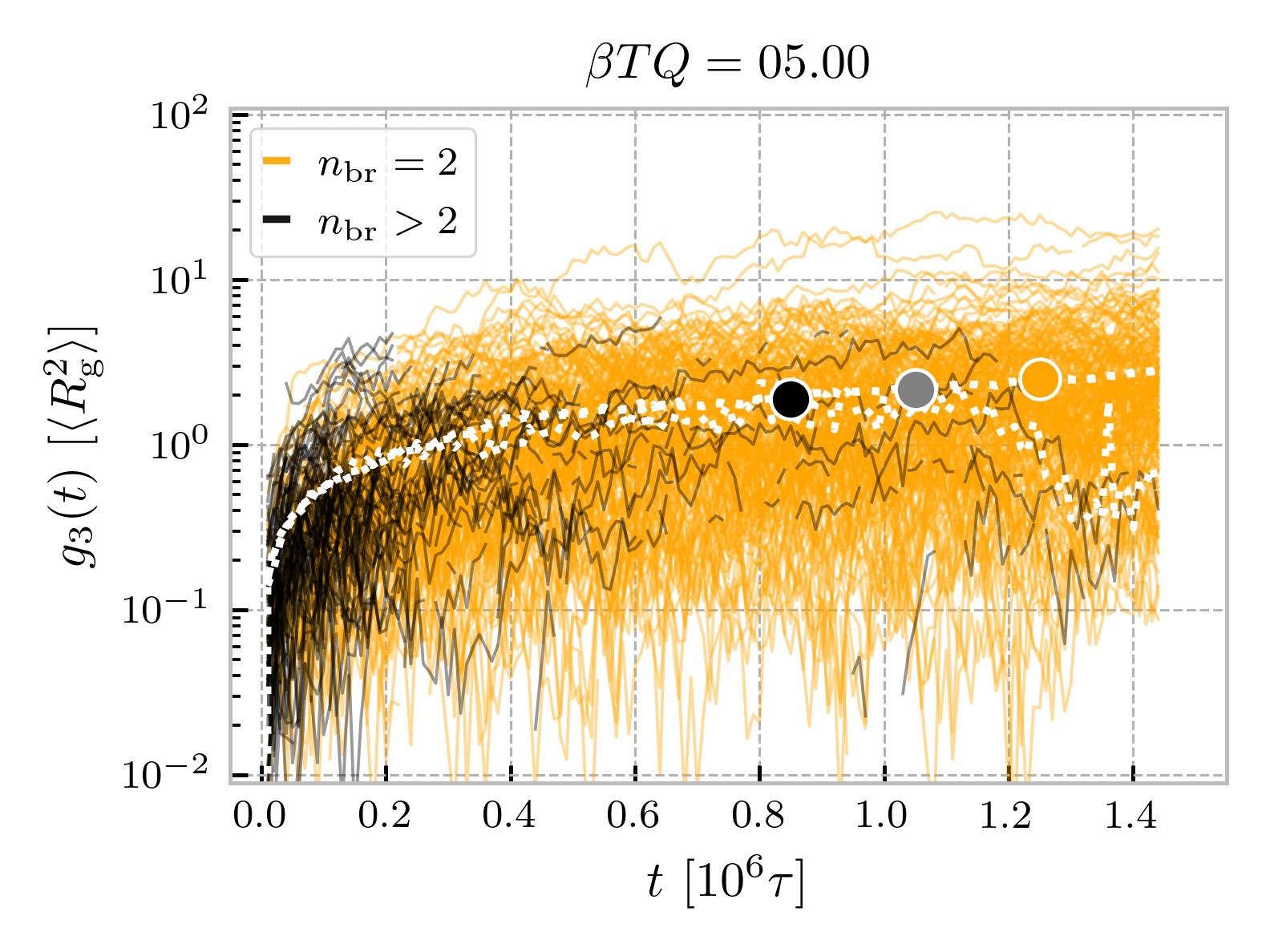} \hfill \includegraphics[scale=0.80]{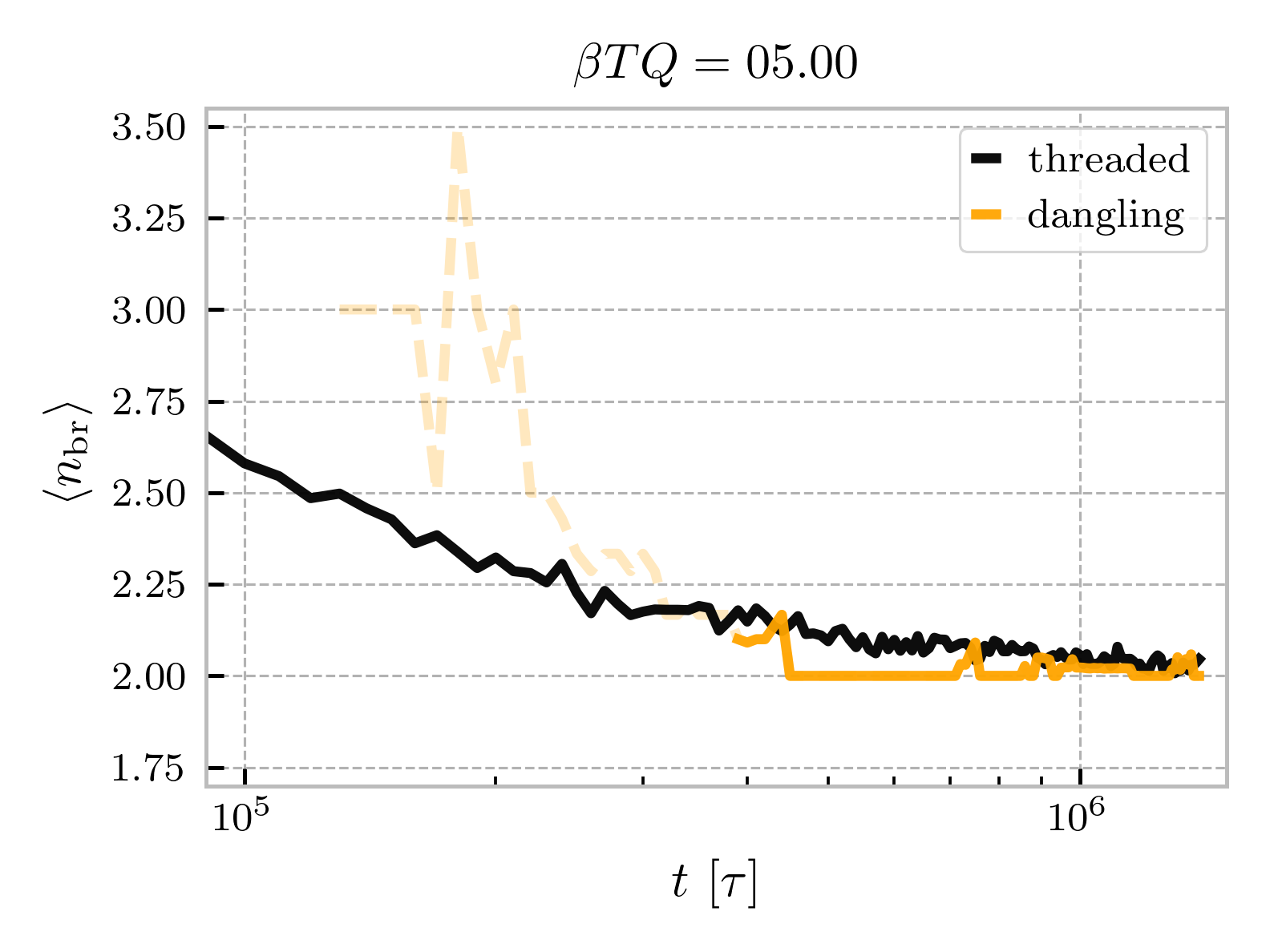} \\
    \includegraphics[scale=0.80]{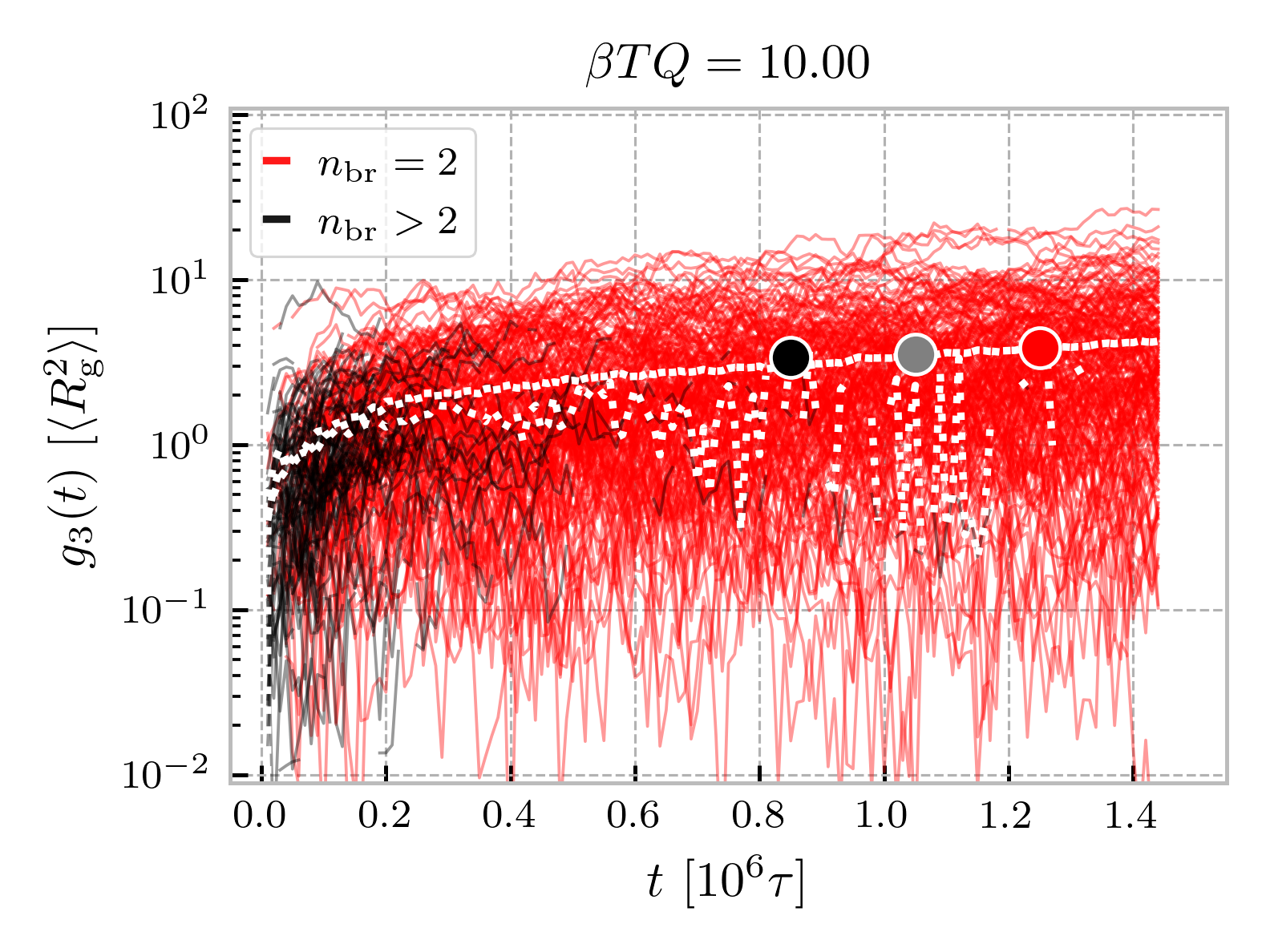} \hfill \includegraphics[scale=0.80]{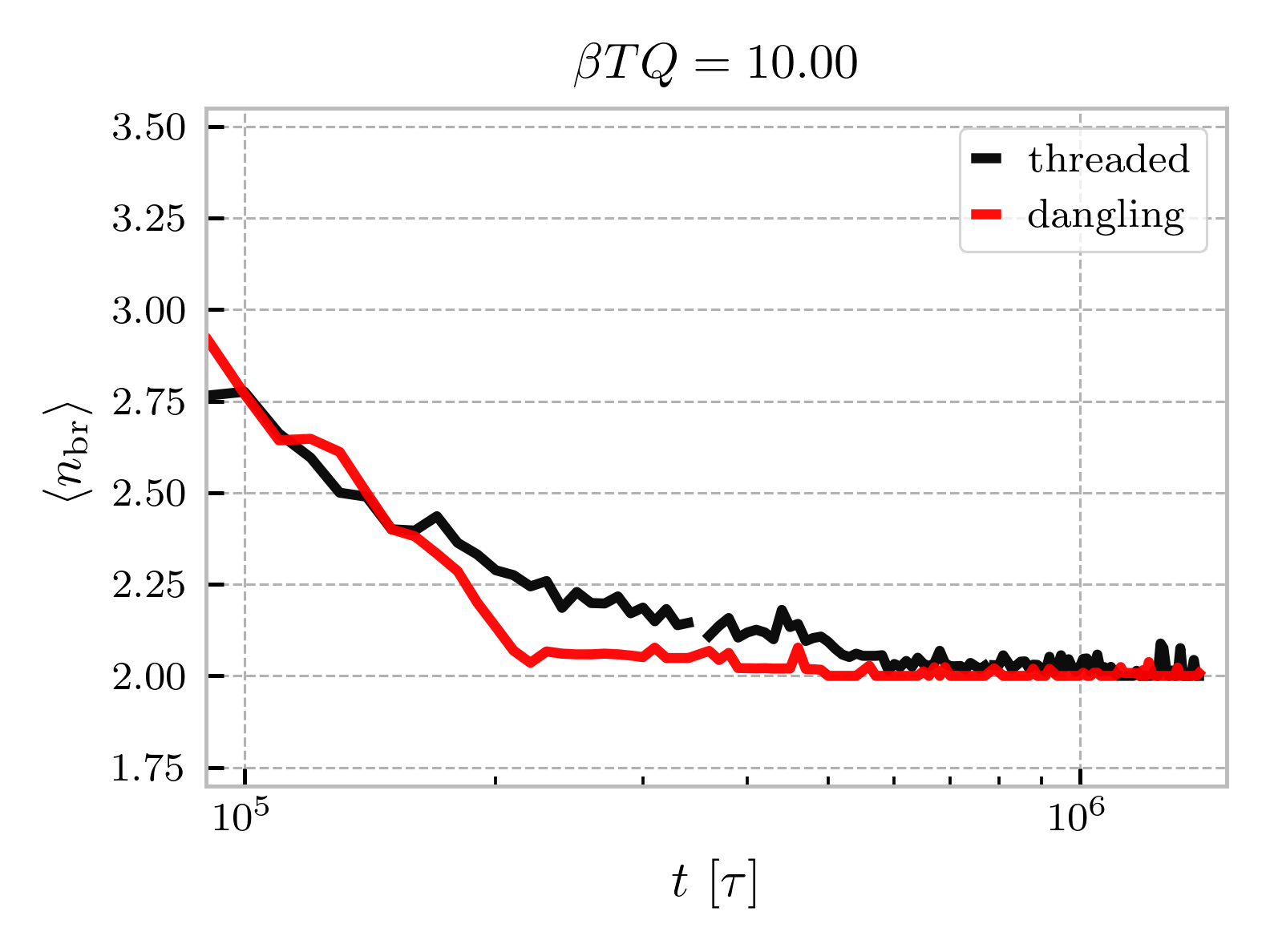} \\
    \caption{
    Left: The mean square displacement of all individual rings, with the point at time $t$ colored black if the ring has more than two branches. Three white dotted lines show means over three subpopulations of rings – over the rings with two branches, more than two branches and over all of the rings.
    Right: The mean branching averaged over two populations of rings -- threaded and dangling. Partially translucent dashed lines signify that less then ten rings contribute to the mean causing large fluctuations, solid full bodied line is considered to be statistically more significant. The few breaks in black solid lines are snapshots that were not analyzed due to threading detection method (see \refsec{esi:sec:model:threading}).
    }
    \label{esi:fig:results:g3_vs_branching}
\end{figure}

%%%%%%%%%%%%%%%%%%%%%%%%%%%%%%%%%%%%%%%%%%%%%%%%%%
\clearpage
\subsection{Separation Length Distributions}
\label{esi:sec:results:separation}
%%%%%%%%%%%%%%%%%%%%%%%%%%%%%%%%%%%%%%%%%%%%%%%%%%

\begin{figure*}[htbp]
    \centering
    \includegraphics[scale = 0.90]{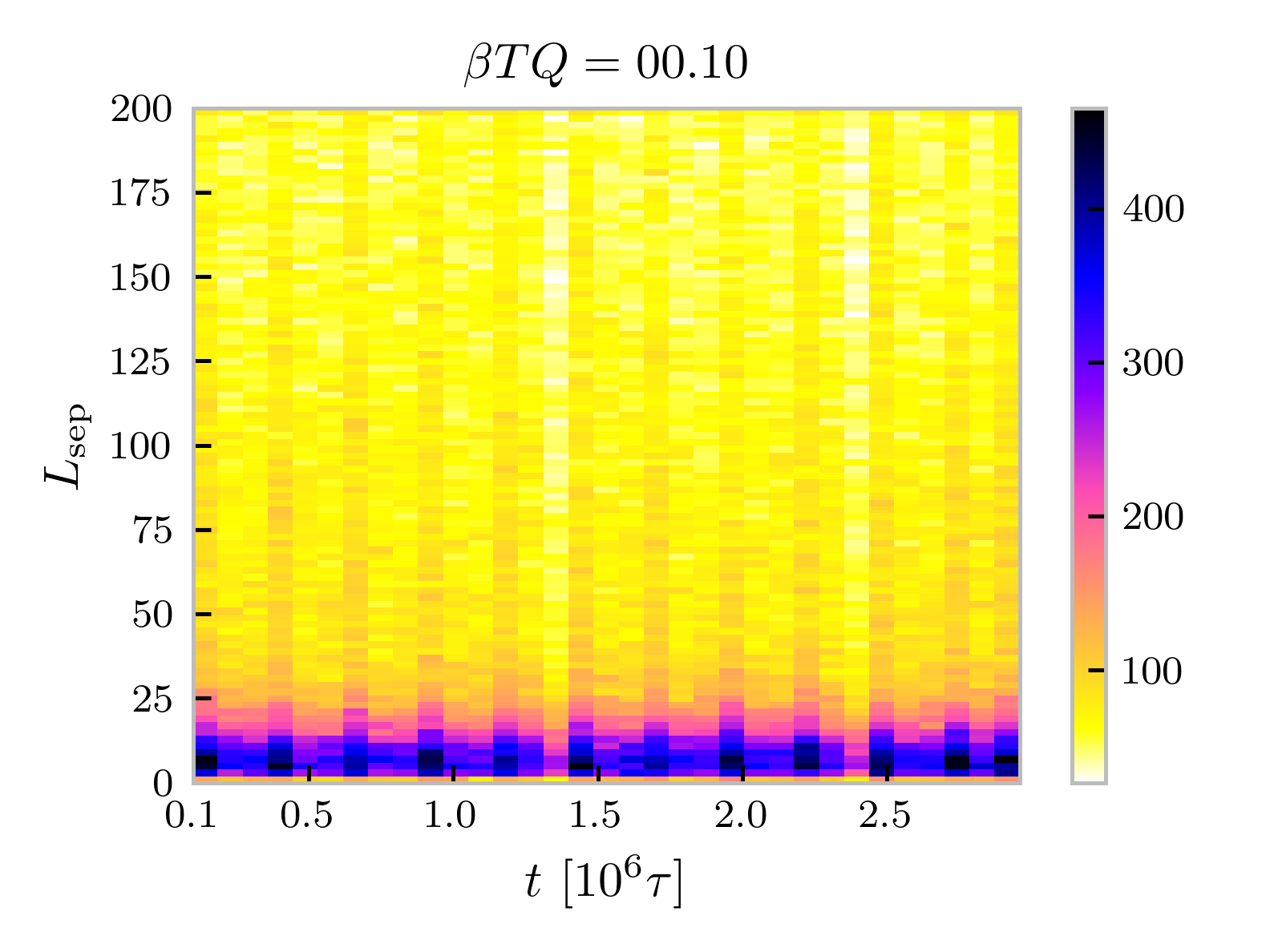}
    \includegraphics[scale = 0.90]{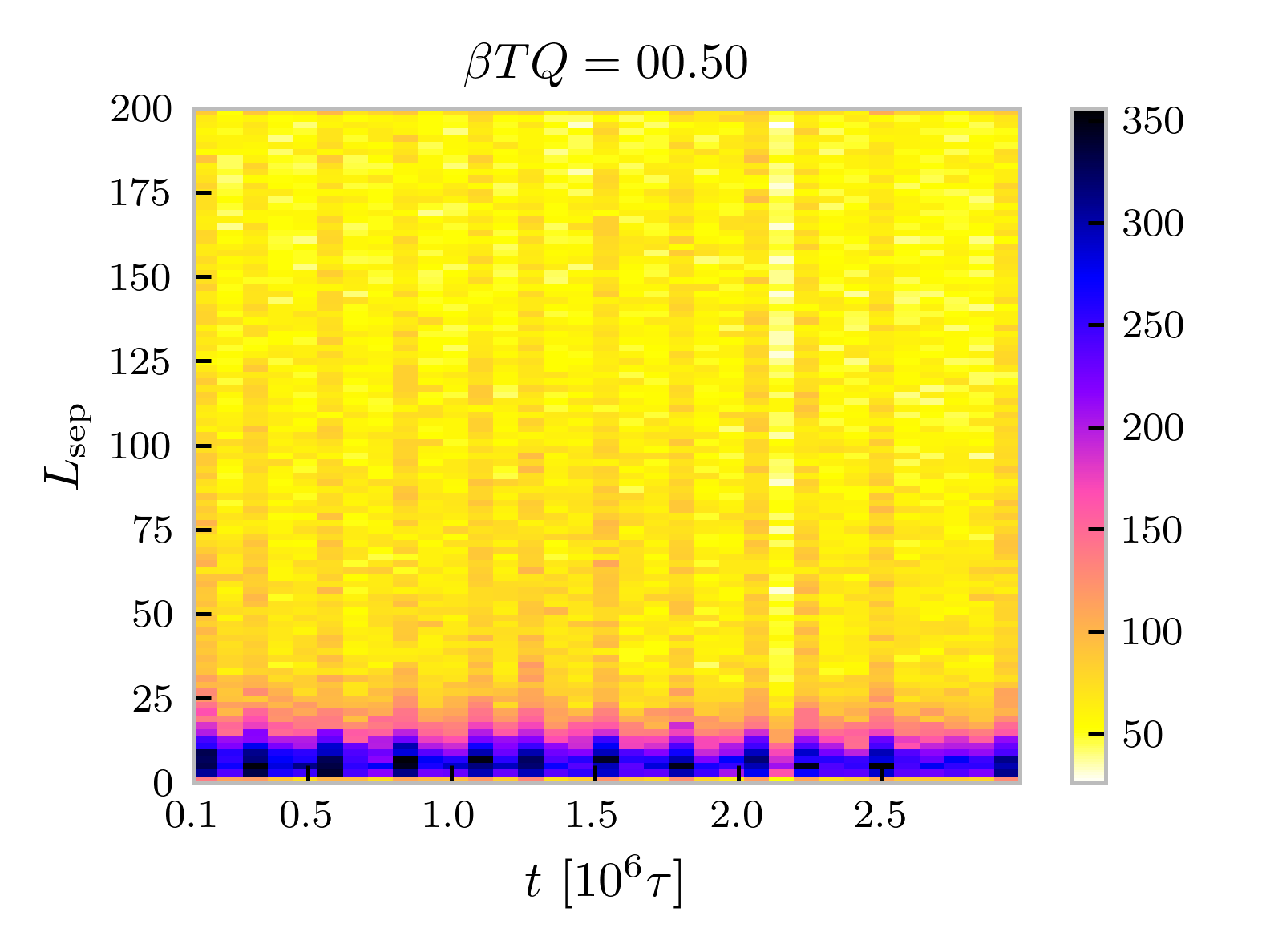}
    \includegraphics[scale = 0.90]{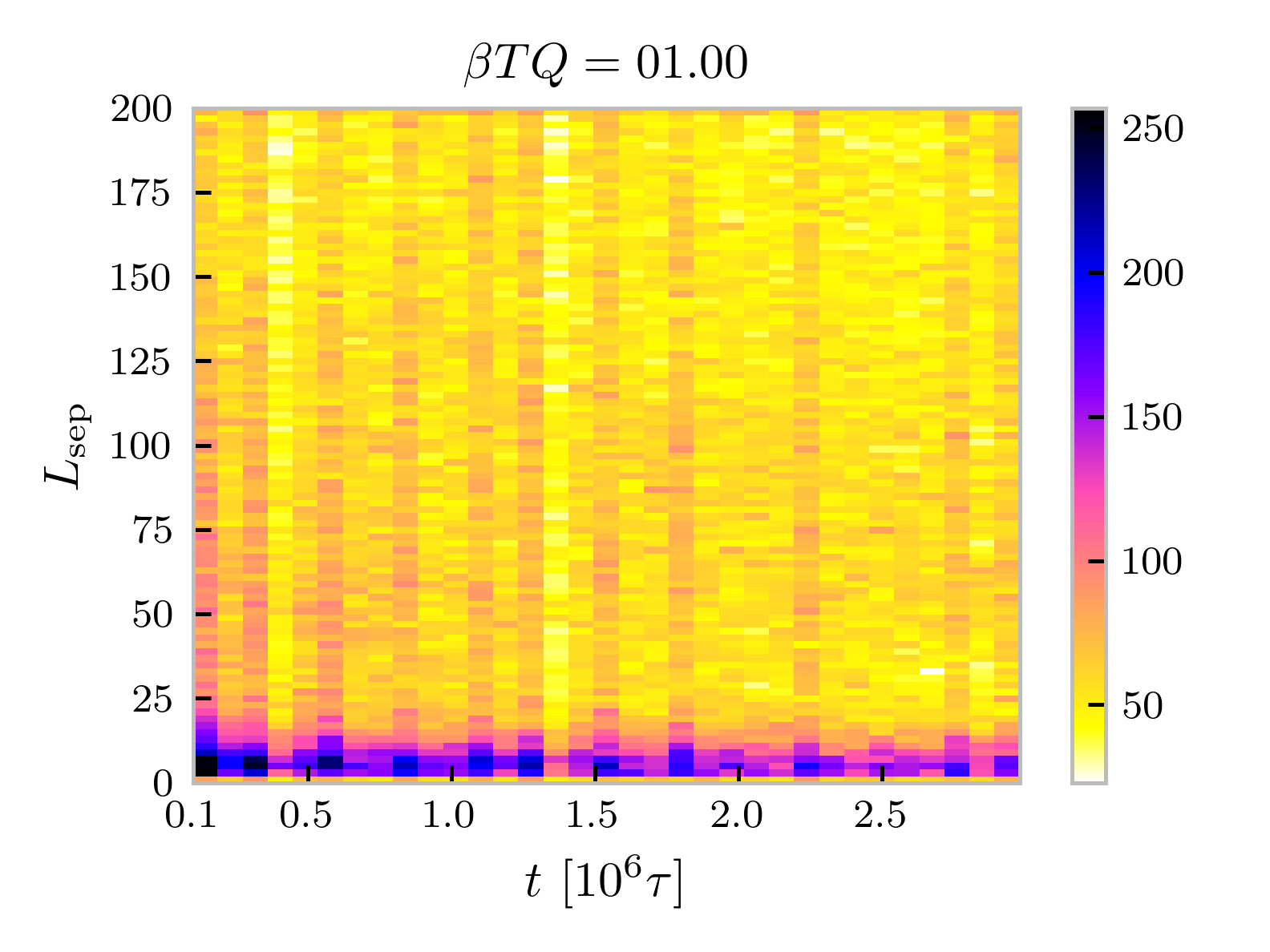}
    \includegraphics[scale = 0.90]{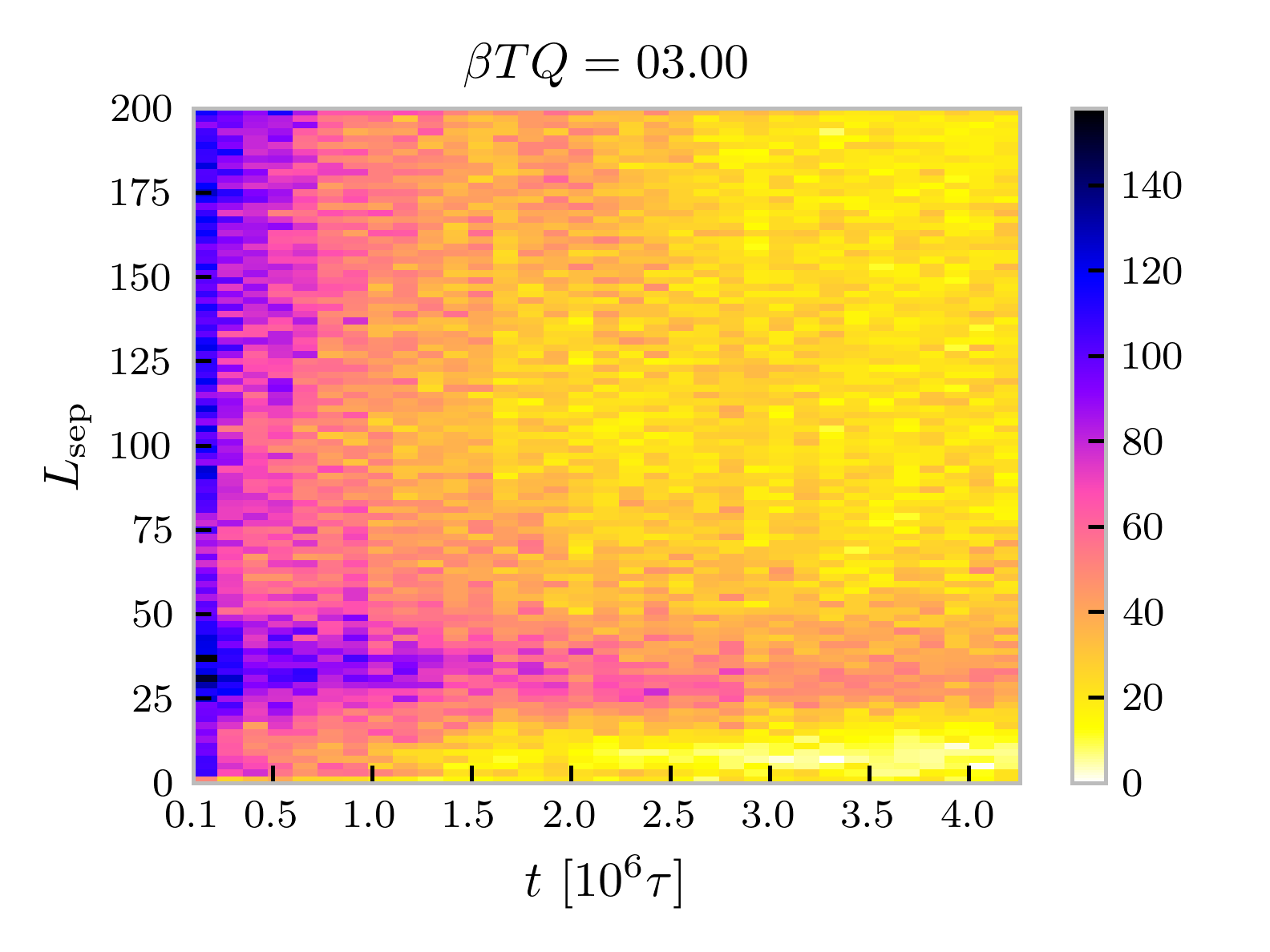}
    \includegraphics[scale = 0.90]{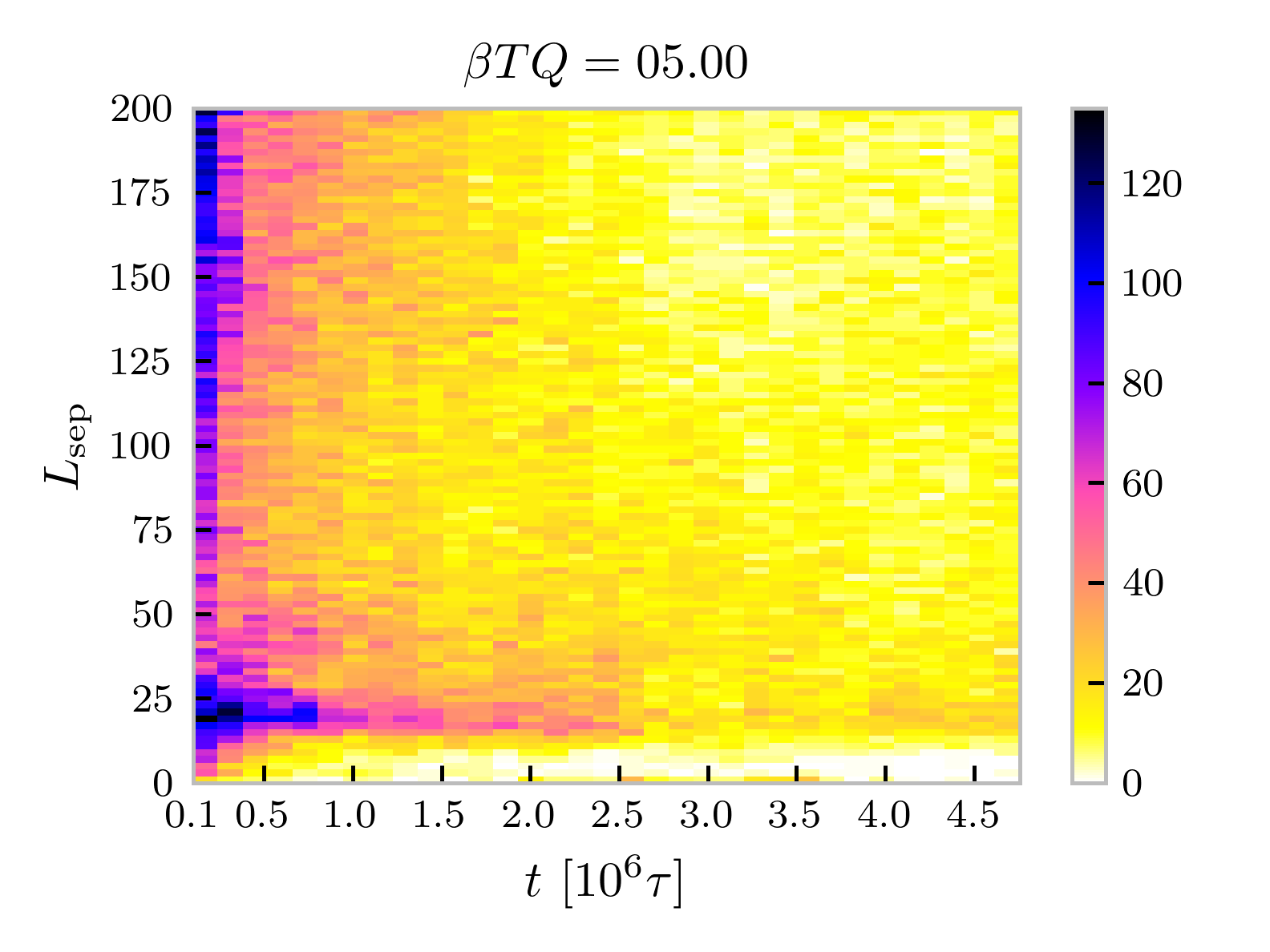}
    \includegraphics[scale = 0.90]{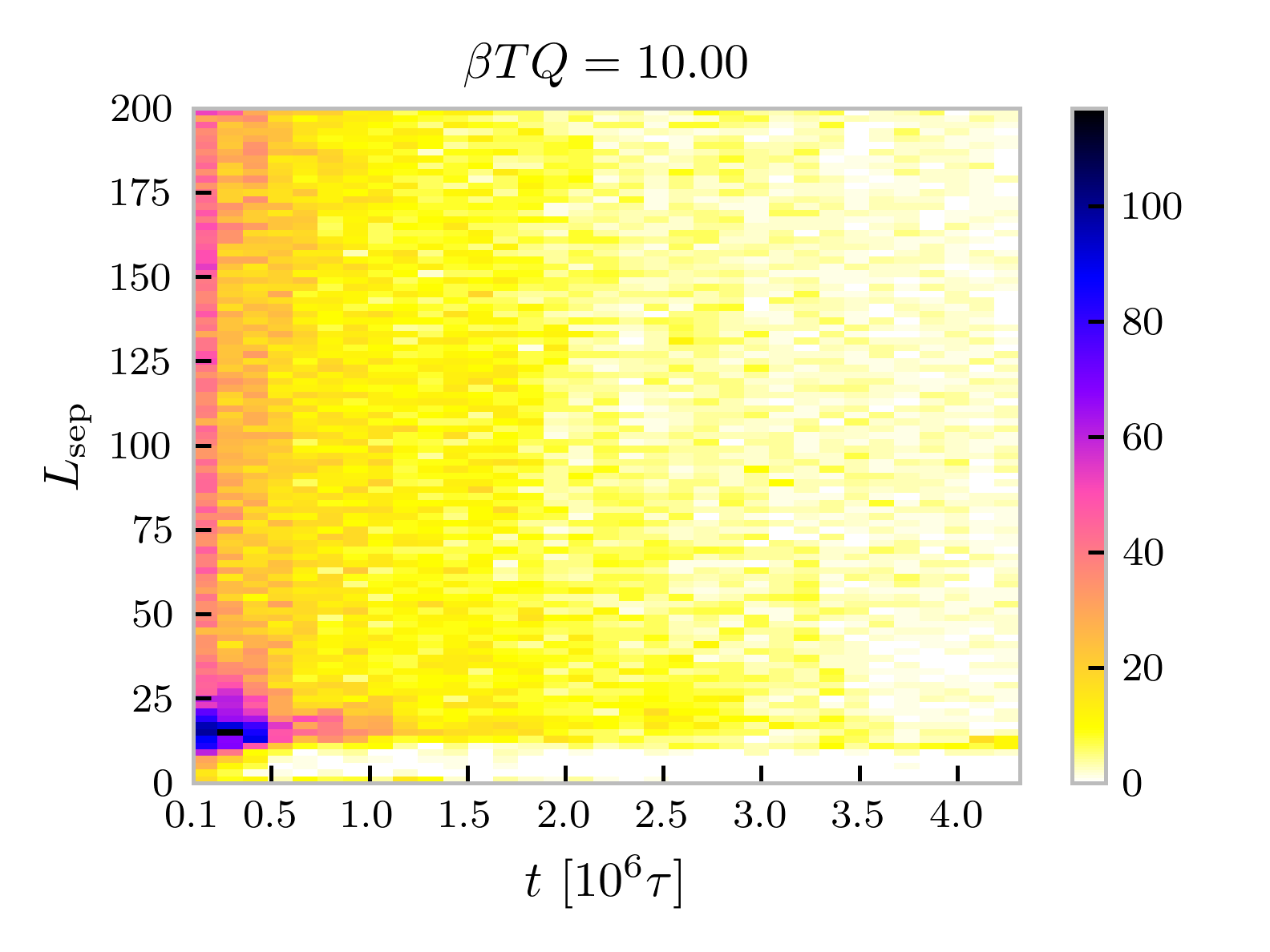}
    \caption{
    Two dimensional probability distributions of separation length, $L_\mathrm{sep}$, Eq.~\eqref{eq:Lsep}, in time.
    The color code corresponds to the number of threadings of a given separation length in the a given time.
    The time axis starts at $10^5\tau$, only after the initial stage of rapid supercoiling.
    }
    \label{esi:fig:results:lsep}
\end{figure*}

\begin{figure*}[htbp]
    \centering
    \includegraphics[scale = 0.90]{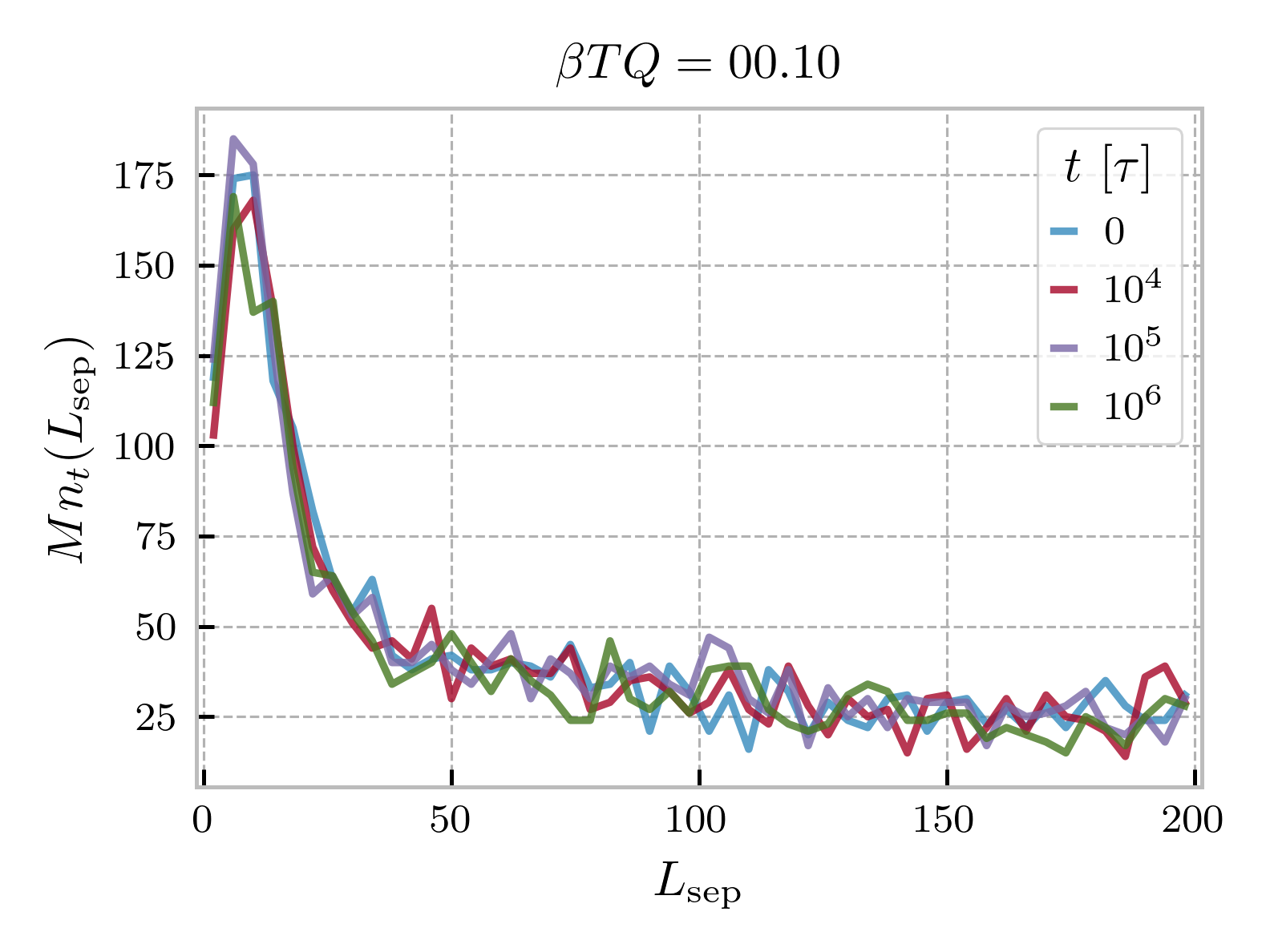}
    \includegraphics[scale = 0.90]{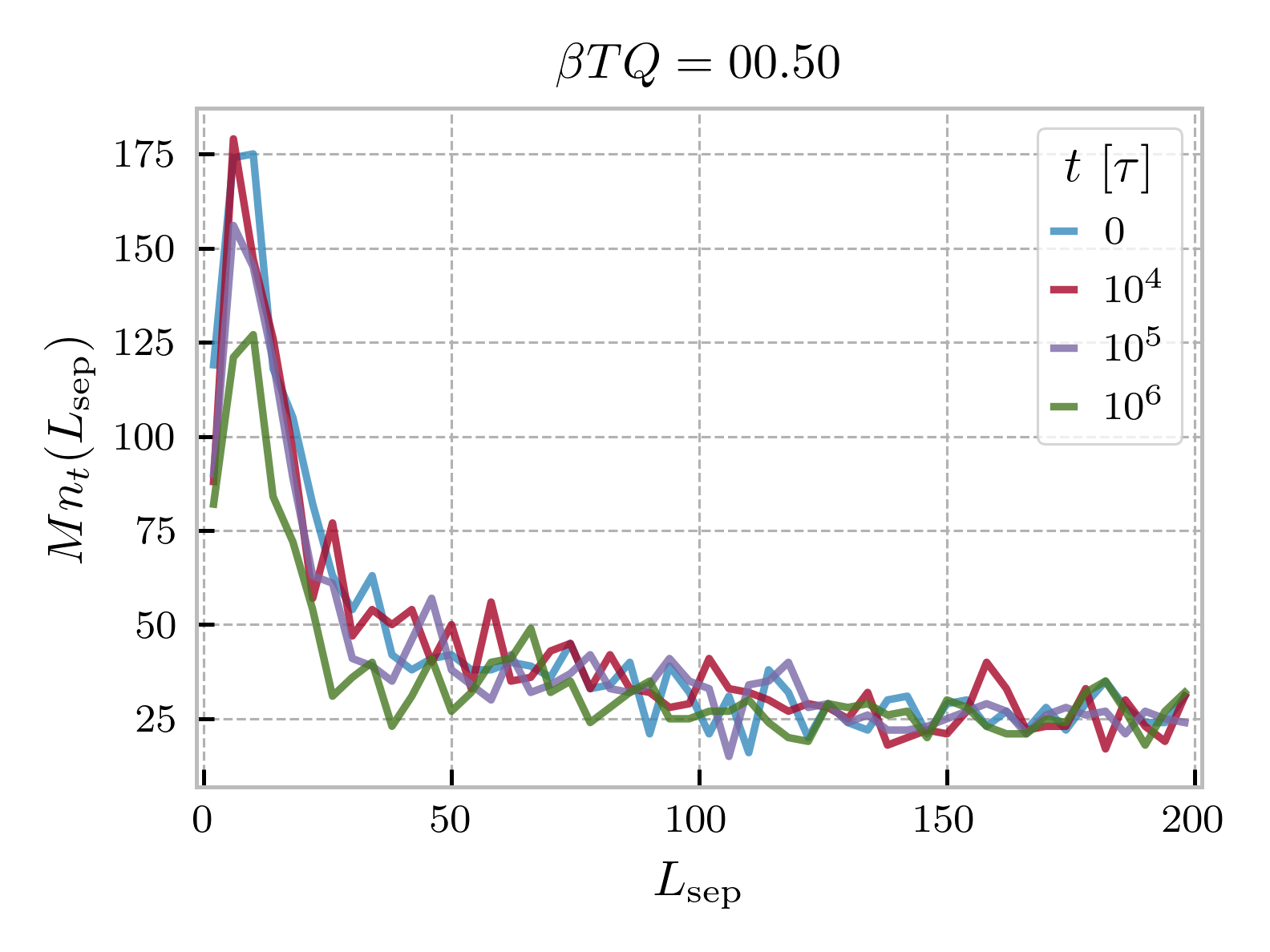}
    \includegraphics[scale = 0.90]{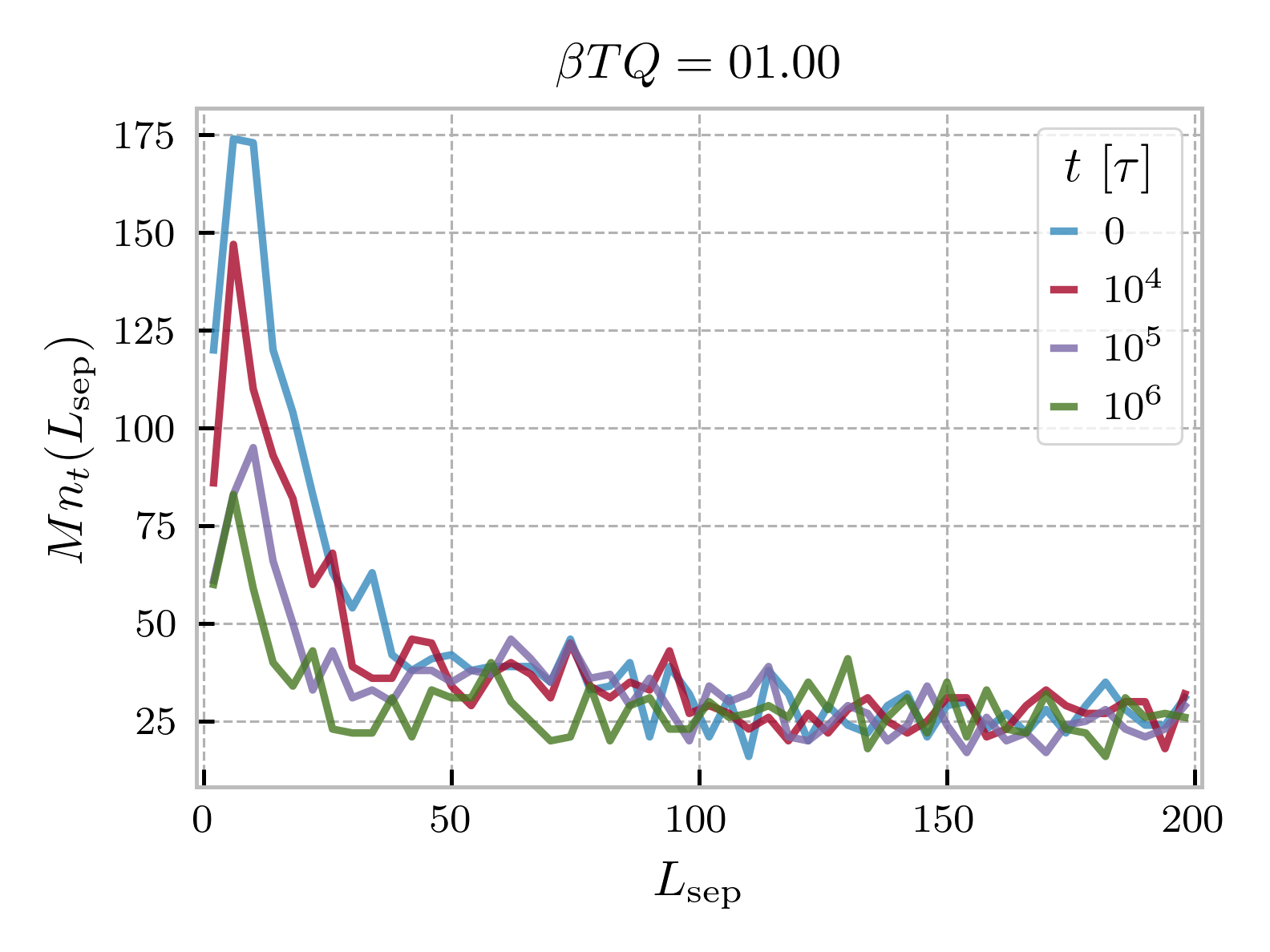}
    \includegraphics[scale = 0.90]{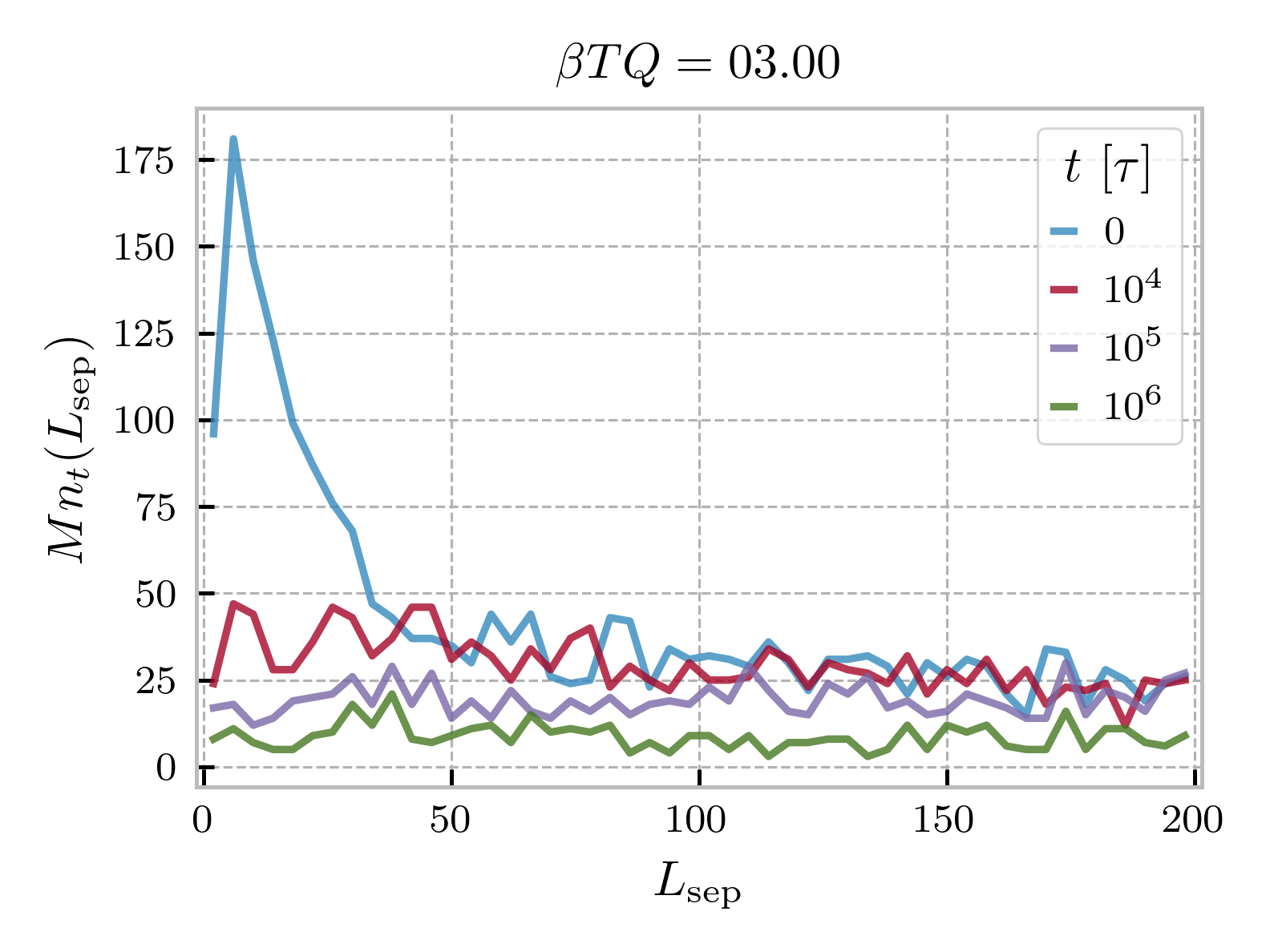}
    \includegraphics[scale = 0.90]{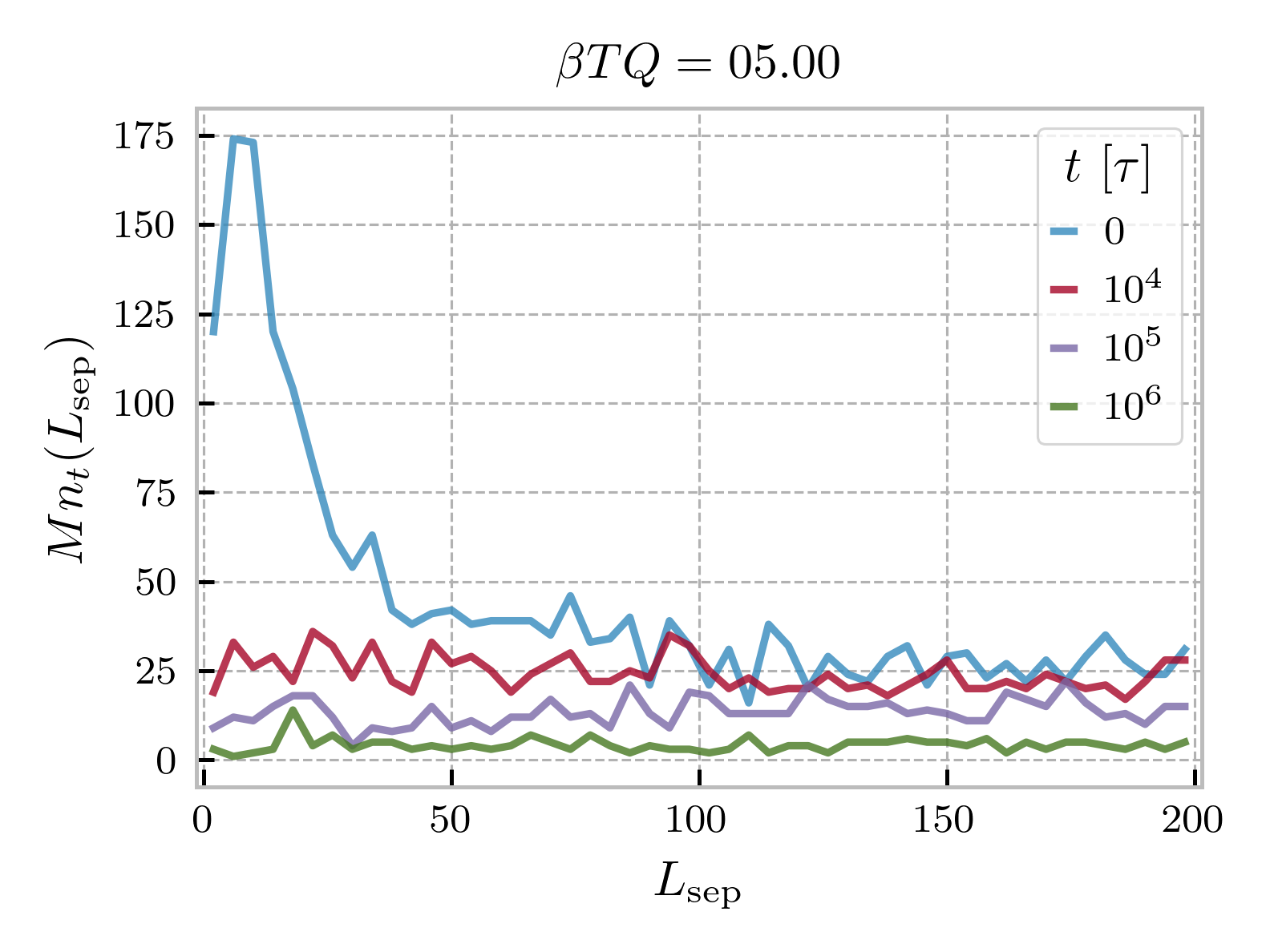}
    \includegraphics[scale = 0.90]{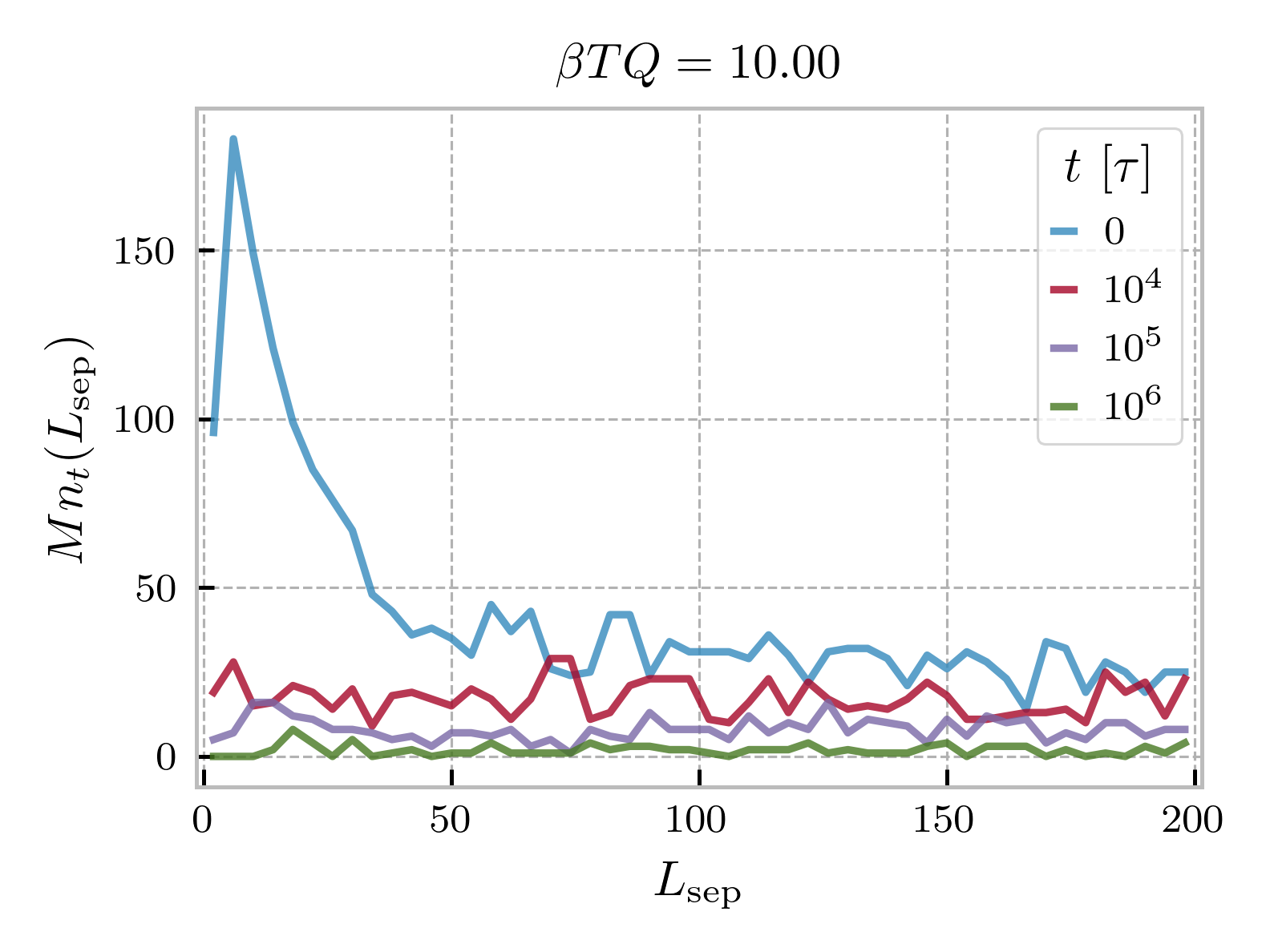}
    \caption{
    Total number of threadings of separation length $L_\mathrm{sep}$ observed in the system of $M = 200$ rings plotted for three different times -- during the stage of rapid supercoiling; during the transition from rapid supercoiling to the slow tightening regime; during the regime of slow aging -- in the order of increasing time.
    Time $t = 0$ corresponds to the initial (equilibrium) configuration.
    Presented histograms are essentially vertical slices of the \reffig{esi:fig:results:lsep}.
    }
    \label{esi:fig:results:lsep:histogram}
\end{figure*}

%%%%%%%%%%%%%%%%%%%%%%%%%%%%%%%%%%%%%%%%%%%%%%%%%%
\clearpage
\subsection{Network Topology Analysis}
\label{esi:sec:results:network}
%%%%%%%%%%%%%%%%%%%%%%%%%%%%%%%%%%%%%%%%%%%%%%%%%%

\begin{figure*}[htbp]
    \centering
    \includegraphics[scale = 0.90]{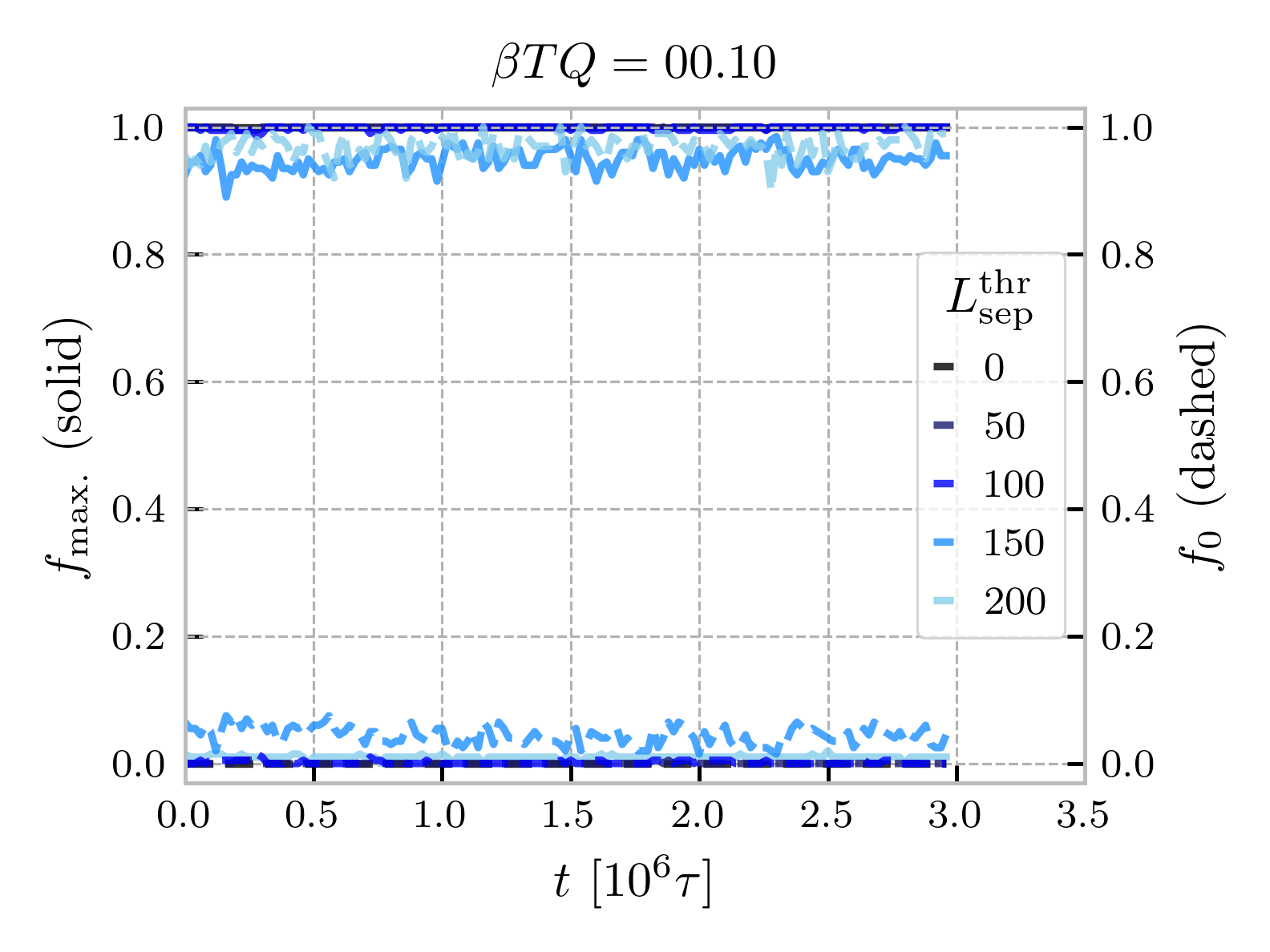}
    \includegraphics[scale = 0.90]{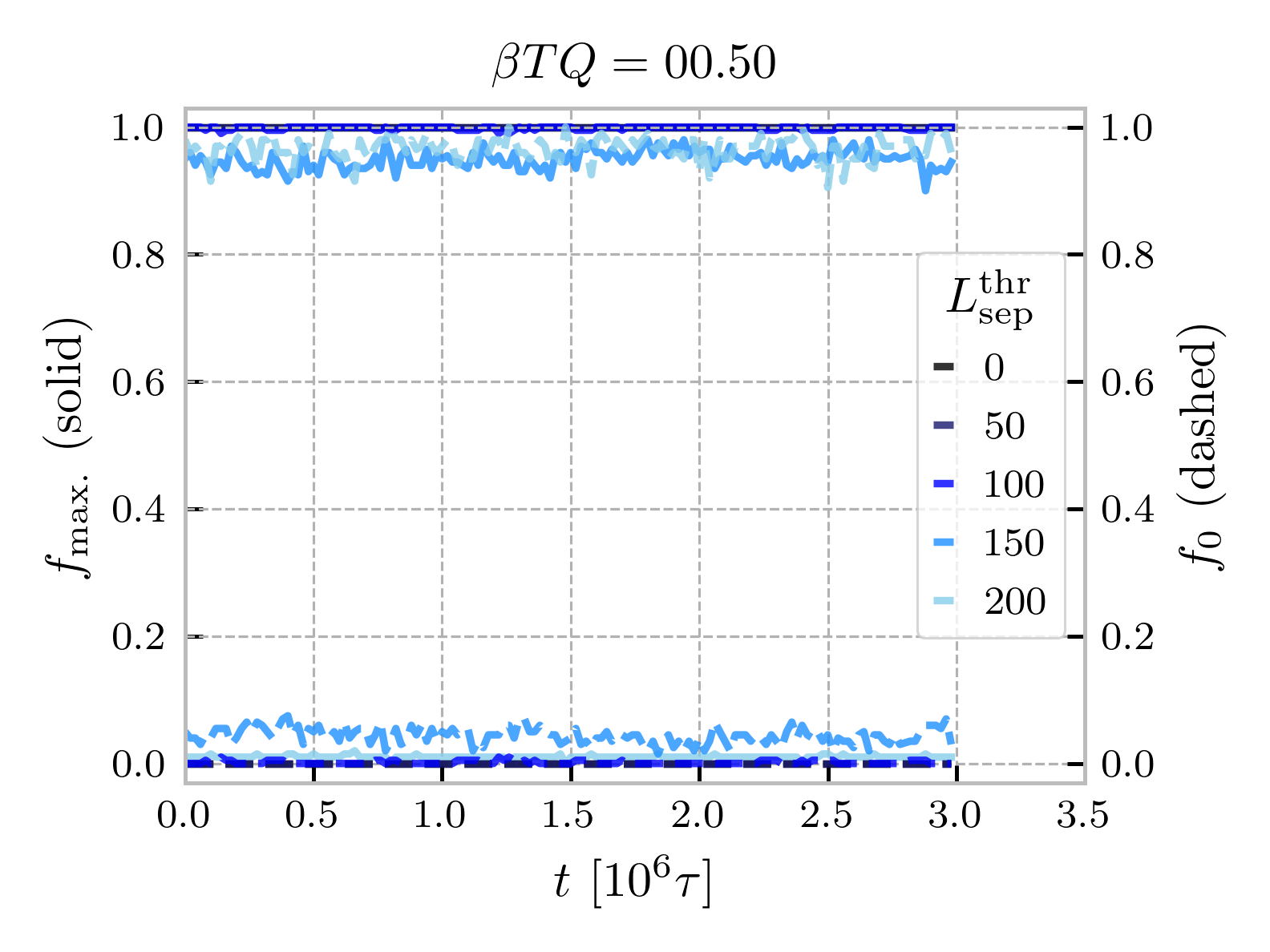}
    \includegraphics[scale = 0.90]{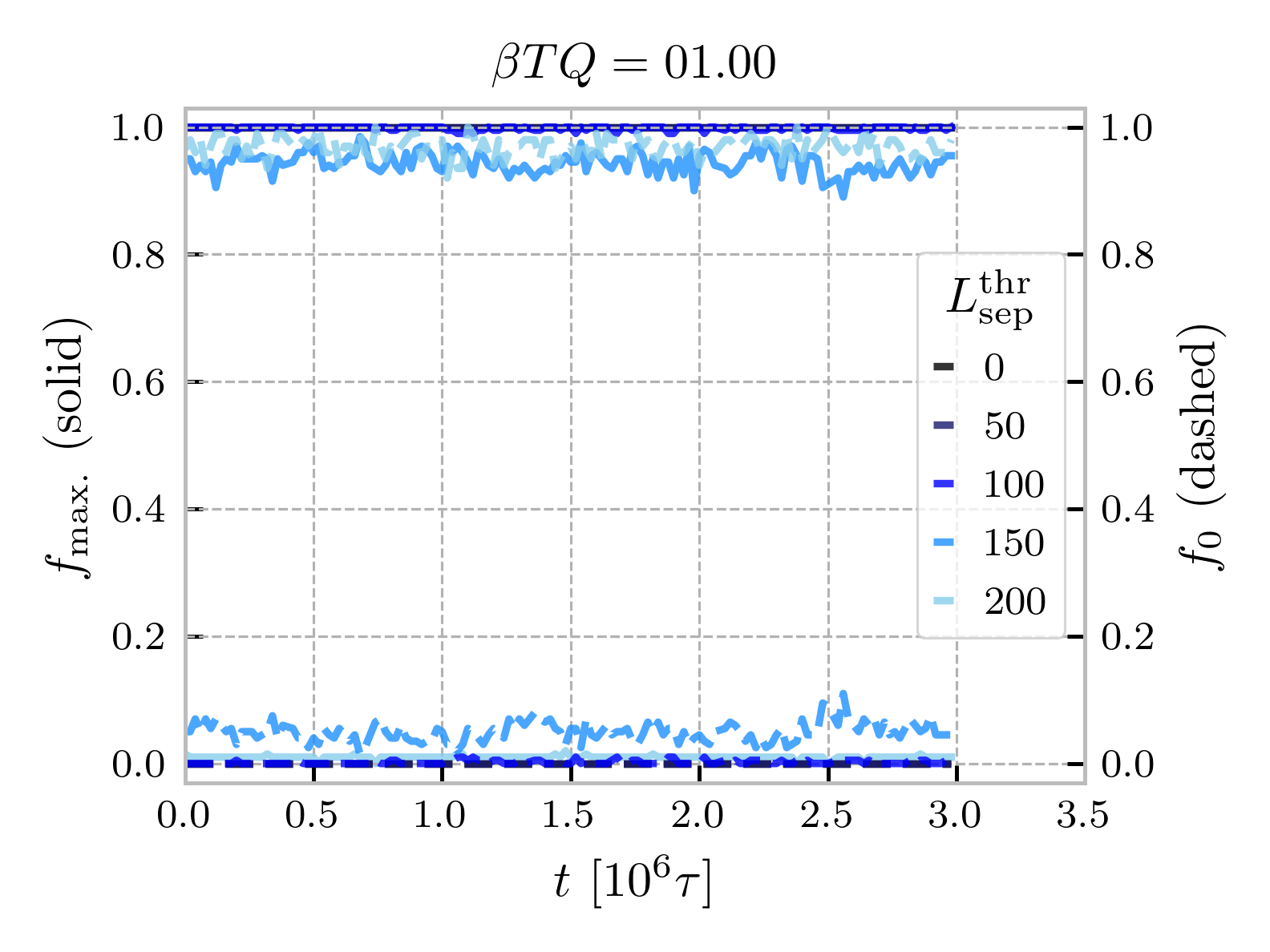}
    \includegraphics[scale = 0.90]{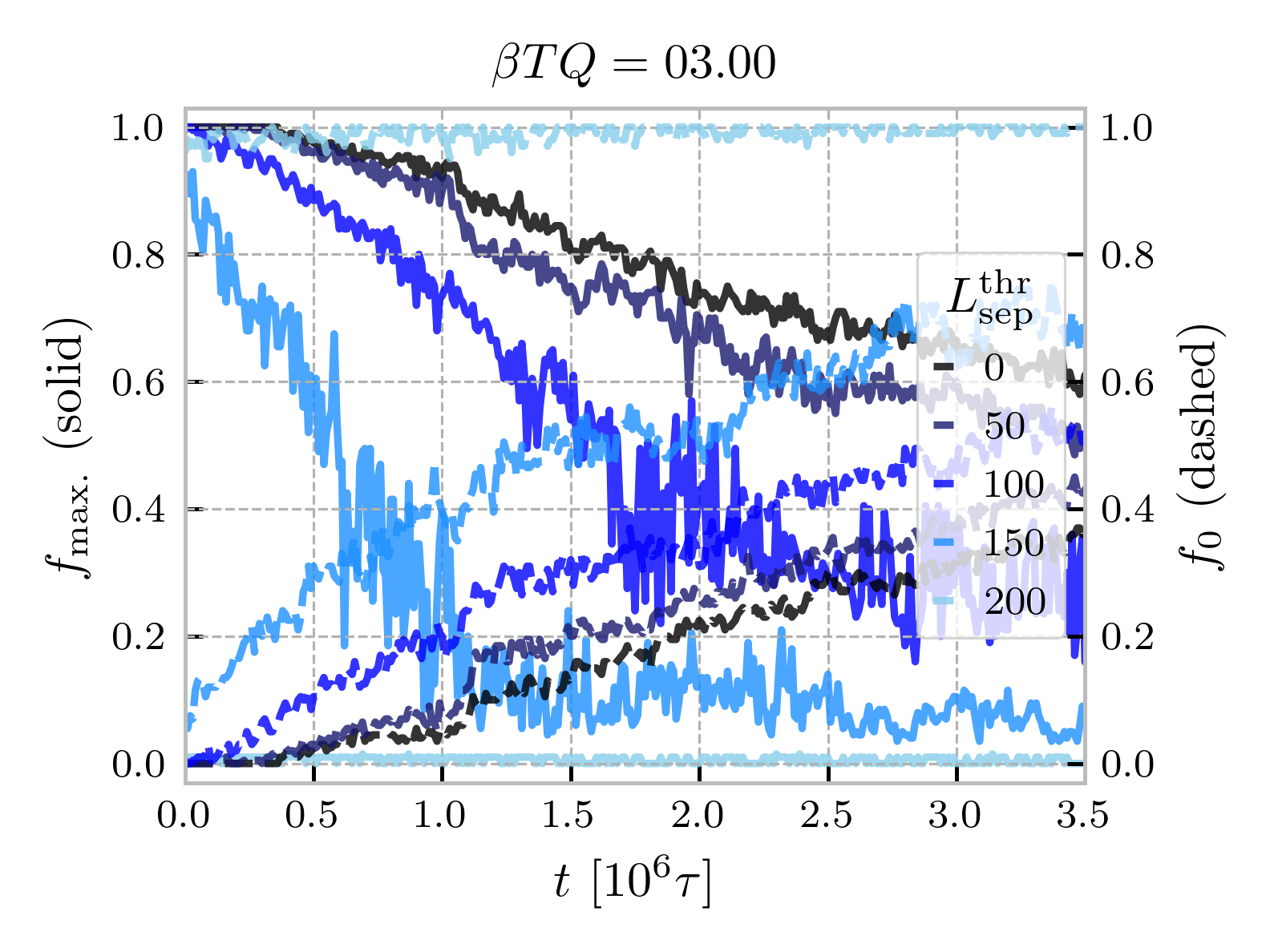}
    \includegraphics[scale = 0.90]{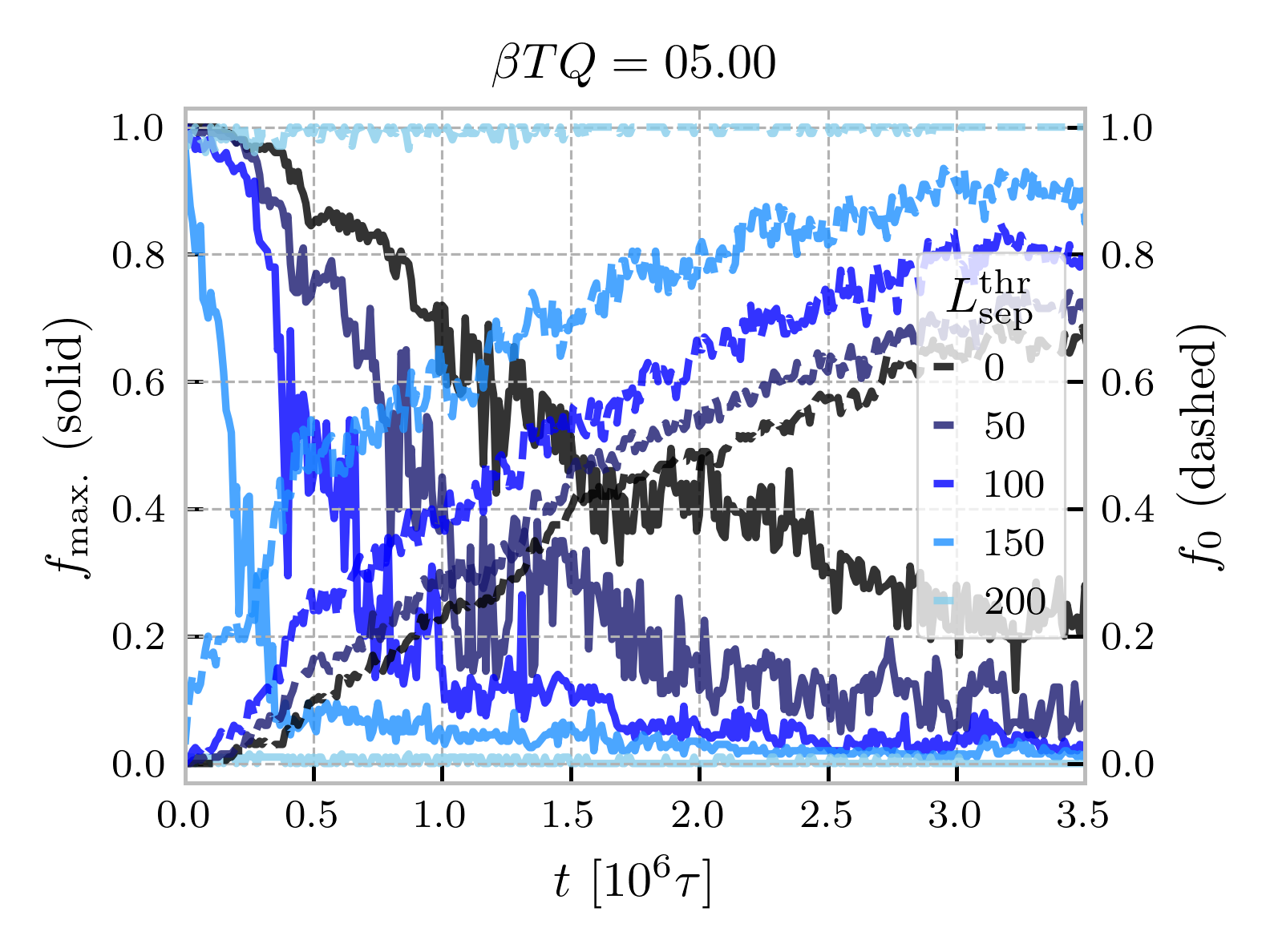}
    \includegraphics[scale = 0.90]{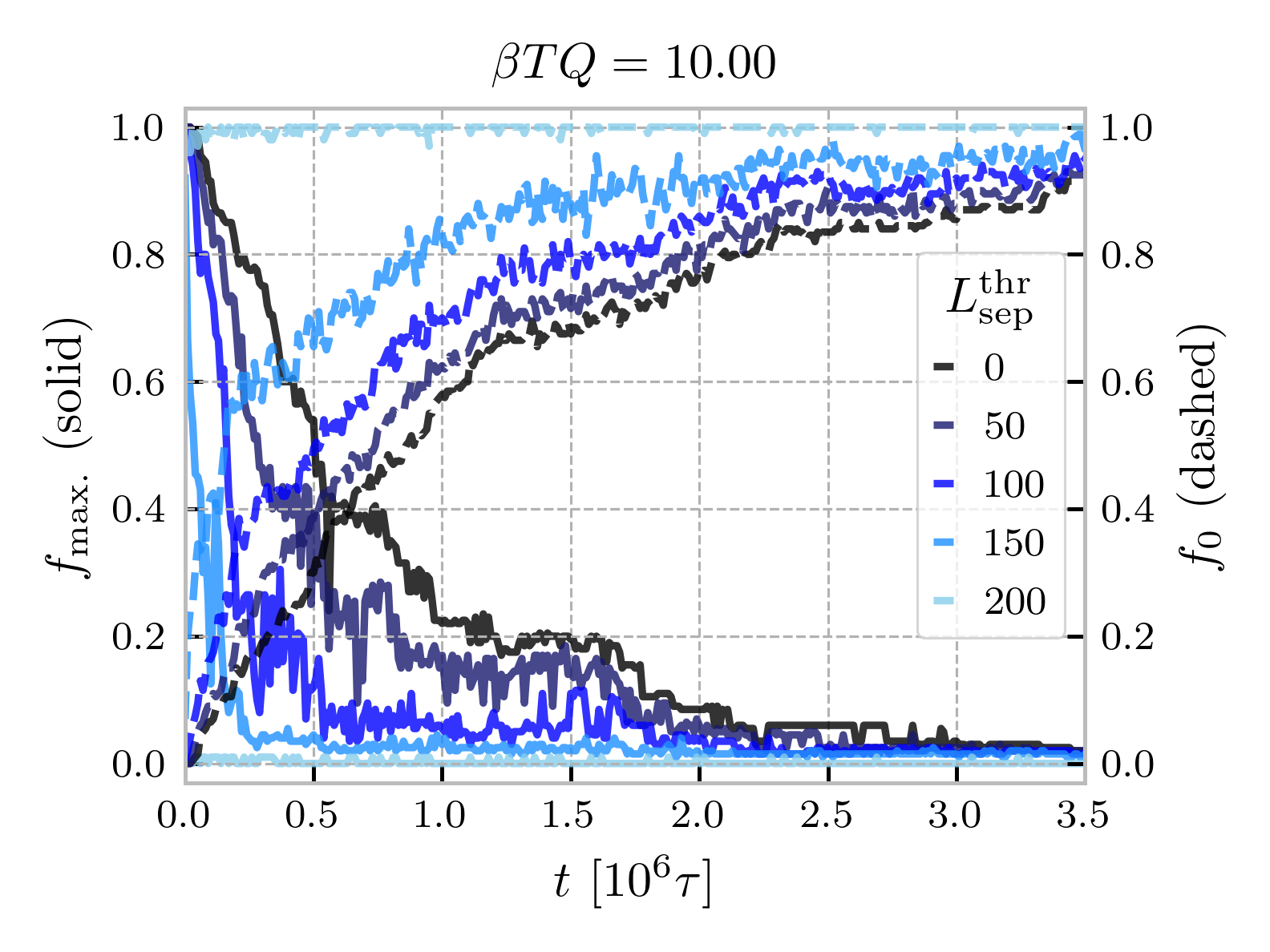}
    \caption{
    Time series of network properties for different torques and different threshold values $L^\mathrm{thr}_{sep}$ used in the definition of the network following the \refsec{esi:sec:model:threading}.
    }
    \label{esi:fig:results:threshold}
\end{figure*}

%%%%%%%%%%%%%%%%%%%%%%%%%%%%%%%%%%%%%%%%%%%%%%%%%%
\clearpage
\subsection{Dynamics of Deadlocked Rings}
\label{esi:sec:results:dynamics}
%%%%%%%%%%%%%%%%%%%%%%%%%%%%%%%%%%%%%%%%%%%%%%%%%%

\begin{figure*}[htbp]
    \centering
    \includegraphics[scale = 0.90]{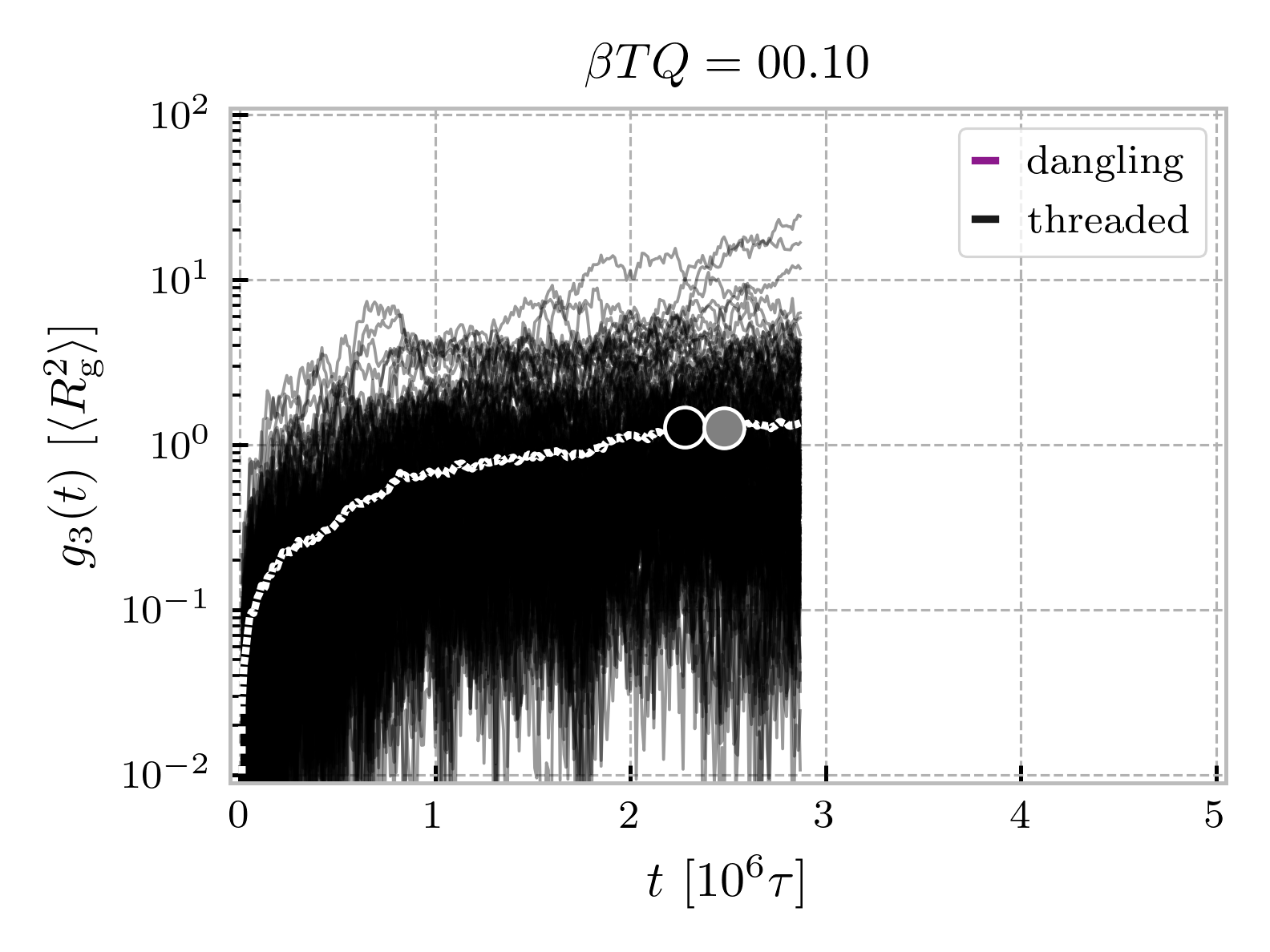}
    \includegraphics[scale = 0.90]{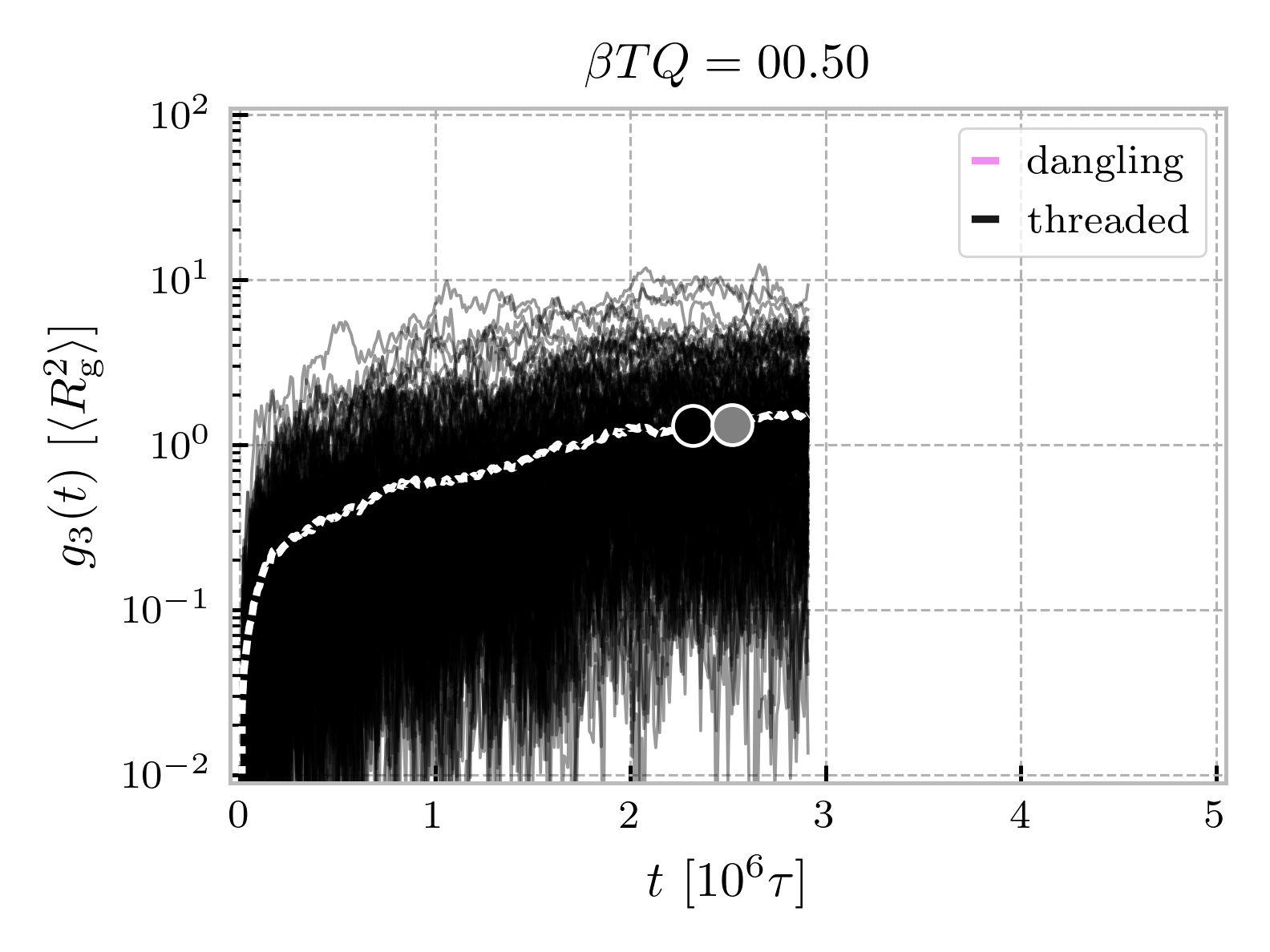}
    \includegraphics[scale = 0.90]{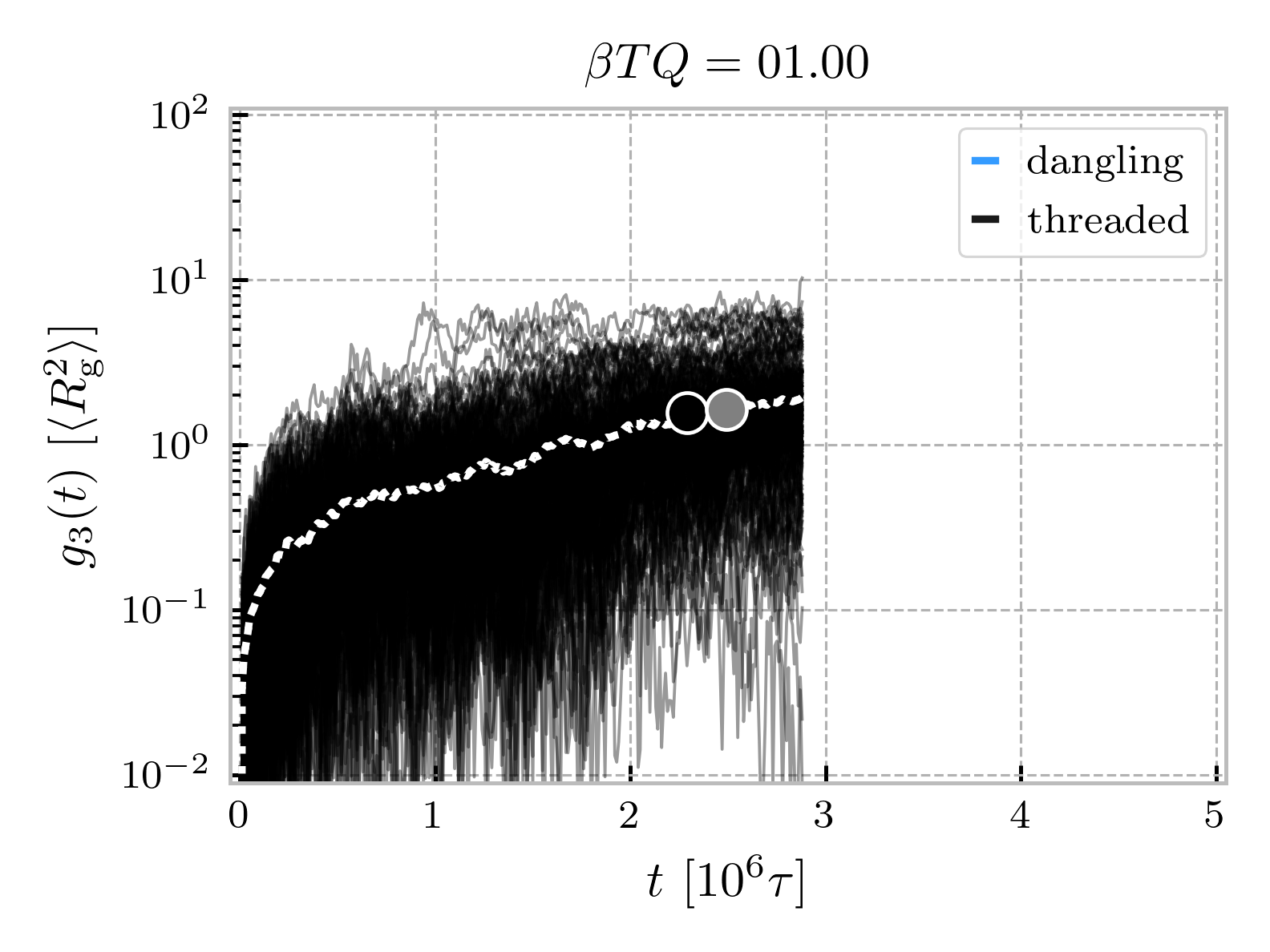}
    \includegraphics[scale = 0.90]{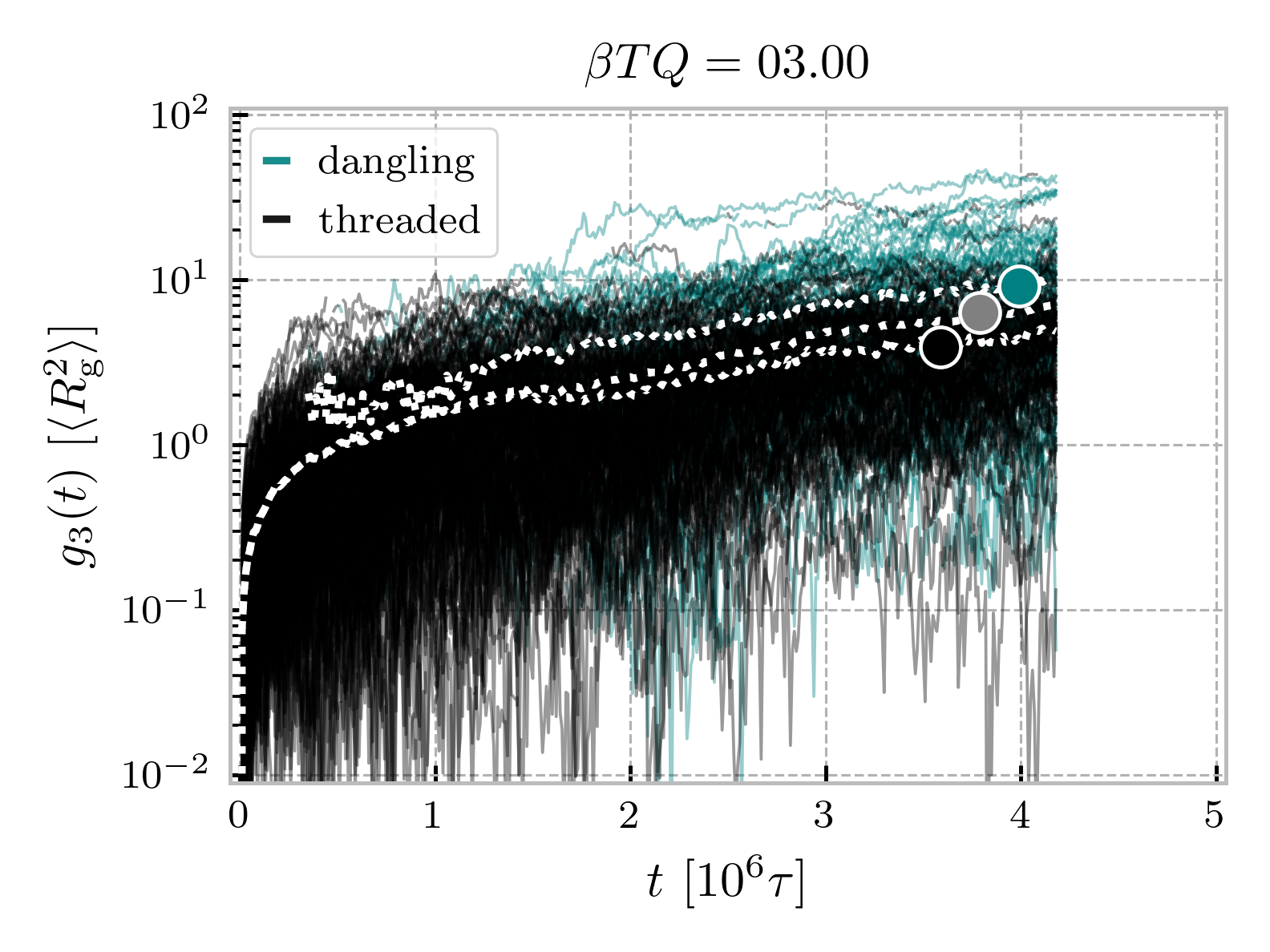}
    \includegraphics[scale = 0.90]{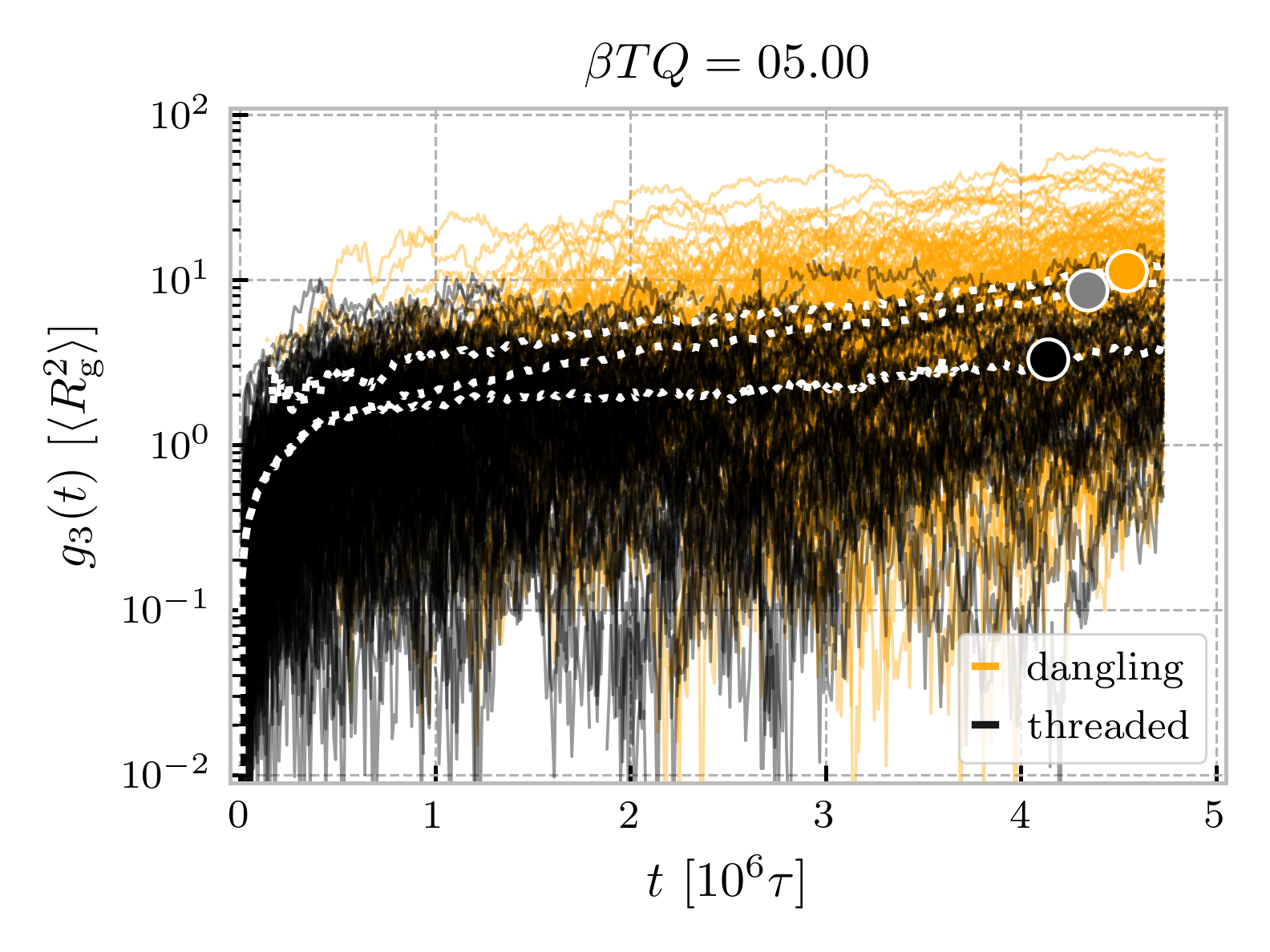}
    \includegraphics[scale = 0.90]{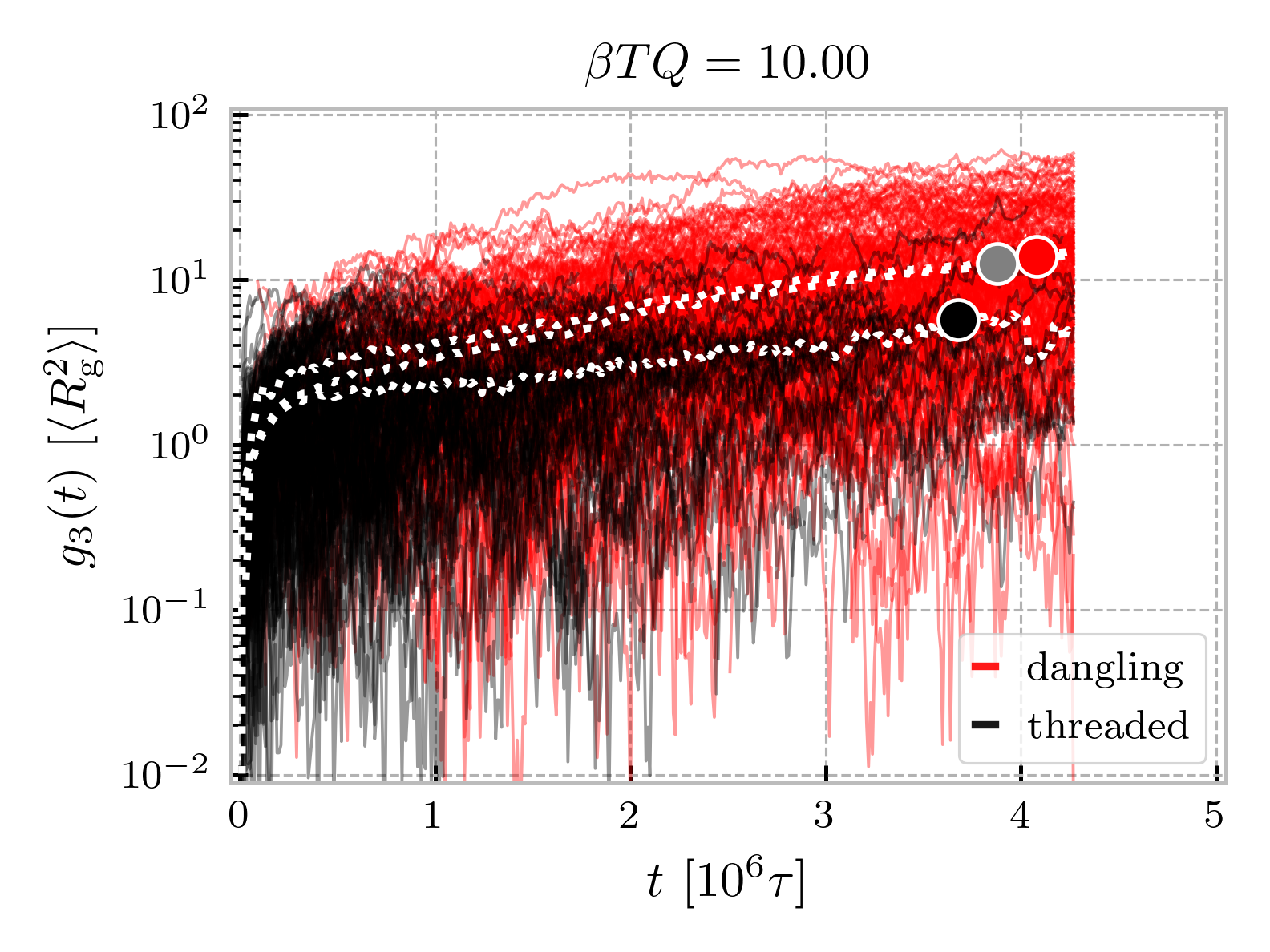}
    \caption{
    The mean square displacement of all individual rings, with the point at time $t$ colored black if the ring is threaded or color if is dangling, plotted for different torques.
    White dotted lines show means over three subpopulations of rings -- over all dangling rings, over all threaded rings and over all of the rings, marked with color, %\RR{color} 
    black and gray circular marker respectively. 
    In some of the cases, we have no dangling rings.
    }
    \label{esi:fig:results:deadlock}
\end{figure*}

\begin{figure*}[htbp]
    \centering
    \includegraphics[scale = 0.90]{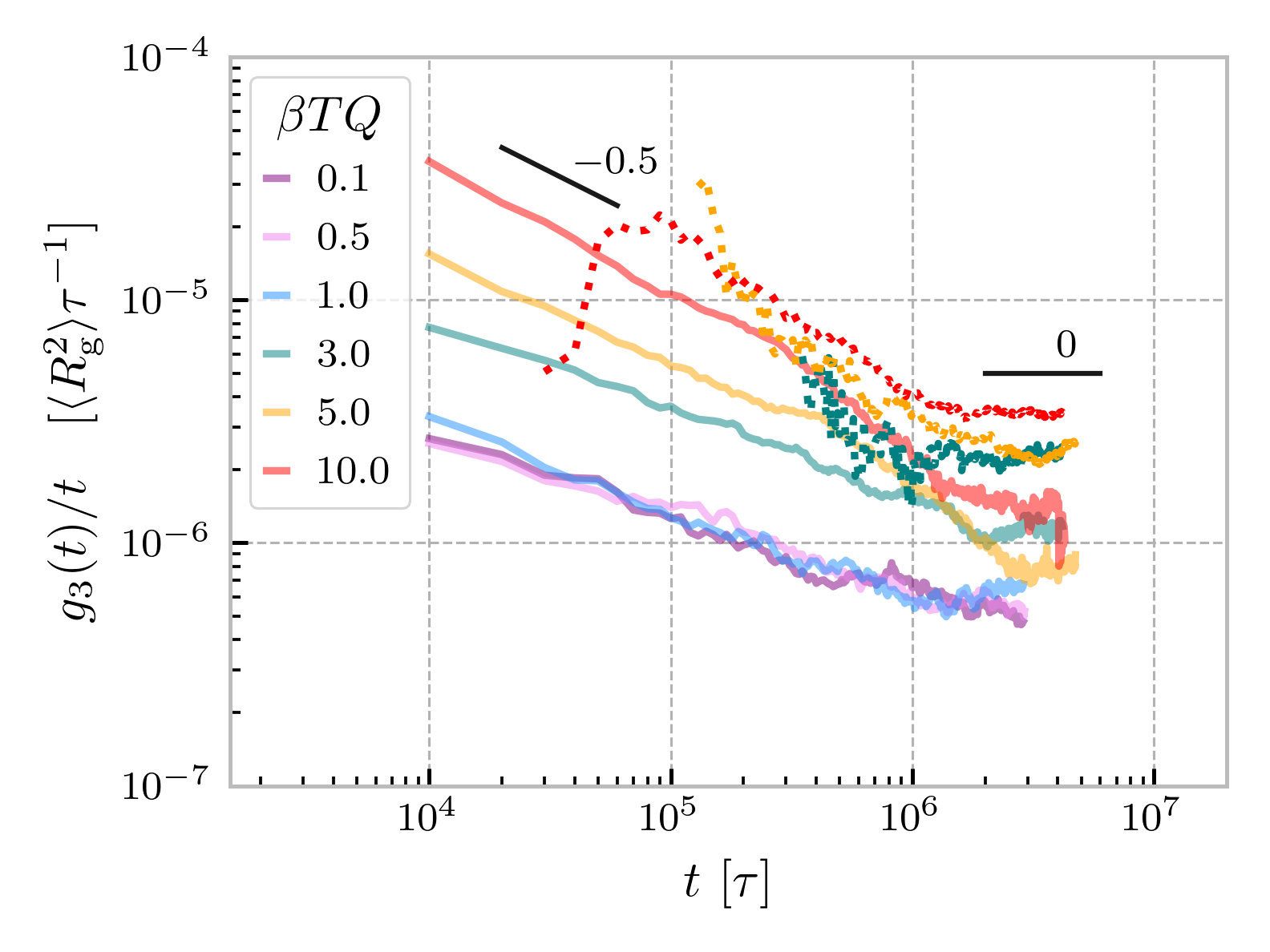}
    \includegraphics[scale = 0.90]{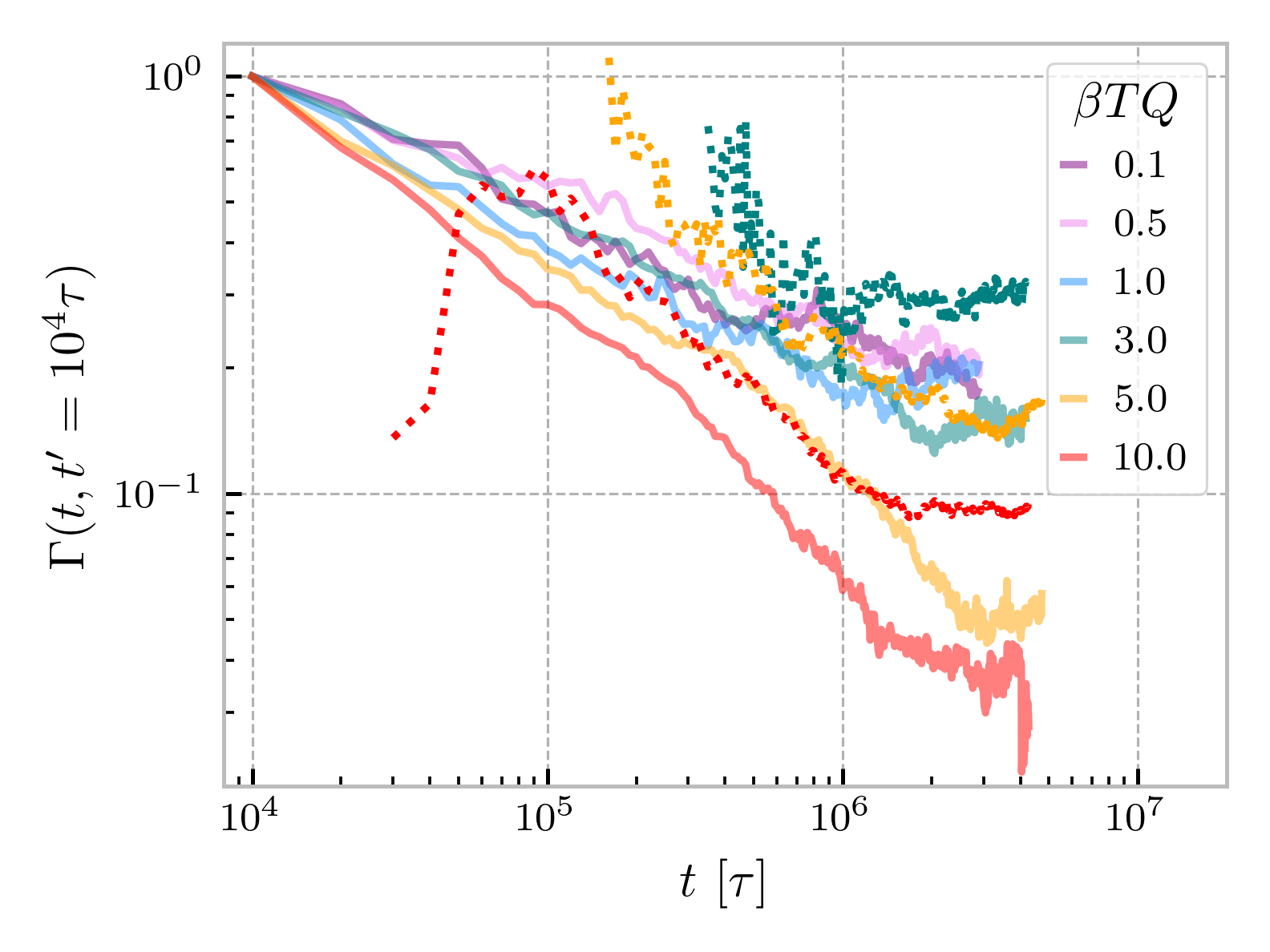} 
    \caption{
    The mean square displacement divided by time (left) and relaxation functions (right) plotted for different torques.
    Solid lines denote the averages over the threaded rings only, while the dotted lines are averaged over the dangling rings only.
    Note that the dangling rings emerge only for the three largest values of $\beta TQ$.
    Since dangling rings emerge only in the long time limit, the normalization factor of the relaxation function $\Gamma(t, t')$ is strictly defined only for the threaded rings.
    To plot the functions for the dangling rings as well, we just use the normalization of the threaded ones at the same $TQ$, hence the dotted lines in $\Gamma(t,t')$ do not start at one.
    }
    \label{esi:fig:results:dynamics2}
\end{figure*}

\begin{figure*}[htbp]
    \centering
    \includegraphics[scale = 0.90]{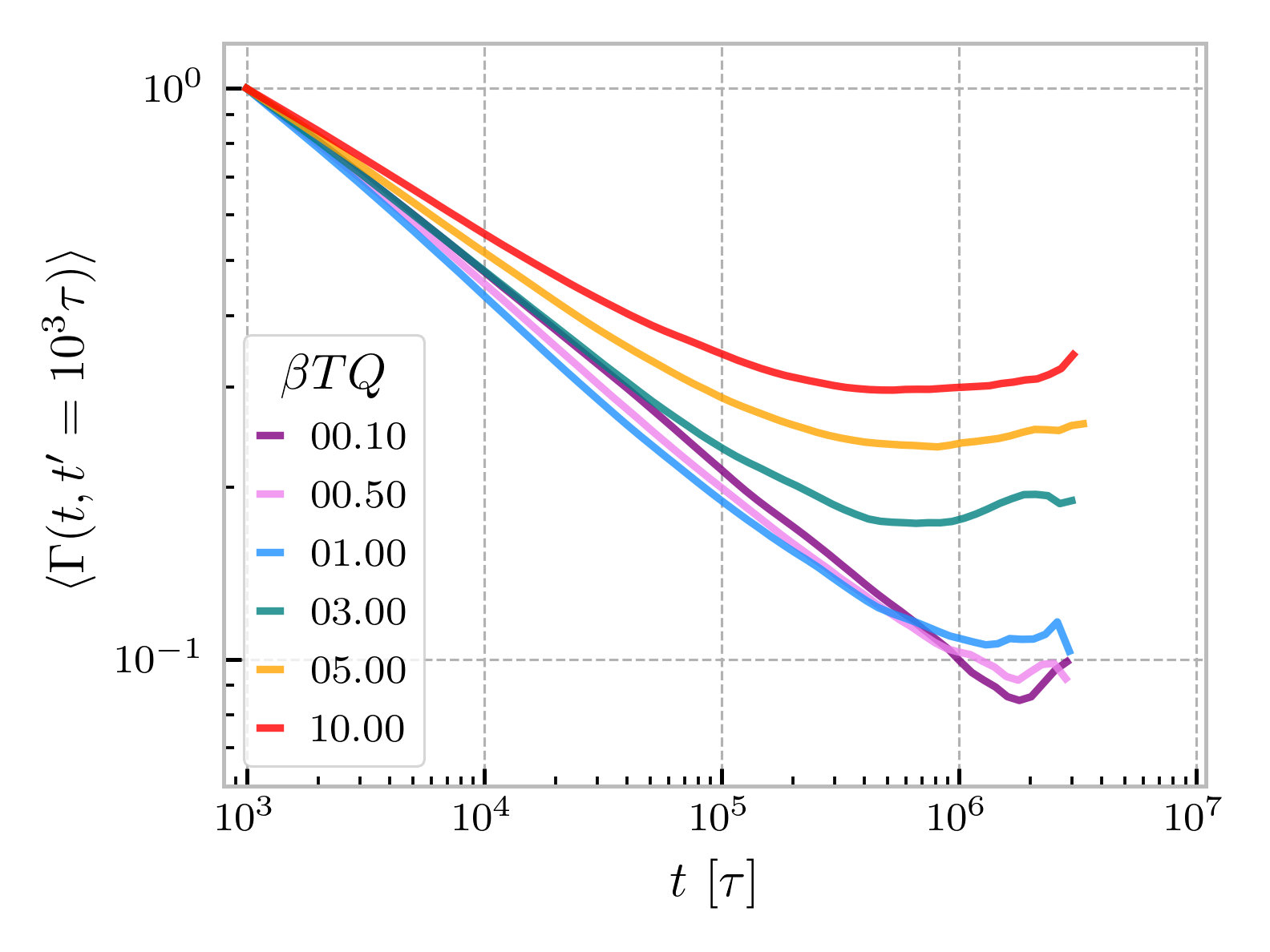}
    \caption{
    The averaged relaxation function $\Gamma$ plotted for different torques in the limit of long times.
    In this case, we redefine the time zero in the corresponding $g_3(t)$ as $t_0 = 10^6\tau$ as opposed to $t_0 = 0$ in all other $\Gamma$ plots. This change means that the function presented here probes only the relaxation in the regime where the supercoiling saturates (\reffig{fig:1}\textbf{d}), long after the initial rapid supercoiling period. For every lag time $t$, we present $\langle \Gamma \rangle$ which is $\Gamma$ averaged over all possible origins after $t_0 = 10^6 \tau$. This relaxation function is hence equivalent to equilibrium relaxation of rings with the supercoiling degree corresponding to its saturated value at the given $\beta TQ$.
    }
    \label{esi:fig:results:dynamics3}
\end{figure*}